\documentclass[12pt]{article}
\usepackage{jheppub}
\pdfoutput=1
\usepackage{epstopdf}

\usepackage{jheppub}
\usepackage{mathrsfs}
\usepackage{psfrag}
\usepackage{color}
\usepackage{hyperref}

\usepackage{slashed}
\usepackage{feynmp-auto}
\usepackage{simplewick}
\usepackage{cancel}

\usepackage[table]{xcolor}

\usepackage{amsmath,bbm,array,amsfonts,graphicx,wrapfig,arydshln,lscape,float,multirow,longtable,rotating,makecell}
\usepackage{url}

\newcommand{\smallminus}{{\rm\rule[2.4pt]{6pt}{0.65pt}}}

\newcommand{\ab}[1]{\langle #1 \rangle}
\newcommand{\sqb}[1]{[ #1 ]}
\newcommand{\sab}[1]{s_{#1}}

\def\dlog{d\log}
\def\A{\mathcal{A}}

\def\N{\mathcal{N}}
\def\M{\mathcal{M}}
\def\I{\mathcal{I}}
\def\J{\mathcal{J}}

\def\R{\mathcal{R}}

\def\MHVbar{\overline{\text{MHV}}}
\def\eps{\epsilon}
\def\sm{\smallminus}

\newcommand{\tw}[1]{\widetilde{#1}}
\newcommand{\lam}[1]{\lambda_{#1} }
\newcommand{\lamt}[1]{\widetilde \lambda_{#1} }

\def\MHVbar{\overline{\text{MHV}}}

\definecolor{airforceblue}{rgb}{0.36, 0.54, 0.66}
\definecolor{bananayellow}{rgb}{1.0, 0.88, 0.21}
\definecolor{bittersweet}{rgb}{1.0, 0.44, 0.37}
\definecolor{blue(ncs)}{rgb}{0.0, 0.53, 0.74}
\definecolor{bole}{rgb}{0.47, 0.27, 0.23}
\definecolor{brass}{rgb}{0.71, 0.65, 0.26}
\definecolor{bronze}{rgb}{0.8, 0.5, 0.2}
\definecolor{brgreen}{rgb}{0.0, 0.26, 0.15}
\definecolor{burgundy}{rgb}{0.5, 0.0, 0.13}
\definecolor{cherry}{rgb}{1.0, 0.72, 0.77}
\definecolor{cocao}{rgb}{0.82, 0.41, 0.12}
\definecolor{citrine}{rgb}{0.99, 0.82, 0.07}

\newcommand{\tblue}[1]{\textcolor{blue}{#1}}

\newcommand{\twhite}[1]{\textcolor{white}{#1}}

\title{UV cancelations in gravity loop integrands}

\author[1]{Enrico Herrmann}
\author[2]{Jaroslav Trnka}

\affiliation[1]{ SLAC National Accelerator Laboratory, Stanford University, Stanford, CA 94039, USA}
\affiliation[2]{Center for Quantum Mathematics and Physics (QMAP),\\ 
Department of Physics, University of California, Davis, CA 95616, USA}

\emailAdd{eh10@stanford.edu, trnka@ucdavis.edu}

\preprint{\begin{flushright} \end{flushright}}

\abstract{In this work we explore the properties of four-dimensional gravity integrands at large loop momenta. This analysis can not be done directly for the full off-shell integrand but only becomes well-defined on cuts that allow us to unambiguously specify labels for the loop variables. The ultraviolet region of scattering amplitudes originates from poles at infinity of the loop integrands and we show that in gravity these integcrands conceal a number of surprising features. In particular, certain poles at infinity are absent which requires a conspiracy between individual Feynman integrals contributing to the amplitude. We suspect that this non-trivial behavior is a consequence of yet-to-be found symmetry or hidden property of gravity amplitudes. We discuss mainly amplitudes in $\N=8$ supergravity but most of the statements are valid for pure gravity as well.}

\begin{document}

\maketitle

\section{Introduction}
\label{sec:intro}

The precise nature of the ultraviolet~(UV)~structure of gravity theories has been a longstanding area of research starting with the classic result about the one-loop UV finiteness of pure gravity~\cite{tHooft:1974toh} due to the topological nature of the Gauss-Bonnet term in four spacetime dimensions. Increasing the loop order, explicit computations have confirmed that pure gravity indeed diverges at two loops~\cite{Goroff:1985sz, Goroff:1985th, vandeVen:1991gw, Bern:2015xsa} consistent with power-counting expectations. Adding supersymmetry softens the UV-behavior of scattering amplitudes owing to the usual cancellations between fermions and bosons so that the loop order at which amplitudes are expected to diverge increases. For the maximally supersymmetric theory in four dimensions, $\N=8$ supergravity, the critical loop order is $L=7$. Whether the $\N=8$ amplitudes indeed diverge at this loop order is still an open question. A number of attempts have been made to predict or rule out certain divergences, based on supersymmetry (see e.g.~\cite{Bjornsson:2010wm}) and duality symmetry (see e.g.~\cite{Kallosh:2008rr,Beisert:2010jx}) only to name a few. At the moment only the direct amplitudes calculation unambiguously resolve the problem. The standard procedure consists of the construction of the full integrand and the expansion around the UV-region which reduces the problem to the computation of certain vacuum integrals. Performing the direct evaluation at seven-loop order for the four point amplitude is currently out of reach, but recently the five-loop result was obtained after a heroic computation and making use of a number of clever tricks \cite{Bern:2018jmv}. The amplitude diverges in the expected critical dimension $D=24/5$ consistent with a $D^8\R^4$ counterterm which gives an indirect hint that the seven-loop and higher amplitudes would indeed diverge in four dimensions barring further miracles. On the other hand, there are results for certain $\N<8$ supergravity amplitudes that show a surprising UV behavior due to {\it enhanced cancelations}, defined as cancelations of UV divergences between individual Feynman integrals that render the amplitude finite in a certain dimension \cite{Bern:2007xj,Bern:2012cd,Bern:2017lpv}. In light of these comments, we can claim that the UV structure of gravity amplitudes is still deserves further study.


In parallel, there has been great progress in the effort to understand amplitudes as differential forms on various geometric spaces. This was first seen in the context of the planar $\N=4$ sYM theory where tree-level amplitudes and the loop integrands correspond to logarithmic forms on the Amplituhedron space \cite{Arkani-Hamed:2013jha,Arkani-Hamed:2013kca}, more recently similar structures have been seen for tree-level amplitudes in $\phi^3$ theory via the Associahedron \cite{Arkani-Hamed:2017mur,delaCruz:2017zqr,Frost:2018djd,He:2018pue} and in cosmological correlation functions \cite{Arkani-Hamed:2015bza, Arkani-Hamed:2017fdk}. The important input in the Amplituhedron construction is the uniqueness of tree-level amplitudes and loop integrands -- they are fully fixed by a list of certain homogenous conditions: logarithmic singularities, no poles at infinity, and absence of unphysical singularities. These conditions are then reformulated geometrically as positivity constraints defining the Amplituhedron geometry and an associated logarithmic volume form which gives rise to the scattering amplitude. The natural question is if the geometry can underly the scattering amplitudes in other quantum field theories. The problem naturally splits into two steps: find the constraints which fix amplitudes uniquely in a given theory, and then search for the geometry and corresponding ``amplitude forms". The recent topological reformulation of the Amplituhedron shows that these forms can be defined directly in the kinematical space without the need to introduce any auxiliary variables \cite{Arkani-Hamed:2017vfh}. On the other hand, it was observed in \cite{Arkani-Hamed:2014via, Bern:2014kca,Bern:2015ple} that full $\N=4$ sYM amplitudes share the same properties as their planar counterparts: logarithmic singularities, no poles at infinity, and demanding the absence of unphysical singularities were sufficient to fix the amplitude uniquely. The loss of planarity does not allow yet to formulate the associated geometry because of the absence of right kinematical variables. The search for the right variables and the unique non-planar integrand is an ongoing process.


In this paper, we focus on gravity amplitudes and our goal is to find non-trivial properties of the loop integrands in the UV which can later be used as defining conditions in a possible geometric formulation. Instead of taking the full non-planar integrands which suffer from labeling ambiguities we can study their \emph{cuts}. According to the principles of unitarity, cuts of loop amplitudes are given by the product of tree-level amplitudes. Therefore, cuts are well-defined functions whose analytic properties can be analyzed without any difficulties. Iterative use of the cut procedure leads to elementary objects called on-shell diagrams \cite{ArkaniHamed:2012nw} which were also studied in the gravity context \cite{Herrmann:2016qea,Heslop:2016plj}. In \cite{Herrmann:2016qea} we found that gravity on-shell diagrams exhibit some surprising features in the collinear regions and as a application we found a direct link to the mild IR singularities in gravity amplitudes. If we write the loop amplitude as a sum of basis integrals these collinear conditions can not be made manifest term-by-term and require non-trivial cancelations between individual pieces in the sum. Unlike in $\N=4$ sYM theory the gravity on-shell diagrams have poles at infinity as a consequence of gravity power-counting and they point to the non-trivial UV singularities. The standard maximal-cut argument relates these poles at infinity to the loop-momentum dependence in the numerators of the associated Feynman integrals which become UV divergent starting at 7-loops in $\N=8$ supergravity and at lower loops for $\N<8$ theories. Naively, this implies that while the gravity integrands are very constrained in the IR through the collinear conditions they are unbounded in the UV because of the presence of poles at infinity, losing any hope to fix them uniquely using homogeneous constraints. We show that this is not the case, and there is some very surprising behavior at infinity within the gravity cuts which requires massive cancelations between individual Feynman integrals contributing to the amplitude. While there is not a direct link to the UV divergence of the amplitude there seems to be a very non-trivial structure associated with the UV region which is not a consequence of standard symmetries. 

While we discuss mainly $\N=8$ supergravity amplitudes all statements about the cancelations are also true for $\N<8$ supergravity including pure GR. The only difference is the overall scaling at infinity but the cancelations are still there, and our observations seem to be general features of four-dimensional gravity amplitudes. The practical advantage of working in $\N=8$ supergravity is the use of on-shell superspace. More importantly, we make a conjecture that the improved UV behavior on cuts together with certain IR conditions are enough to fully fix $\N=8$ amplitudes. If true this would be certainly special for $\N=8$ as for lower $\N$ there will be additional UV poles. 


The paper is organized as follows: In the end of this section we review some basic material about loop integrands and cuts. Section~\ref{sec:uv_poles_at_infinity} is the main result of this paper where we discuss the absence of certain poles at infinity. In section~\ref{sec:ir_to_uv} we review some IR properties of gravity amplitudes and address the ultimate goal of this program of fixing gravity integrands uniquely using a combination of IR and UV constraints as a prelude for a possible geometric interpretation. We close with final remarks in section~\ref{sec:conclusions}.

\subsection{Loop integrands}
\label{subsec:loop_integrands}

We usually refer to the \emph{loop integrand} as a rational, gauge invariant function written in terms of a sum of Feynman integrals, which still have to be integrated over the loop momenta $\ell_i$ to get the final \emph{amplitude}. In general, the loop integrand as a \emph{single} function is not well-defined and what we really mean is the collection of integrands $\I_k$ which correspond to individual Feynman integrals,

\begin{equation}
{\cal A} = \sum_k \int d^4\ell_1\,d^4\ell_2\dots d^4\ell_L\,\I_k\,. 
\label{int1}
\end{equation}

It seems suggestive to exchange the summation and the integration symbol in order to define the loop integrand $\I$ as

\begin{equation}
{\cal A} =\int d^4\ell_1\,d^4\ell_2\dots d^4\ell_L\, {\cal I}\qquad \mbox{, where} \qquad {\cal I} = \sum_k {\cal I}_k\,.
\end{equation}

Typically, $\I$ is not uniquely defined because there is no global invariant meaning of the loop momenta $\ell_j$ - different choices for $\ell_j$ in Feynman integrals give us a different integrand function $\I$
\begin{align}
\begin{split}
& \hspace{1.1cm}
\raisebox{-55pt}{\includegraphics[scale=.6]{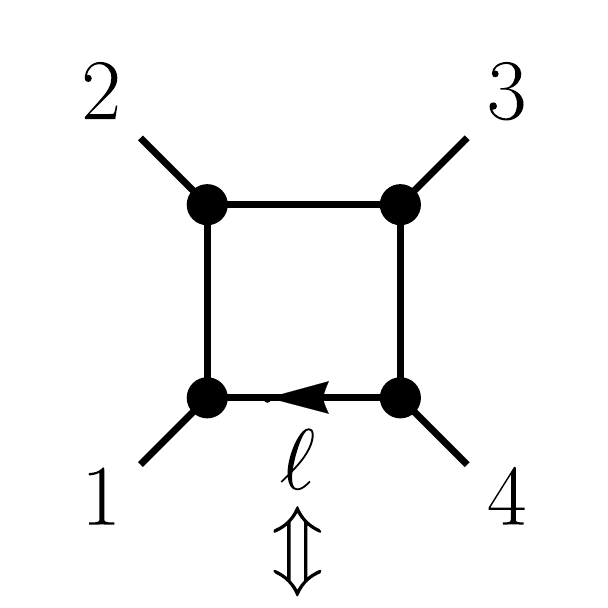}}
\hspace{3.1cm}
\raisebox{-55pt}{\includegraphics[scale=.6]{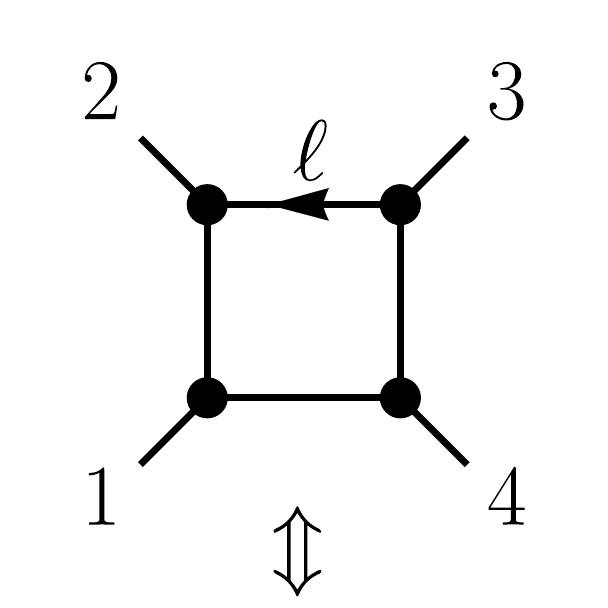}}  	\,.		\\
& \frac{d^4\ell}{\ell^2 (\ell-p_1)^2(\ell-p_1-p_2)^2(\ell+p_4)^2}
\qquad
 \frac{d^4\ell}{\ell^2 (\ell-p_2)^2(\ell-p_1-p_2)^2(\ell+p_3)^2}
\end{split}
\end{align}

\vspace{0.3cm}

However, in planar theories there exists a preferred choice of loop variables given by dual coordinates: instead of defining momenta as the flow along edges $p_k$ (external) and $\ell_j$ (loop) we define dual momenta associated with the faces $x_k$ and $y_j$ of the graph
\begin{align}
\raisebox{-45pt}{\includegraphics[scale=.6]{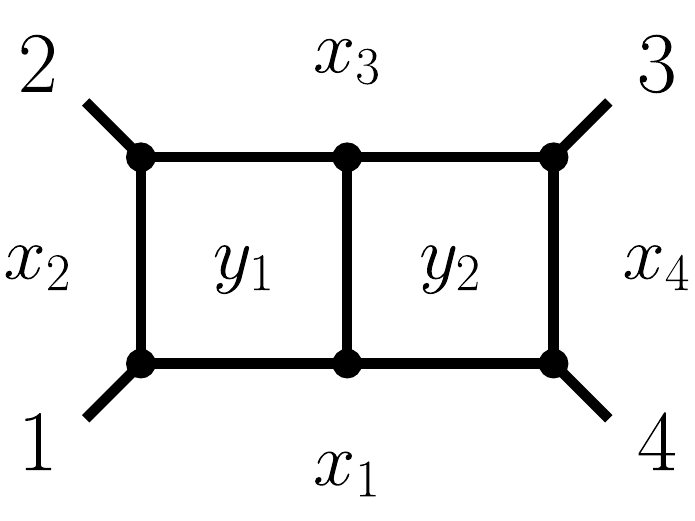}}\,.
\end{align}
Then the amplitude can be written as 
\begin{equation}
\A =\int d^4y_1\,d^4y_2\dots d^4y_L\,\I \,,
\end{equation}
where the integrand ${\cal I}$ is now uniquely defined, and one does not have to refer to the sum of Feynman integrals (\ref{int1}) anymore. This allowed to find BCFW recursion relations for the loop integrand in planar ${\cal N}=4$ sYM \cite{ArkaniHamed:2010kv} which were then reformulated in terms of on-shell diagrams \cite{ArkaniHamed:2012nw}. Naively even with good global coordinates the integrand is still not uniquely defined because we can add terms proportional to total derivatives 
\begin{equation}
\I \sim\I + \frac{\partial}{\partial \ell} \widetilde{\I}\,.
\end{equation}
This is true if we are interested in amplitudes, $\A$, directly (as the total derivatives integrate to zero). However, if we want to obtain the integrand $\I$ in the context of generalized unitarity as the function which satisfies all field theory cuts then no total derivatives can be added as they would spoil matching the cuts. In other words, any function which is a total derivative would change the value of the cuts which are already matched by $\I$ or introduce unphysical poles.

In contrast, for non-planar theories the above set of unique labels in terms of dual face variables is not available and we are forced to think about the integrand in the context of (\ref{int1}) as sum of individual Feynman integrals. Let us demonstrate this for the one-loop four-point amplitude in ${\cal N}=8$ supergravity first calculated by Brink, Green and Schwarz \cite{Green:1982sw} as low energy limit of string amplitudes. The amplitude can be written in terms of three scalar box integrals,
\begin{align}
\label{eq:one-loop-gravity-amp}
-i\M^{(1)}_4 = stu\ \M^{(0)}_4\Big[ I^{\text{box}}_4(s,t) +  I^{\text{box}}_4(u,t)  + I^{\text{box}}_4(s,u)\Big]\,,
\end{align}
where $\M^{(0)}_4$ is the tree-level amplitude and the individual Feynman integrals
\begin{align}
I^{\text{box}}_4(s,t) = \hskip -.4cm \raisebox{-33pt}{\includegraphics[scale=.4]{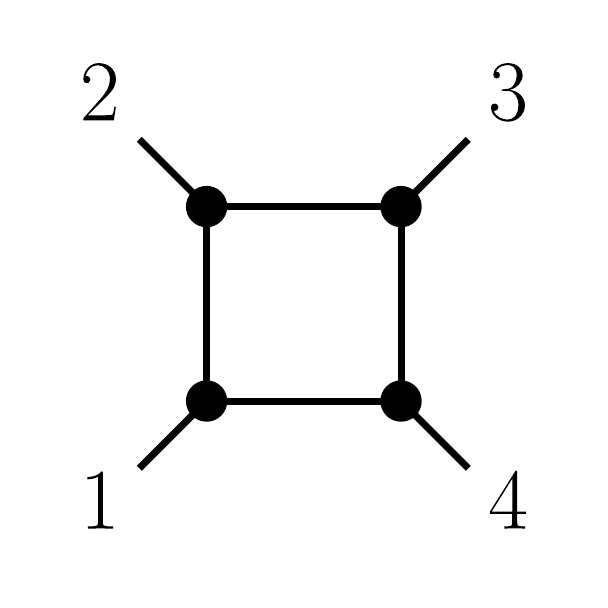}} \hskip -.7cm \,, \qquad
I^{\text{box}}_4(u,t) = \hskip -.4cm \raisebox{-33pt}{\includegraphics[scale=.4]{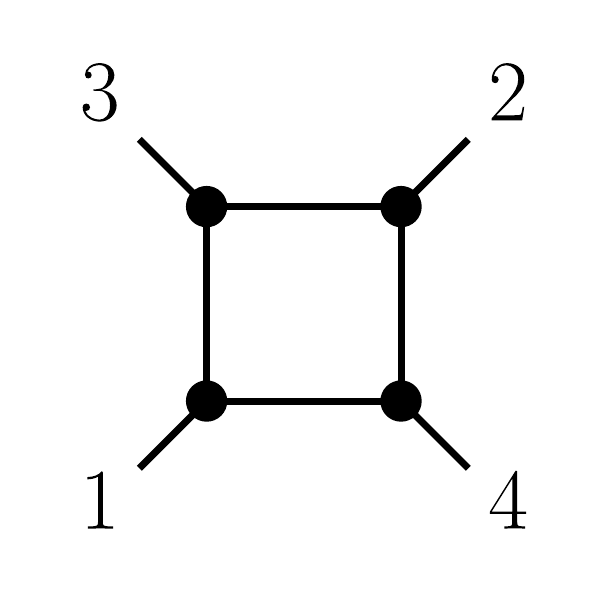}}\hskip -.7cm \,, \qquad
I^{\text{box}}_4(s,u) = \hskip -.4cm  \raisebox{-33pt}{\includegraphics[scale=.4]{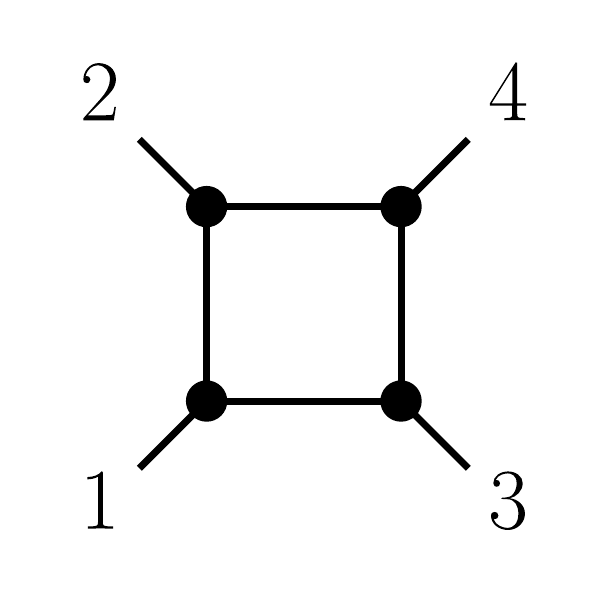}}\hskip -.7cm \,.
\end{align}
are defined with unit numerators. In (\ref{eq:one-loop-gravity-amp}), the usual $\sab{ij}$-dependent box-normalization is included in the totally crossing-symmetric $stu\ \M^{(0)}_4$ prefactor.

The question here is again how to choose the loop variables $\ell$ in individual diagrams. One natural instruction is to sum over all choices of labeling an edge by $\ell$. While this gives a unique function there is some intrinsic over-counting in this prescription. The other suggestion appeared in the context of Q-cuts \cite{Baadsgaard:2015twa} and ambitwistor strings \cite{Geyer:2015jch,Geyer:2016wjx} where the non-planar integrand was written using terms with linearized propagators. For the case above the integrand would be written as a sum of 24 terms of the form,
\begin{equation}
\label{eq:qcut_1loop}
{\cal I} = \sum_\sigma \frac{1}{\ell^2(\ell\cdot p_1)(\ell\cdot p_{12})(\ell\cdot p_4)}\,,
\end{equation}
where $p_{ij} = (p_i+p_j)$ and $\sigma$ labels the $4!=24$ permutations of external legs. While this gives a unique prescription there is a problem with spurious poles as ${\cal I}$ does not vanish on the residue $(\ell\cdot p_1)=0$. However, the representation (\ref{eq:qcut_1loop}) reproduces the known answer after integration. While both proposals seem promising and the ultimate solution to finding good loop coordinates for non-planar loop integrands might involve some of the ideas involved there if we demand that the integrand is absent of spurious poles or over-counting of singular regions then no such function exists. 

\subsection{Cuts of integrands}
\label{subsec:unitarity_cuts}

The problem of the non-planar integrand disappears if we consider unitarity cuts of the integrand by putting some of the propagators on shell. In particular, if we cut sufficiently many propagators, the cut defines natural coordinates and makes the {\it cut integrand} $\I_{cut}$ well-defined. The most extreme example is the \emph{maximal cut} when all propagators in a corresponding Feynman integral are set on-shell
\begin{align}
\label{fig:max_cut_matching}
\raisebox{-57pt}{
\includegraphics[scale=.62]{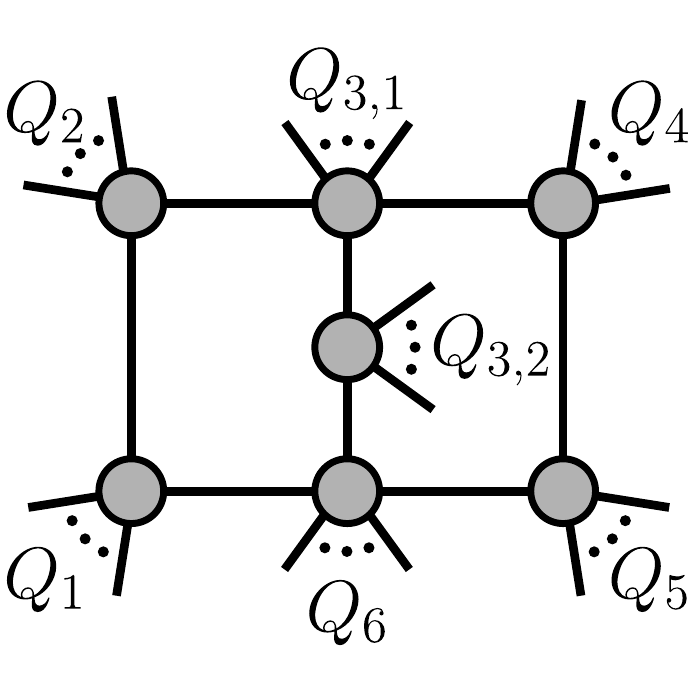}}
\qquad
\Leftrightarrow 
\qquad
\raisebox{-57pt}{
\includegraphics[scale=.62]{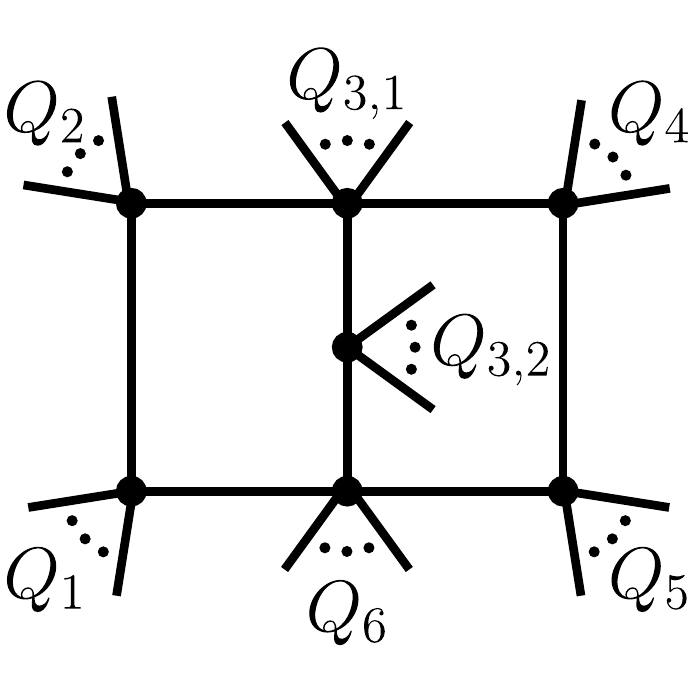}}\,.
\end{align}
In this particular case, the integral has $4L$ propagators so that the maximal cut localizes all degrees of freedom and constitutes a \emph{leading singularity} \cite{Cachazo:2008vp}. The on-shell conditions localize all internal degrees of freedom, so that the residue of the loop-integrand is a rational function of external kinematics. In a diagrammatic representation of the amplitude where we only introduce Feynman integrals with at most $4L$ propagators, such a maximal cut isolates a single term, and its coefficient is directly given by this on-shell function. Note that sometimes it is wise to go beyond diagrams with $4L$ propagators to expose special features of a given theory, see e.g. the representations of $\N=4$ sYM amplitudes in \cite{ArkaniHamed:2010gh,Bourjaily:2015jna,Bourjaily:2017wjl}. In a recently developed prescriptive approach to unitarity, the basis of Feynman integrals is explicitly tailored so that each term is matched exactly by one unitarity cut picture \cite{Bourjaily:2017wjl}. In the next step, one can go beyond the leading singularities and cut fewer than $4L$ propagators. Generically, we get a contribution of multiple integrals to a given cut,
\begin{align}
\label{fig:cut_matching}
\raisebox{-40pt}{
\includegraphics[scale=.55]{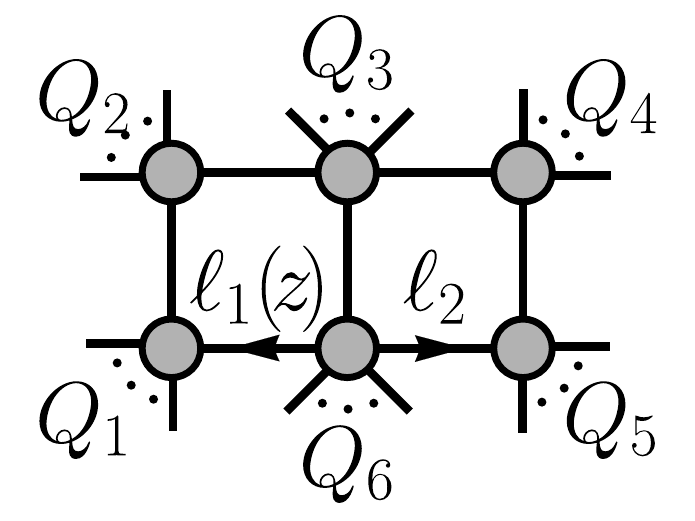}}
\hskip -.3cm
\Leftrightarrow 
\hskip -.3cm
\raisebox{-40pt}{
\includegraphics[scale=.55]{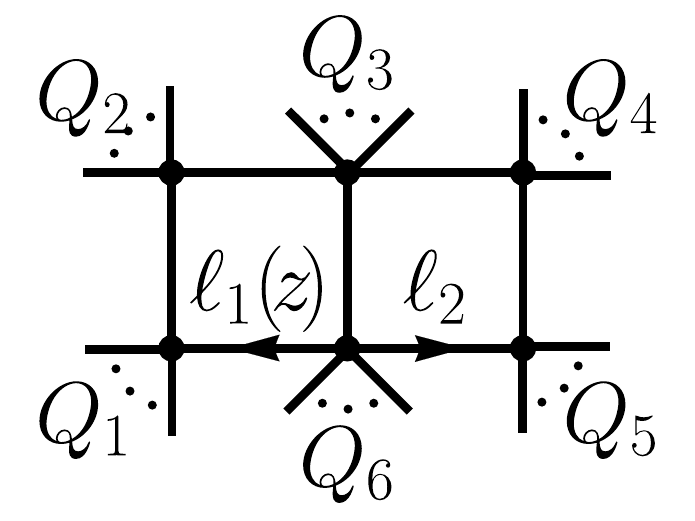}}
\hskip -.2cm
+
\hskip-.3cm
\raisebox{-40pt}{
\includegraphics[scale=.55]{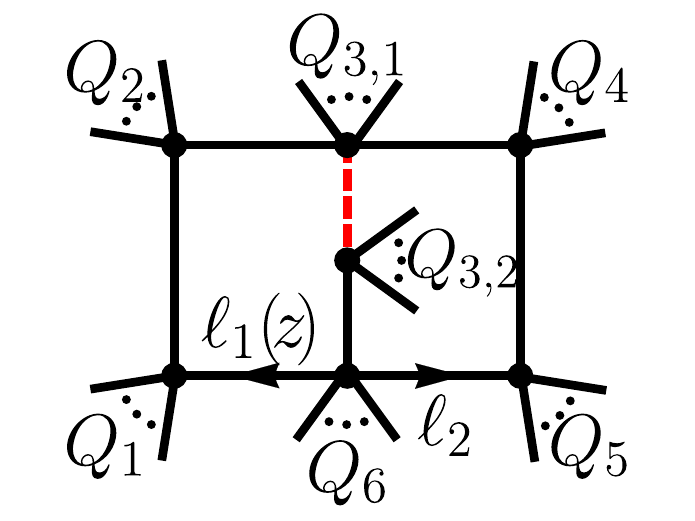}}
\hskip -.4cm
+\ \cdots
\,.
\end{align}
Here, all internal momenta $P_i$ are on-shell $P_i^2=0$, except the red, dashed propagator in the double-pentagon integral which is left uncut. The two momenta $\{Q_{3,1},Q_{3,2}\}$ in the double-pentagon sum to the corresponding momentum $Q_3=Q_{3,1}\!+\!Q_{3,2}$ of the cut. This particular cut contains only one unfixed parameter, $z$, in the solution to the on shell conditions. The main statement of \emph{generalized unitarity} is that the residue of the integrand on the cut is equal to the product of the corresponding tree-level amplitudes (left figure in (\ref{fig:cut_matching})). We will refer to these residues as {\it cut} or {\it on-shell functions}.

On the other hand, in a Feynman integral representation of the amplitude, on this cut we can directly pair the coordinates of the cut with the ones in the Feynman integrals and match the double-box coefficients by taking into account that the eight-propagator integrals have been fixed previously. This hierarchical cut-matching procedure described here is known as the \emph{method of maximal cuts}~\cite{Bern:2007ct}. 

In order to have well defined loop coordinates it is unnecessary to go all the way to the maximal cut as good labels are already available for lower cuts. In fact, even cutting a single propagator per loop suffices and the set of contributing integrals would correspond to the symmetric sum over labeling all possible edges by $\ell$ ((\ref{fig:multi_unitarity_cut}) left),

\vspace{-0.5cm}

\begin{align}
\label{fig:multi_unitarity_cut}
\raisebox{-40pt}{
\includegraphics[scale=.5]{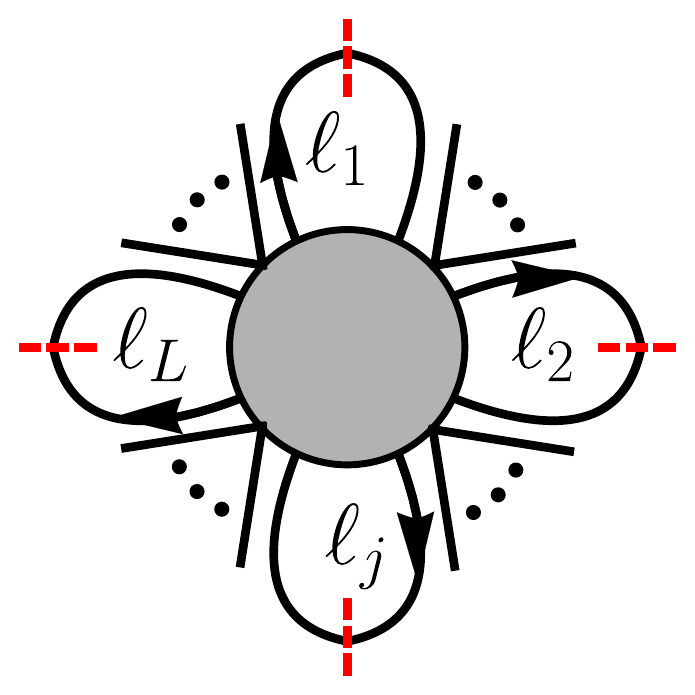}}\,, \qquad
\raisebox{-57pt}{
\includegraphics[scale=.57]{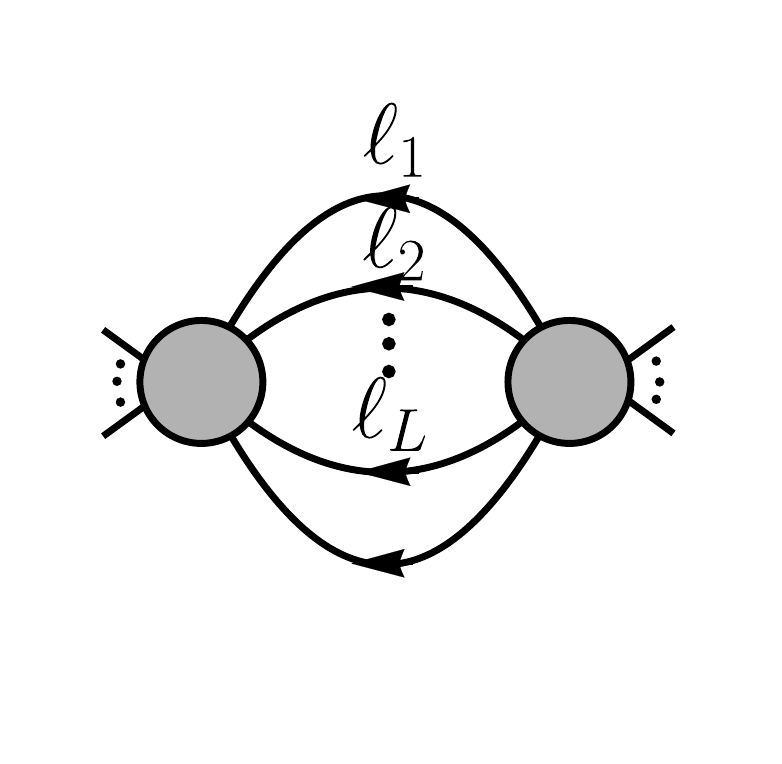}}\,.
\end{align}

\vspace{-0.5cm}

The value of this cut is not well-defined in a general QFT due to forward limit issues, see e.g.~\cite{Benincasa:2015zna}. The best starting point therefore is the traditional \emph{unitarity cut} at one loop, where two propagators are put on shell, and its higher-loop \emph{multi-unitarity cut} generalization ((\ref{fig:multi_unitarity_cut}) right).

This is not the only cut that yields good coordinates and any other cuts between the multi-unitarity cut and the maximal cuts are acceptable for this purpose. Let us discuss one concrete one-loop example to see how on-shell functions are calculated. If we set three propagators to zero, $(\ell-p_3)^2=\ell^2=(\ell+p_4)^2=0$, these on-shell conditions have two parity conjugate solutions which localize $\ell^\ast=\gamma \lam{4}\lamt{3}$ or $\tw{\ell}^\ast = \delta \lam{3}\lamt{4}$ with $\gamma,\delta$ unfixed. The residue of the amplitude on the first cut solution is equal to

\vspace{-0.5cm}
\begin{align*}
\label{fig:one_loop_triple_cut}
\raisebox{-50pt}{\includegraphics[scale=0.65]{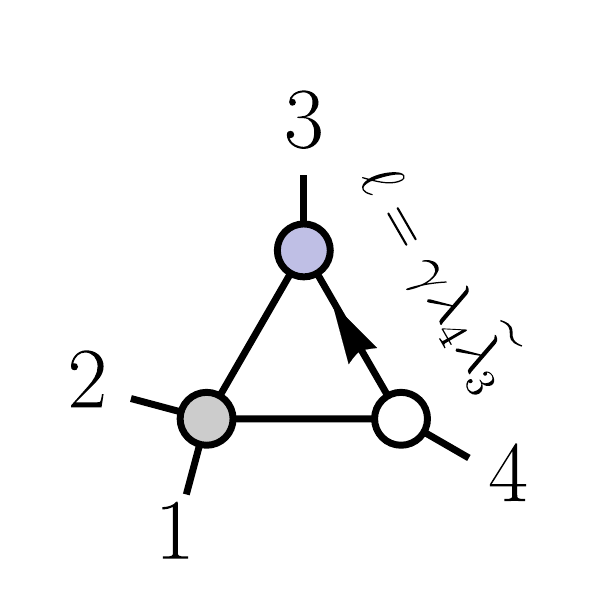}}
\hskip -.5 cm
 = & \int\!\! \frac{d\gamma}{\gamma \ \sab{34}}\!\! \int\!\! d \tw{\eta}\, \M^{(0)}_{4}(1,2,\sm\ell_1,\sm\ell_2) \  \M^{\text{MHV}}_{3}(\ell_1,3,\sm\ell) \M^{\MHVbar}_{3} (\ell,4,\ell_2)\,.
\end{align*}

\vspace{-0.2cm}

In $\N=8$ supergravity this is written in terms of products of tree-level (super-) amplitudes, $d \tw{\eta}$ schematically denotes the supersymmetric state sum, and $\gamma \sab{34}$ is the Jacobian of the three on-shell conditions. For pure gravity, one would have to specify the helicity states of the external gravitons and then sum over all allowed helicity configurations of internal on-shell legs. The $\N=8$ four-point amplitude is given by
\begin{align}
\begin{split}
\raisebox{-28pt}{\includegraphics[scale=.6]{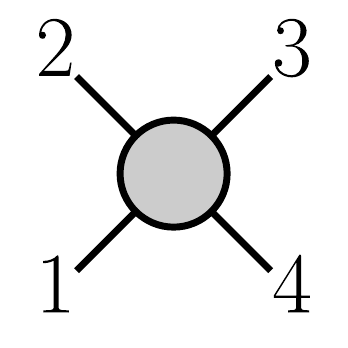}}  \hskip -.3cm = 
 \M^{(0)}_{4} \! =\! \frac{\sqb{12}}{\ab{12}\ab{13}\ab{14}\ab{23}\ab{24}\ab{34}^2}\,\delta^{(16)}(\lam{}\cdot \widetilde \eta).
\end{split}
\end{align}
The three-point on-shell kinematics is special and it allows only for two solutions: (i) $\widetilde{\lambda}_1\sim\widetilde{\lambda}_2\sim\widetilde{\lambda}_3$ which corresponds to an MHV amplitude, and (ii) ${\lambda}_1\sim{\lambda}_2\sim{\lambda}_3$ which corresponds to an $\overline{\rm MHV}$ amplitude,
\begin{align}
\begin{split}
&\raisebox{-35pt}{\includegraphics[scale=.47]{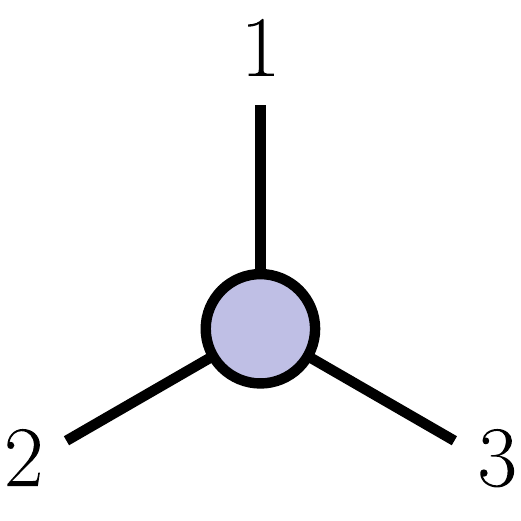}} \hskip -.3cm =
  \M^{\text{MHV}}_{3} = \frac{\delta^{(16)}(\lam{1}\widetilde{\eta}_1+\lam{2}\widetilde{\eta}_2+\lam{3}\widetilde{\eta}_3)}{\ab{12}^2\ab{23}^2\ab{31}^2}\,, \\
&\raisebox{-35pt}{\includegraphics[scale=.47]{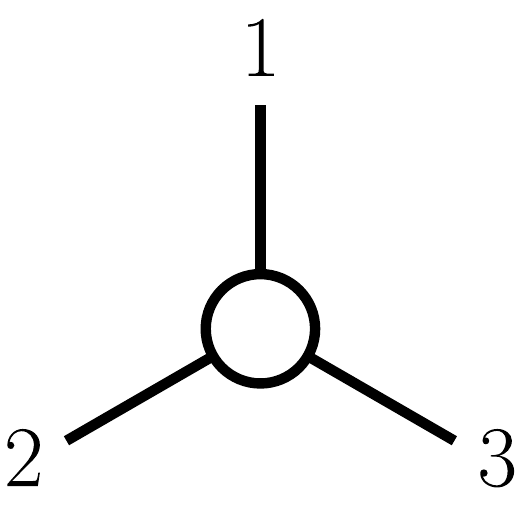}} \hskip -.3cm =
   \M^{\MHVbar}_{3} = \frac{\delta^{(8)}(\sqb{23}\widetilde{\eta}_1+\sqb{31}\widetilde{\eta}_2+\sqb{12}\widetilde{\eta}_3)}{\sqb{12}^2\sqb{23}^2\sqb{31}^2}
 \end{split}
\end{align}
We always use these colored vertices for three-point amplitudes to distinguish between the MHV degree $k=2$ and $k=1$ respectively. General unitarity cuts (or \emph{on-shell functions}), such as (\ref{fig:max_cut_matching}), contain higher point amplitudes denoted by gray blobs where the $k$-charge must be specified. On the other hand, we define \emph{on-shell diagrams} to only involve the bi-colored three-point vertices. 

After inserting the building blocks and performing the state sum, we obtain a cut function of the remaining degree of freedom, $\gamma$, 
\vspace{-0.4cm}
\begin{align}
\label{eq:one_loop_triple_cut_os_function}
\raisebox{-48pt}{\includegraphics[scale=0.65]{./figures/1_loop_4pt_BCFW_MHV_cut_non_hel.pdf}}
\hskip -.5 cm
 = & \int\!\! \frac{d\gamma \ \sab{34}\sqb{12}\  \delta^{(16)}(\lam{}\!\cdot\! \widetilde \eta) }{\gamma \ \ab{12}(\gamma \ab{14}-\ab{13})(\gamma\ab{24}-\ab{23})\ab{14}\ab{24}\ab{34}^2}\,.
\end{align}

\vspace{-0.3cm}

In the generalized unitarity approach, this on-shell function (\ref{eq:one_loop_triple_cut_os_function}) is to be compared to the cut of the integral expansion of the amplitude. Only basis integrals which contain at least the cut propagators contribute. For the $\N=8$ one-loop example, these are only two of the box integrals in (\ref{eq:one-loop-gravity-amp}),
\vspace{-0.5cm}
\begin{align}
\label{fig:one_loop_BCFW_cut}
\raisebox{-50pt}{\includegraphics[scale=0.7]{./figures/1_loop_4pt_BCFW_MHV_cut_non_hel.pdf}}
\hskip .1cm
 = 
 \hskip .1cm
\raisebox{-55pt}{\includegraphics[scale=0.52]{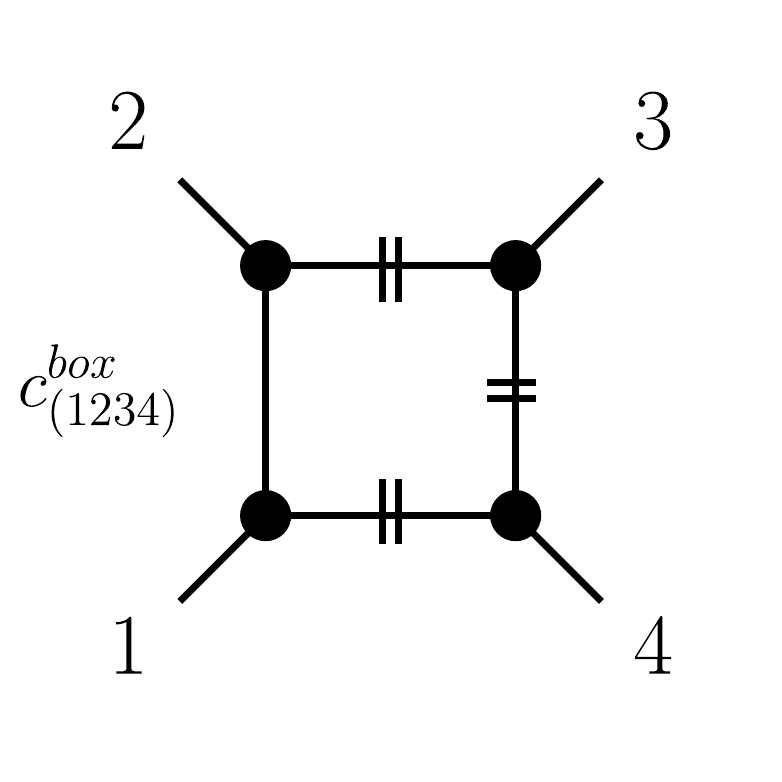}} 
\hskip -.5cm
+ 
\raisebox{-55pt}{\includegraphics[scale=0.52]{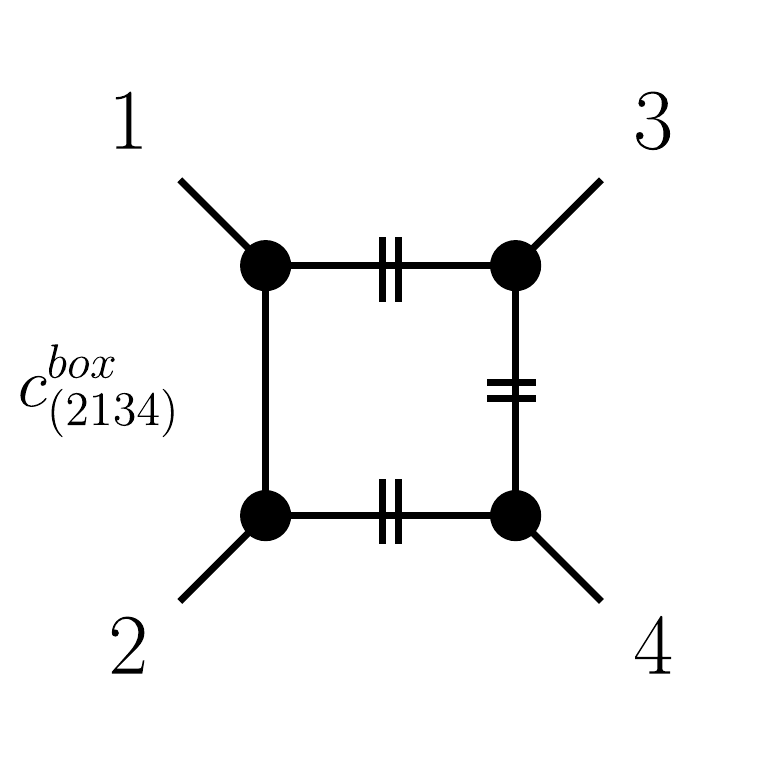}}\,.
\end{align}

\vspace{-0.5cm}

To calculate the box coefficients $c^{box}_{i}$ we can take further residues of (\ref{eq:one_loop_triple_cut_os_function}) at $\gamma=\ab{23}/\ab{24} \leftrightarrow c^{box}_{(1234)}$ and $\gamma=\ab{13}/\ab{14}\leftrightarrow c^{box}_{(2134)}$ which correspond to two of the remaining factorization channels of the on-shell function and gives formulae for the respective box coefficients
\begin{align}
c^{box}_{(1234)} = c^{box}_{(2134)} = stu\, \M^{(0)}_4\,.
\end{align}
Here we would like to point to a different feature of the gravity on-shell function (\ref{eq:one_loop_triple_cut_os_function}) and that is the $\ell\rightarrow\infty$ limit by sending $\gamma\rightarrow\infty$. In the large $\gamma$ limit, (\ref{eq:one_loop_triple_cut_os_function}) scales like $1/\gamma^3$ whereas the individual box integrals behave like $1/\gamma^2$ on the cut
\begin{align}
\label{eq:one_loop_BCFW_cut_infinity}
\begin{split}
\int d\gamma\,{\cal O}\left(\frac{1}{\gamma^3}\right) & = \int \frac{d\gamma}{\gamma} \left[ \frac{c^{box}_{(1234)}}{(\ell^\ast \sm p_3\sm p_2)^2} +  \frac{c^{box}_{(2134)}}{(\ell^\ast \sm p_3\sm p_1)^2} \right] \\
	& = \int \frac{d\gamma}{\tblue{\gamma}}  \frac{c^{box}_{(1234)}(\tblue{\gamma} \ab{14}-\ab{13})\sqb{13} + c^{box}_{(2134)}(\tblue{\gamma} \ab{24}-\ab{23})\sqb{23}}
									     {(\tblue{\gamma}\ab{14}-\ab{13})\sqb{13}(\tblue{\gamma} \ab{24}-\ab{23})\sqb{23}}\,.
\end{split}
\end{align}

In order for the leading $1/\gamma^2$ terms to cancel between the two boxes on the right hand side of (\ref{eq:one_loop_BCFW_cut_infinity}) imposes the correct relation $c^{box}_{(1234)}=c^{box}_{(2134)}$ between the box coefficients and reproduces the precise $1/\gamma^3$ scaling of the cut. In this case the improved scaling of the on-shell function can be understood from the good large-$z$ behavior of the tree-level amplitudes under BCFW shifts. However, we checked that a much stronger statement is true at one-loop: the behavior at infinity on \emph{any} triple-cut is $1/\gamma^3$
\vspace{-1.5cm}
\begin{align}
\label{fig:massive_triple_cut}
\raisebox{-50pt}{\includegraphics[scale=0.78]{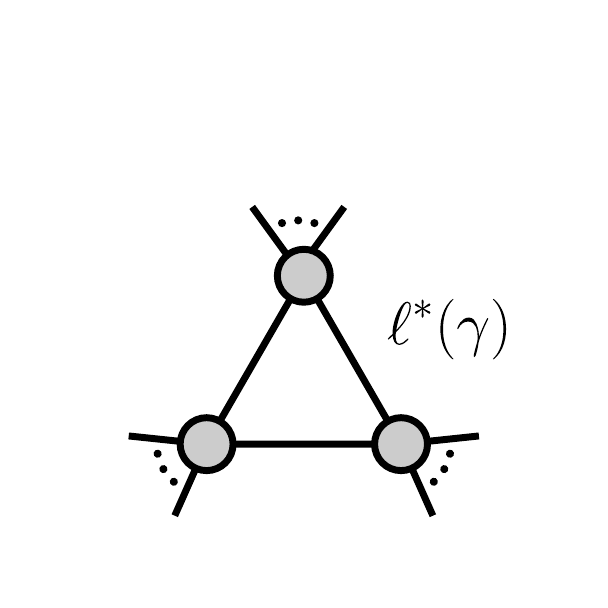}}\,,
\end{align}

\vspace{-0.5cm}

in comparison to the $1/\gamma^2$ behavior of individual box integrals. The cancelation between boxes on the triple cut is then very general and not limited to special cases. This makes it plausible that the surprising behavior at infinity we are going to discuss later also holds for all amplitudes. The one-loop statement described here goes beyond the ``no-triangle hypothesis" \cite{ArkaniHamed:2008gz} which is equivalent to no poles at infinity on the cut (\ref{fig:massive_triple_cut}) and it would only imply $1/\gamma^2$ behavior in contrast to the observed $1/\gamma^3$. Similar type of constraints arose from dual conformal symmetry in planar ${\cal N}=4$ sYM theory.

\section{UV and poles at infinity}
\label{sec:uv_poles_at_infinity}

The purpose of this section is to study the UV region of gravity integrands. We will refer to the $\ell\rightarrow\infty$ singularities  as \emph{poles at infinity}. In line with the previous section, such poles can not be meaningfully approached for the full non-planar integrand $\I$ as there are no canonical loop momenta $\ell_k$ which we can set to infinity. However, we can probe these poles for the cut integrand ${\cal I}_{cut}$ once the $\ell_k$ become well-defined. \twhite{......}
\subsection{Maximal cuts and powercounting}
\label{subsec:max_cuts_power_counting}

Let us start with a simple example. The two-loop four-point amplitude is written using two integral topologies, planar and non-planar double boxes, and permutations thereof,

\vspace{-0.7cm}

\begin{align}
\label{eq:two_loop_four_point}
\M^{(2)}_4 = 
N_{np} \hskip -.6cm \raisebox{-53pt}{
\includegraphics[scale=.5]{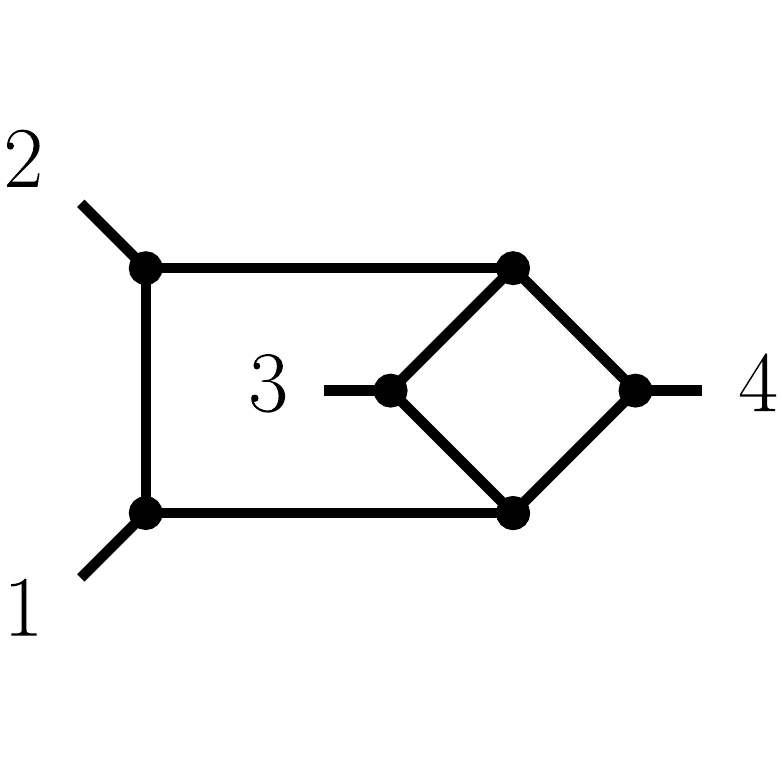}}\hspace{.3cm} + 
N_p \hskip -.6cm
\raisebox{-53pt}{
\includegraphics[scale=.5]{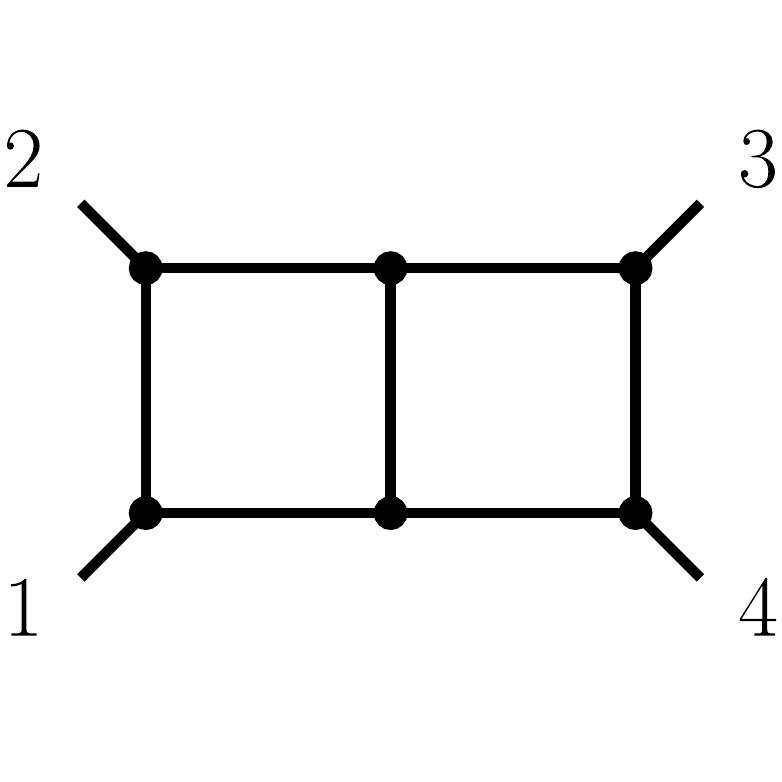}} + \text{perms}
\end{align}

\vspace{-0.7cm}

The $N_{np}$ and $N_p$ in (\ref{eq:two_loop_four_point}) indicate explicit diagram numerators. In the following we will suppress these numerators and take them to be implied for all Feynman integrals. There are two distinct maximal cuts (heptacuts) which select the individual integral topologies. The corresponding on-shell diagrams depend on one remaining parameter $z$ which can be chosen such that the cut loop momentum depends linearly on $z$. Further residues in $z$ are accessible when we consider cutting the Jacobian factor. Generically, these residues are associated with certain soft- or collinear singularities. Besides approaching these singularities we can calculate the behavior at infinity, $z\rightarrow\infty$, 

\vspace{-0.9cm}

\begin{align}
\label{eq:2_loop_4pt_os_funcs}
\raisebox{-53pt}{\includegraphics[scale=.5]{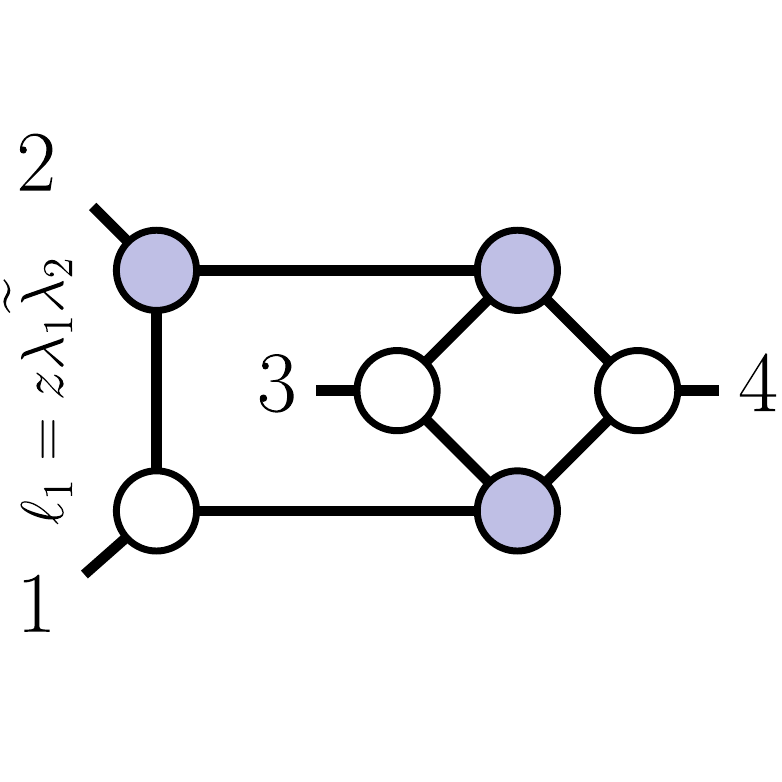}} \hspace{.1cm} \sim {\cal O}\left(\frac{1}{z^2}\right)\,,  \qquad
\raisebox{-53pt}{\includegraphics[scale=.5]{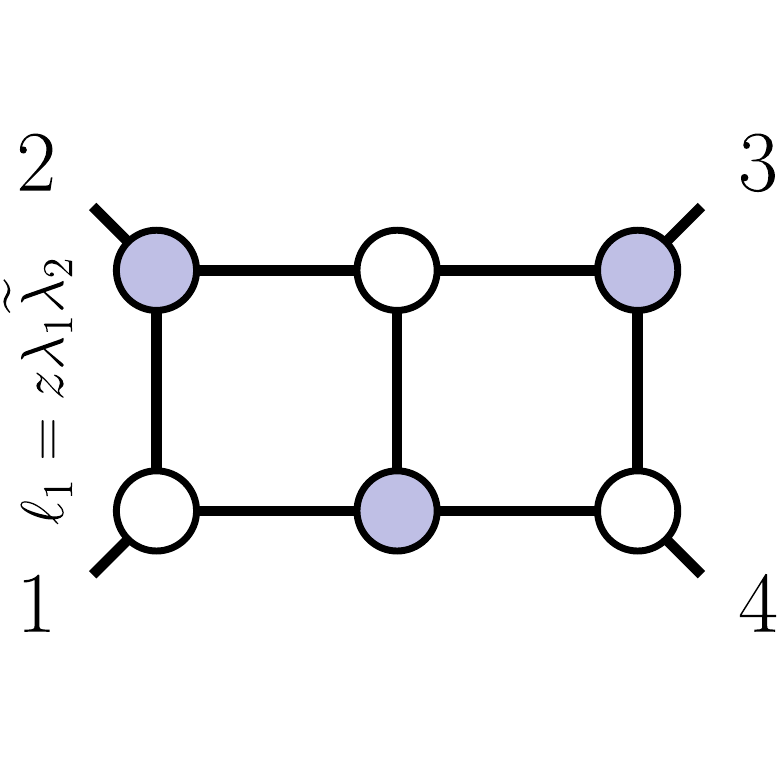}}	 	\hspace{-.5cm}\sim {\cal O}\left(\frac{1}{z^2}\right)\,.
\end{align}

\vspace{-0.6cm}

We see that in both cuts there are no poles at infinity and after comparing the on-shell diagrams (\ref{eq:2_loop_4pt_os_funcs}) to the cuts of the Feynman integrals (\ref{eq:two_loop_four_point}) we find 
\begin{align}
 N_{np} = N_{p} = s t u\, \M^{(0)}_4  \times s^2\,.
\end{align}
These numerators do not depend on the loop momenta \cite{Bern:1998xc} and turn out to be the square of the Yang-Mills numerators in line with the BCJ double-copy structure which is always valid at the level of maximal cuts of cubic three-point diagrams.

At three loops, the local expansion of the amplitude is given by a sum of 12 integral topologies \cite{Bern:2010ue}. We only focus on the planar integrals without dangling trees, but the same analysis can be performed for other terms as well,
\vspace{-0.2cm}
\begin{align}
\label{eq:three_loop_local_partial}
\M^{(3)}_4 = \hskip -1.1cm\raisebox{-33pt}{\includegraphics[scale=.5]{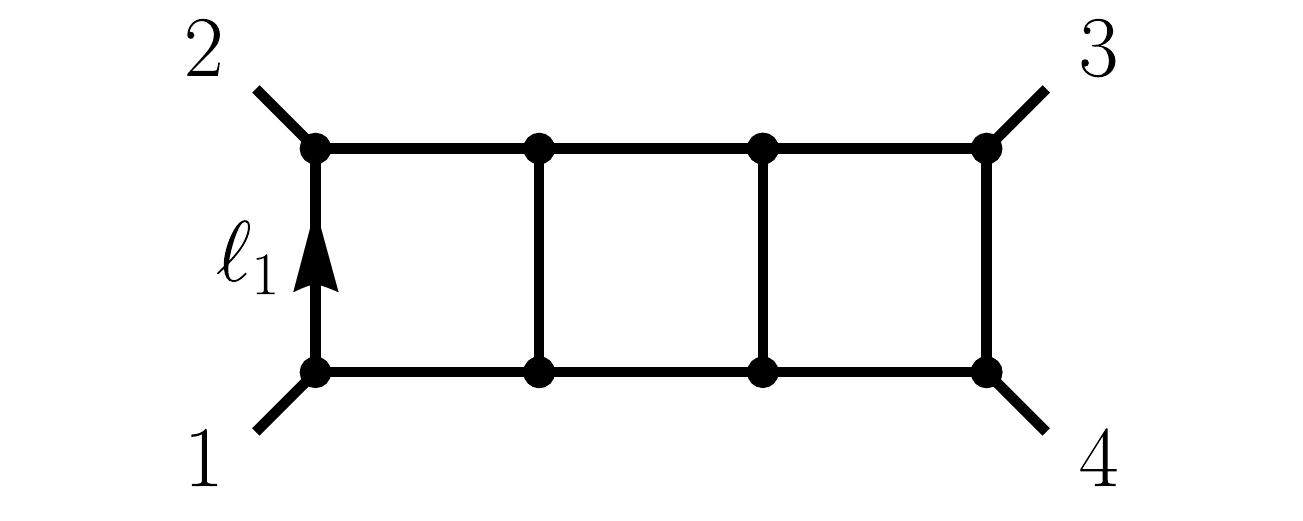}}  \hskip -.7cm+ \hskip -.7cm 
		     \raisebox{-53pt}{\includegraphics[scale=.5]{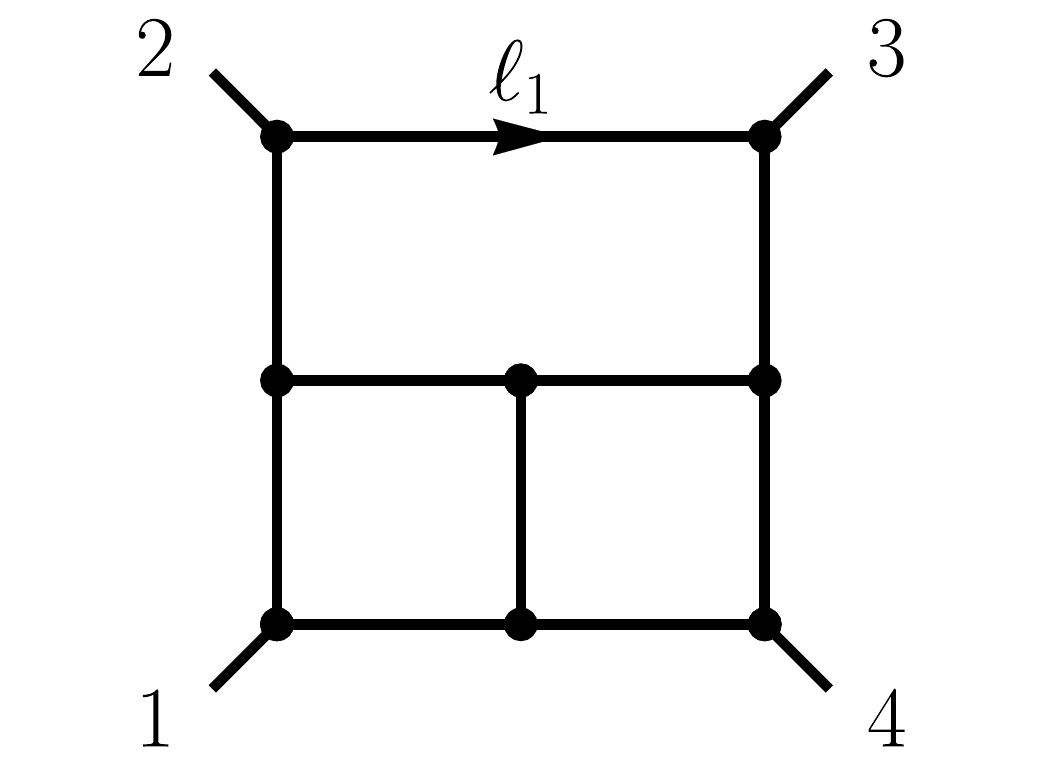}} \hskip -1.1cm+\! \text{nonplanar}\! +\! \text{perms.}
\end{align}

\vspace{-0.2cm}

The associated maximal cuts (\ref{fig:3_loop_4pt_max_cuts}) are functions of two parameters $z,w$ as we only impose 10 on-shell constraints to the $4\times3=12$ off-shell loop variables. It is possible to parameterize the remaining degrees of freedom such that $z$ probes the behavior at infinity. The scaling at infinity is now different for the different on-shell diagrams and consequently their associated Feynman integrals need to reflect this behavior in order to match the cuts. To give an example, the triple-box on-shell function ((\ref{fig:3_loop_4pt_max_cuts}) left) scales like $1/z^2$ as $z$ approaches infinity whereas other cuts, such as  the ``tennis court" ((\ref{fig:3_loop_4pt_max_cuts}) right), display a poorer behavior at infinity,

\vspace{-0.7cm}

\begin{align}
\label{fig:3_loop_4pt_max_cuts}
\raisebox{-33pt}{
\includegraphics[scale=.5]{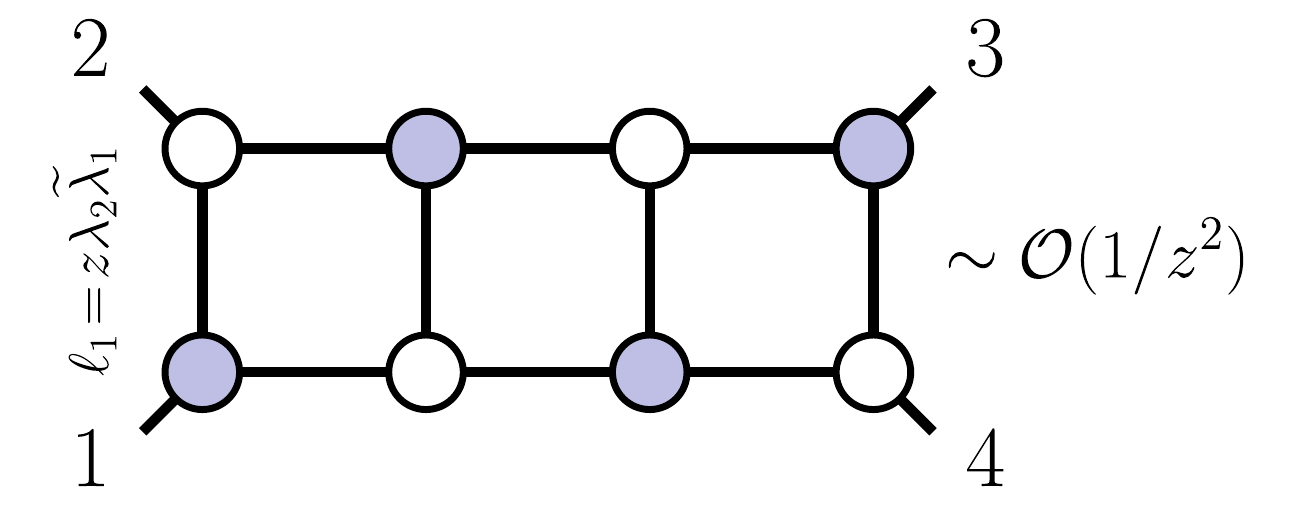}}\,,
\raisebox{-53pt}{
\includegraphics[scale=.48]{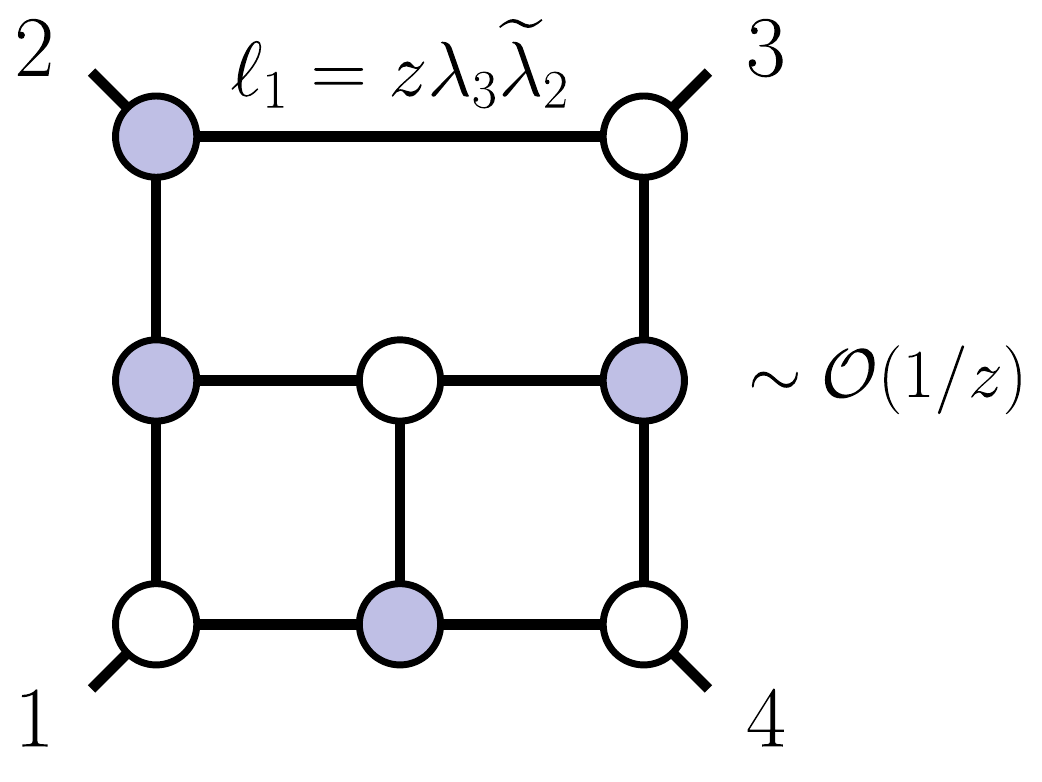}}\,.
\end{align}

The scaling behavior of the on-shell function as $z\to\infty$ has important implications for the numerators of the corresponding Feynman integrals: the tennis court integral in (\ref{eq:three_loop_local_partial}) must contain sufficiently many powers of loop momenta in order to match the large $z$ scaling behavior of the cut. In fact, the numerator in ${\cal N}=8$ supergravity is $N \sim [(\ell_1\! +\! p_1\!+\!p_2)^2]^2$ modulo contact terms which vanish on the maximal cut. 

It is obvious that this trend continues, and therefore on certain maximal cuts the supergravity amplitude has higher poles when we chose a collective coordinate $z$ for $z_1,z_2,\dots$ to approach infinity. The degree of the pole grows with the loop order. One class of maximal gravity cuts where this happens is

\vspace{-0.6cm}
\begin{align}
\label{fig:os_func_uv_scaling_example}
\raisebox{-63pt}{\includegraphics[scale=.52]{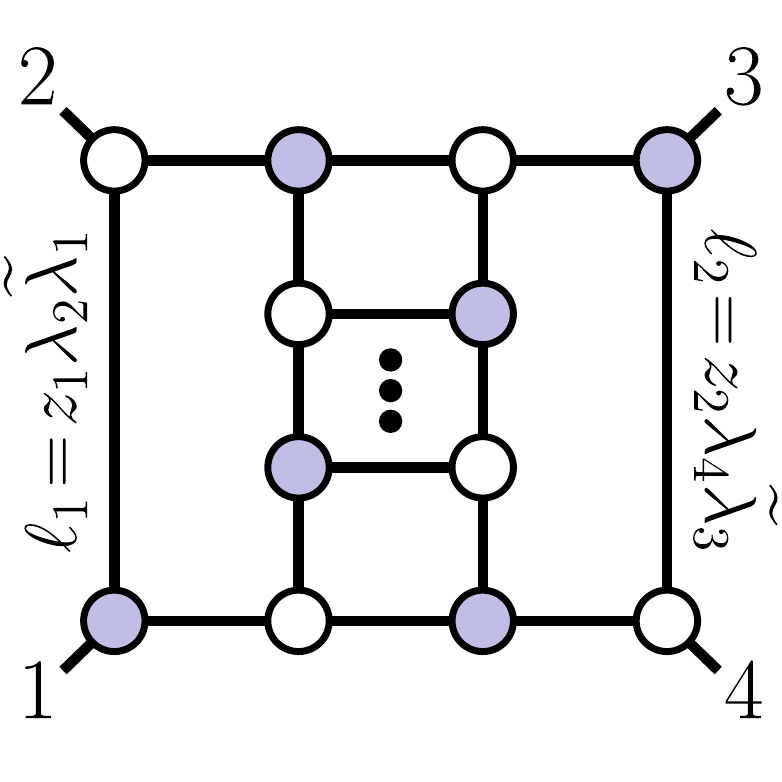}}
\quad \sim {\cal O}\left(\frac{1}{z^{4-L}}\right)\,,
\end{align}
\vspace{-0.35cm}

in contrast to $1/z^2$ for the same $L$-loop on-shell function in $\N=4$ sYM. As a result, the $L$-loop integral with the same topology as (\ref{fig:os_func_uv_scaling_example}) must have a strong loop-momentum dependence in the numerator in order to match this large $z$ scaling of (\ref{fig:os_func_uv_scaling_example}),

\vspace{-0.6cm}
\begin{align}
\label{bad integral}
\raisebox{-63pt}{\includegraphics[scale=.52]{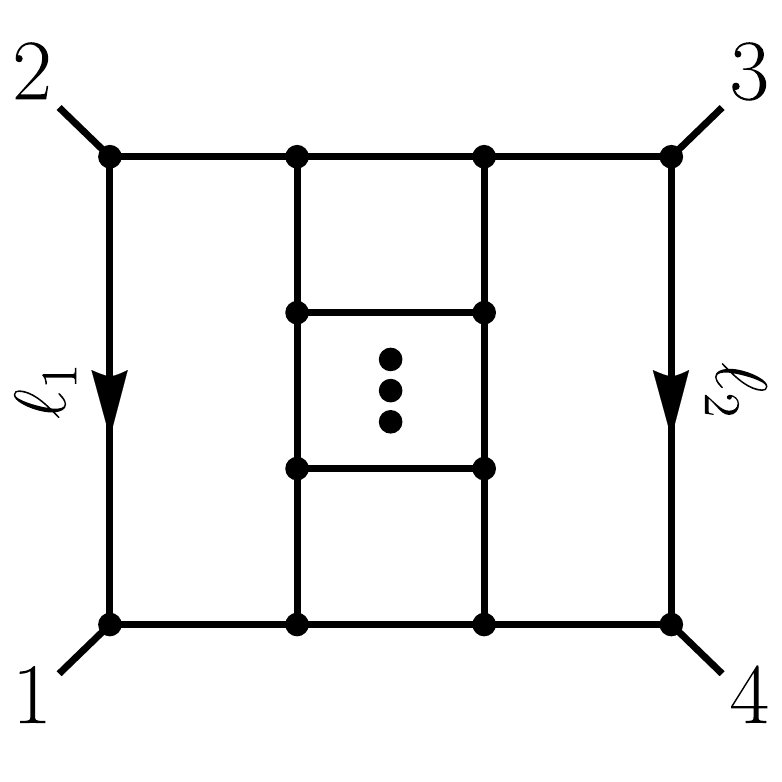}}
\quad \Leftrightarrow  \quad N \sim \left[(\ell_1+\ell_2-p_1-p_4)^2\right]^{2(L-3)}\,.
\end{align}
\vspace{-0.5cm}

The $\ell\rightarrow\infty$ scaling of the numerator can subsequently be related to the UV behavior of the Feynman integral. When integrating (\ref{bad integral}) we can extract the leading UV divergence from sending all loop momenta to infinity,

\begin{equation}
I = \int \frac{d\ell}{\ell^{15-2L}} = \,\mbox{divergent for $L\geq 7$}\,.
\end{equation}

Following this line of reasoning, for maximal cuts there is a direct relation between the degree of the pole at infinity, the loop-momentum dependence of the numerator of the corresponding integral and the degree of the UV divergence. Up to this point everything seems very predictable and unsurprising. One might expect that once poles at infinity are present in maximal cuts, they also appear for lower cuts as well. Furthermore, without relying on any surprises, the naive expectation is that all Feynman integrals that appear in the expansion of the amplitude have the same (or lower) degree poles at infinity as the cut integrand. However, we already saw at the end of section~\ref{sec:intro} that this is not true even at one-loop, where we found cancelations between box integrals on the triple cut (\ref{eq:one_loop_BCFW_cut_infinity}). As we will show below, the unexpected behavior also appears at higher loops.

\subsection{Multi-particle unitarity cut}
\label{subsec:multi_unitarity_cut}

As we have seen in the previous subsection, no surprising features were found on maximal cuts mainly because they isolate individual Feynman integrals in the expansion and there is no room for cancelations. We learned in the one-loop example (\ref{eq:one_loop_BCFW_cut_infinity}) that surprises appear when multiple diagrams contribute on a given lower cut. To this end, we now consider the opposite to maximal cuts: we only cut a minimal number of propagators which still gives us unique loop labels and allows us to approach infinity. The particular cut of our interest is the multi-unitarity cut
\vskip -1.3cm
\begin{align}
\label{fig:multi_unitarity_cut_2}
\raisebox{-40pt}{
\includegraphics[scale=.6,trim={1cm 1.8cm 1cm 0cm},clip]{./figures/multi_unitarity_cut_labels.pdf}}\,.
\end{align}

We will probe cancelations of the large loop momentum behavior between different Feynman integrals that contribute to (\ref{fig:multi_unitarity_cut_2}). The presence of such cancelations and better behavior of the amplitude in comparison to individual integrals then points to some novel mechanism or symmetry we have not yet unraveled in the context of gravity amplitudes.

In the cut (\ref{fig:multi_unitarity_cut_2}) we only put $L+1$ propagators on-shell and are left with an on-shell function $F(z_k)$ which depends on $3L-1$ parameters $z_k$. On this cut surface there are numerous ways to approach infinity, and we will have to make a particular choice. 

\subsubsection*{One-loop}

Let us first discuss the simplest case of the traditional unitarity cut of a one-loop four-point gravity amplitude which already exhibits some surprising behavior.

\vspace{-0.6cm}
\begin{align}
\raisebox{-50pt}{
\includegraphics[scale=.57]{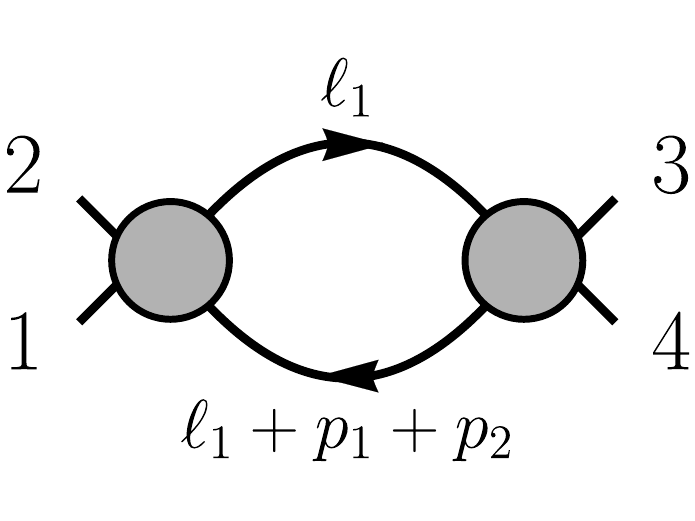}}
\end{align}
\vspace{-0.5cm}

We parametrize the loop momentum using a basis of spinor-helicity variables,
\begin{align}
\ell_1 = \alpha_1 \lam{1}\lamt{1} + \beta_1\lam{2}\lamt{2}+\gamma_1 \lam{1}\lamt{2} +\delta_1 \lam{2}\lamt{1}
\end{align}
To get the full external kinematic dependence one has to keep track of the Jacobian from changing the integration measure $d^4\ell_1 =  \sab{12}^2 d\alpha_1d\beta_1d\gamma_1d\delta_1$. Cutting the two propagators $\ell^2_1 = 0 =(\ell_1+p_1+p_2)^2 $ is solved by,
\begin{align}
\label{eq:two_particle_cut_parameterization}
\begin{split}
 \ell^\ast_1 & = \frac{1}{\delta_1}\big[\delta_1 \lam{2}-(\beta_1+1)\lam{1}\big]\big[\delta_1\lamt{1}+\beta_1\lamt{2}\big]\,,\\
 (\ell^\ast_1+p_1+p_2) & = \frac{1}{\delta_1} \big[\delta_1 \lam{2} - \beta_1 \lam{1}\big] \big[ \delta_1 \lamt{1} +(\beta_1+1)\lamt{2}\big]\,.
\end{split}
\end{align}
In these coordinates, solving the two cut conditions results in another Jacobian factor $\J= \delta_1$ which appears both in the on-shell function as well as the cut of the Feynman integrals and therefore drops out of the unitarity matching equation. As before, we look at the relative scaling between the on-shell function compared to individual integrals. In the parametrization (\ref{eq:two_particle_cut_parameterization}) we find that for $\beta_1=\hat{\beta}_1z,\delta_1=\hat{\delta}_1z$ and $z\rightarrow\infty$ (while keeping $\hat{\beta}$ and $\hat{\delta}$ fixed) the cut loop momenta are sent to infinity, $\ell_1^\ast, (\ell_1+p_1+p_2)^\ast = {\cal O}(z)$, and the cut behaves  

\vspace{-0.8cm}

\begin{align}
\label{fig:1loop_unitarity_cut_scaling}
\raisebox{-40pt}{
\includegraphics[scale=.56]{./figures/one_loop_unitarity_cut}}
\hskip -.1cm
\stackrel{z \gg 1}{\sim} \frac{1}{z^8} \,,
\qquad
 \raisebox{-45pt}{\includegraphics[scale=.56]{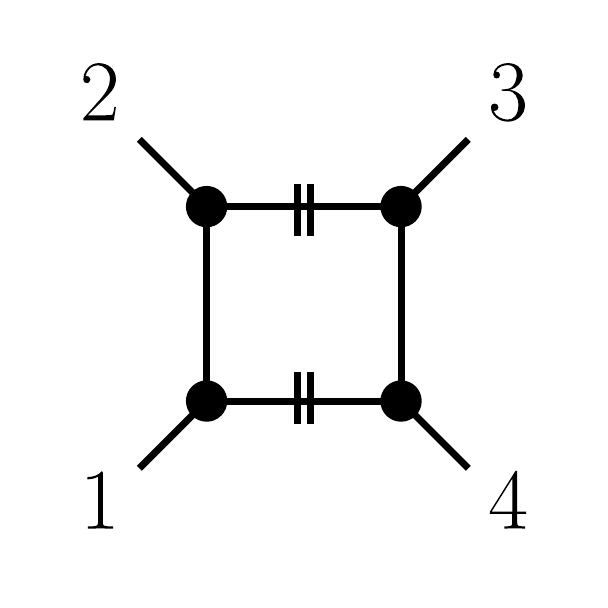}}
 \hskip -1cm
 \stackrel{z \gg 1}{\sim} \frac{1}{z^3}\,.
\end{align}

\vspace{-0.1cm}

The absolute large $z$ scaling powers of (\ref{fig:1loop_unitarity_cut_scaling}) are not really important. What matters is the relative difference between the on-shell function and the individual integral. In this simple one-loop example, there is an easy analytic proof of this enhanced scaling behavior of the complete cut in comparison to individual diagrams. All box integrals have the same crossing-symmetric prefactor $\kappa_8= \left(stA^{4,\text{YM}}_{\text{tree}}(1234)\right)^2 = stu\M^{(0)}_4$, so that the cut is given in a local expansion as the sum of four boxes,
\begin{align}
\raisebox{-32pt}{
\includegraphics[scale=.45]{./figures/one_loop_unitarity_cut}}
 	& = \kappa_8 \Bigg[
				\hskip -.2cm
				\raisebox{-35pt}{\includegraphics[scale=.4]{./figures/box1234_unitarity_cut.pdf}}
				\hskip -.3cm
				+
				\hskip -.3cm
				\raisebox{-35pt}{\includegraphics[scale=.4]{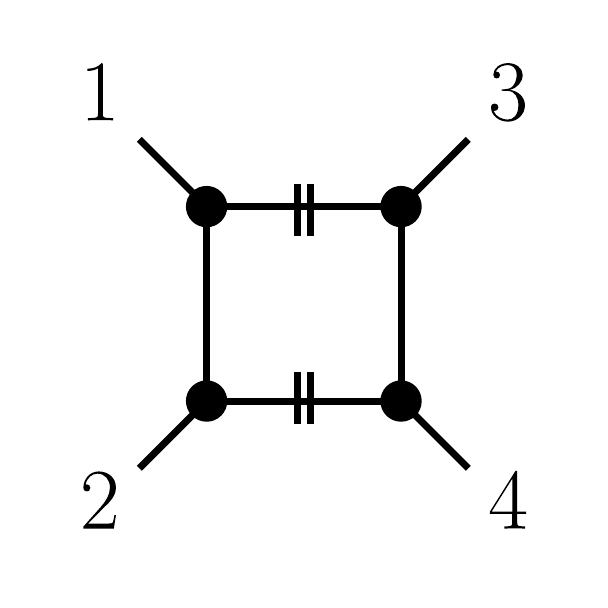}}
				\hskip -.3cm
				+
				\hskip -.3cm
				\raisebox{-35pt}{\includegraphics[scale=.4]{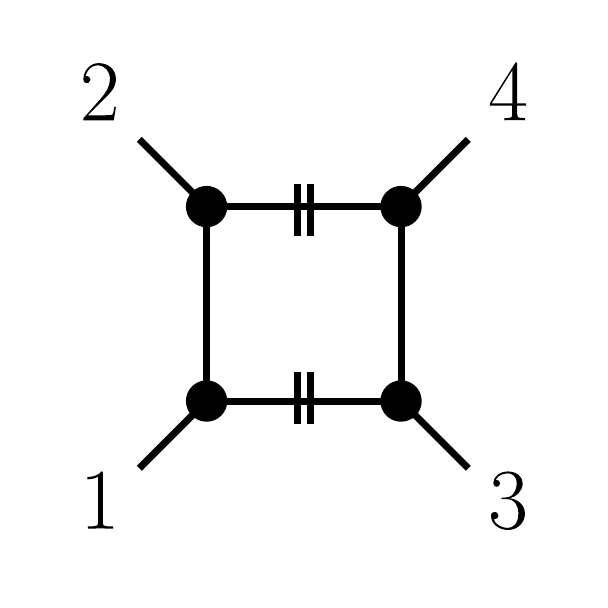}}
				\hskip -.3cm
				+
				\hskip -.3cm
				\raisebox{-35pt}{\includegraphics[scale=.4]{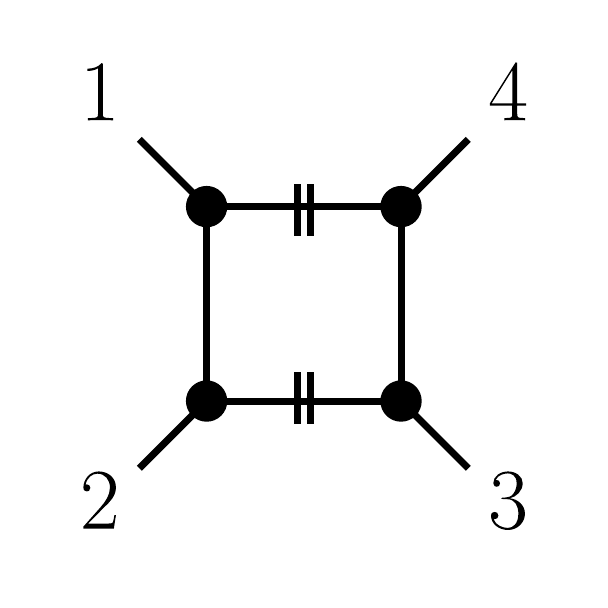}}
				\hskip -.2cm
			     \Bigg]  \\
 		& \sim \frac{1}{(\ell_1\cdot2)(\ell_1 \cdot3)} \!+\! \frac{1}{(\ell_1\cdot1)(\ell_1 \cdot3)}\! +\! 
			\frac{1}{(\ell_1\cdot2)(\ell_1 \cdot4)}\!+\! \frac{1}{(\ell_1\cdot1)(\ell_1 \cdot4)}  \nonumber\\
		& = \frac{\left((\ell_1\cdot1)\!+\!(\ell_1\cdot2)\right)\left((\ell_1\cdot3)\!+\!(\ell_1\cdot4)\right)}{(\ell_1\cdot1)(\ell_1\cdot2)(\ell_1\cdot3)(\ell_1\cdot4)} 
		   = \frac{\sab{12}^2}{(\ell_1\cdot1)(\ell_1\cdot2)(\ell_1\cdot3)(\ell_1\cdot4)}\nonumber
\end{align}
which directly shows the improved behavior of the full cut in comparison to individual local diagrams. For $\N=8$ supergravity, the form of the Feynman integral expansion as well as the structure of the super-amplitudes makes it clear that this is the same result for any helicity configuration of the external states.

\subsubsection*{Two-loops}

The next-to-simplest case is the three-particle cut of two-loop amplitudes. This example is simple enough to keep track of all terms but we already have to choose a particular way how to send the loop momenta to infinity. In our four-dimensional cut analysis, we make such a choice by performing a collective shift on all cut legs. As we will explain below, this appears to be the most uniform choice possible. We point out that this particular limit of approaching infinity is special to $D=4$ where we have spinor-helicity variables at our disposal. This multi-particle cut has been analyzed before~\cite{Bern:1998ug} and was revisited in~\cite{Bern:2017lpv} in an attempt to understand the enhanced cancellations in half-maximal supergravity in $D=5$. The outcome of their analysis was that the improved UV behavior of the amplitude in comparison to individual integrals can not be seen at the integrand level. The authors of \cite{Bern:2017lpv} checked that for some limit $\ell_i\to \infty$ there is no improvement in the large loop-momentum behavior after summing over all terms, compared to the behavior of a single cut integral. For the particular amplitude that was studied, the non-existence of an integrand level cancellation was reduced to the statement that once a certain loop-momentum-dependent, permutation-invariant prefactor is extracted, the remaining sum of diagrams is precisely the same one that appears in the cut of the two-loop four-point amplitude of $\N=8$ \cite{Bern:1998ug}. There are no further cancellations arising at the integrated level from the sum over diagrams so that the UV-divergence of each term coincides with the critical dimension of the full four-point amplitude \cite{Bern:1998ug}.
\begin{align}
\label{fig:three_particle_cut}
 \raisebox{-100pt}{\includegraphics[scale=.5]{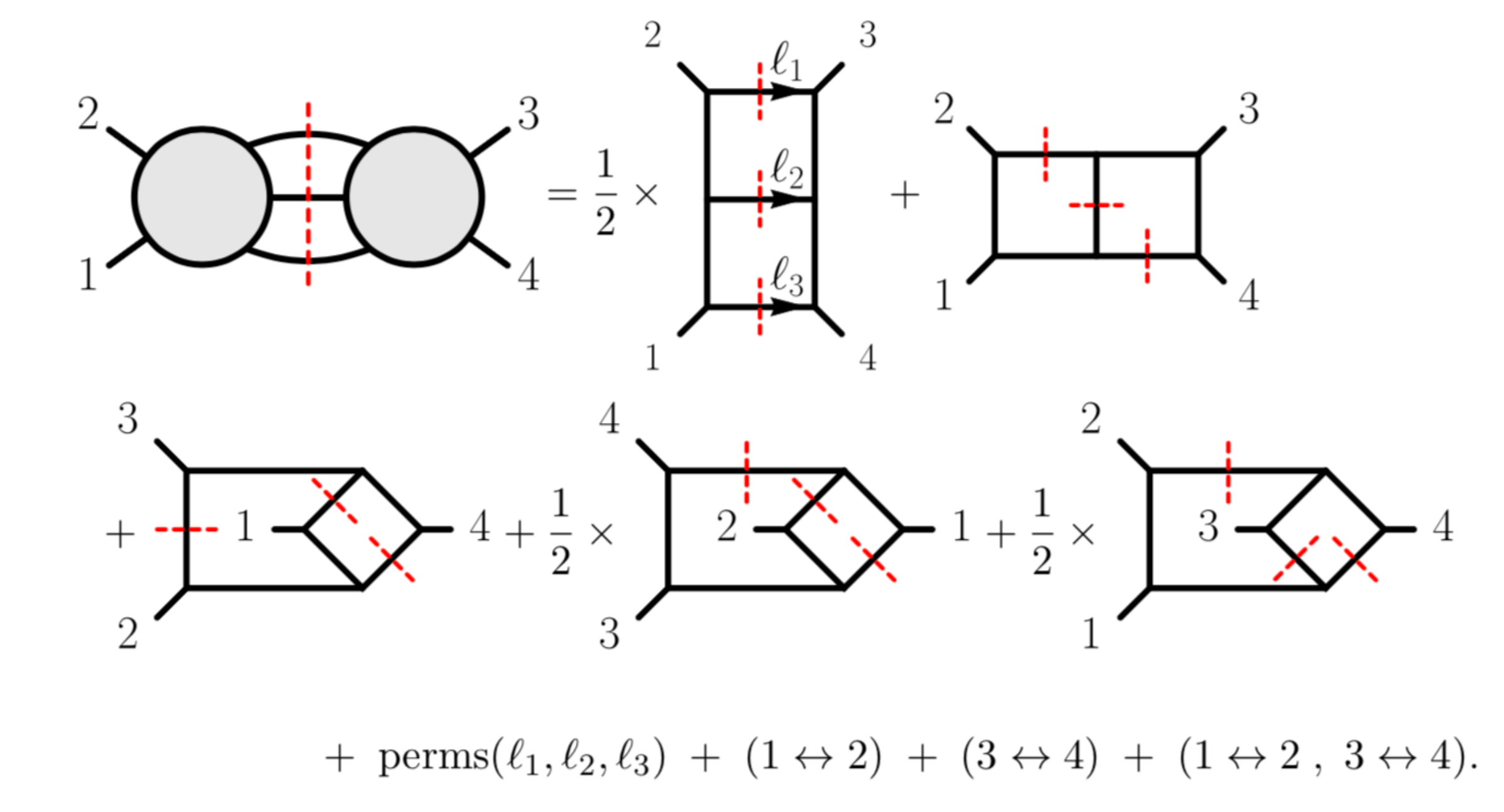}}
\end{align}
Despite the negative result of~\cite{Bern:2017lpv}, here we point out a curious observation at the level of the cut integrand based on the earlier remark that $D=4$ is special and permits powerful spinor-helicity methods. In particular, the factorization of on-shell momenta allows us to approach infinity in a more uniform manner that is common to all loops. Naively, one such prescription to probe infinity comes from a scaling argument, where all loop momenta are scaled simultaneously $\ell_i \to t\,\ell_i$. However, in such an approach terms of the form $(\ell_i\cdot \ell_j) \sim t^2$ are not on the same footing as $(\ell_i\cdot p_k) \sim t$ so that several terms in the diagrammatic sum on the right hand side of (\ref{fig:three_particle_cut}) are decoupled. Instead, we can probe the UV-structure by a holomorphic shift of the on-shell legs by a common reference spinor, $\eta$, such that terms of the form $(\ell_i \cdot \ell_j)$ behave in a uniformly and are on the same footing as $(\ell_i\cdot p_k)$. This is consistent with the origin of these factors from $(\ell_i+\ell_j)^2$ and $(\ell_i+p_k)^2$ which have the same scaling for $\ell_i\rightarrow\infty$. 

In our analysis, we start from some constant numeric parametrization for the momentum-conserving five-particle on-shell kinematics and holomorphically shift the three legs that correspond to the cut loop-momenta. The shift parameter $z$ then allows us to probe the UV-structure by analyzing the $z \to \infty$ behavior,
\begin{align}
\label{eq:holomorphic_shift_2loop}
 \lam{\ell_i} \mapsto \lam{\ell_i} + z\,\sigma_i \eta \,.
\end{align}
Before doing so, we need to make sure that momentum-conservation is satisfied for the deformed kinematic setup. In our parametrization (\ref{eq:holomorphic_shift_2loop}), this is easily implemented by choosing special values for the $\sigma_i$,
\begin{align}
 0=z \eta \left( \sigma_1 \lamt{\ell_1}+\sigma_2 \lamt{\ell_2}+\sigma_3 \lamt{\ell_3}\right) \Leftrightarrow \{\sigma_1 =\sqb{\ell_2\ell_3},\sigma_2=\sqb{\ell_3\ell_1},\sigma_3=\sqb{\ell_1\ell_2}\}
\end{align}
by means of the Schouten identity. In our setup with the uniform chiral parametrization of all internal lines, we actually do find an improvement of the complete cut (\ref{fig:three_particle_cut}), in comparison to individual diagrams,
\begin{align}
\raisebox{-28pt}{\includegraphics[scale=.3]{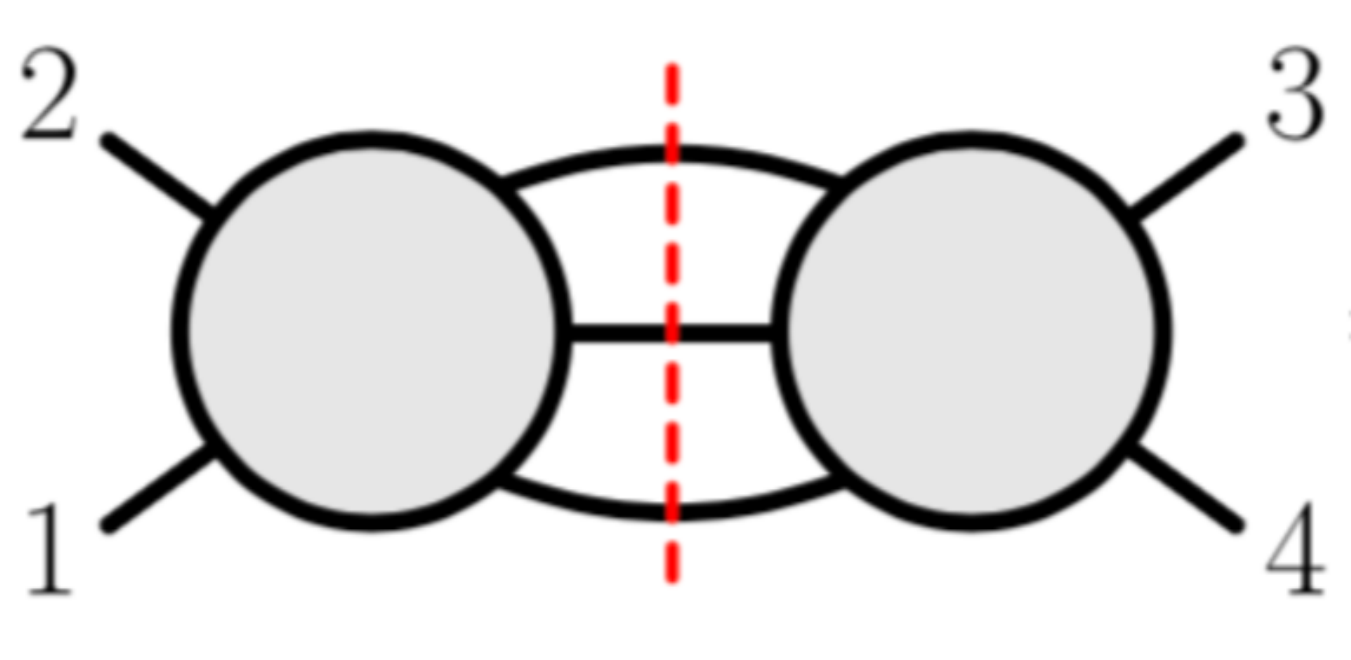}} \sim \frac{1}{z^5} \qquad vs. \qquad\raisebox{-28pt}{\includegraphics[scale=.3]{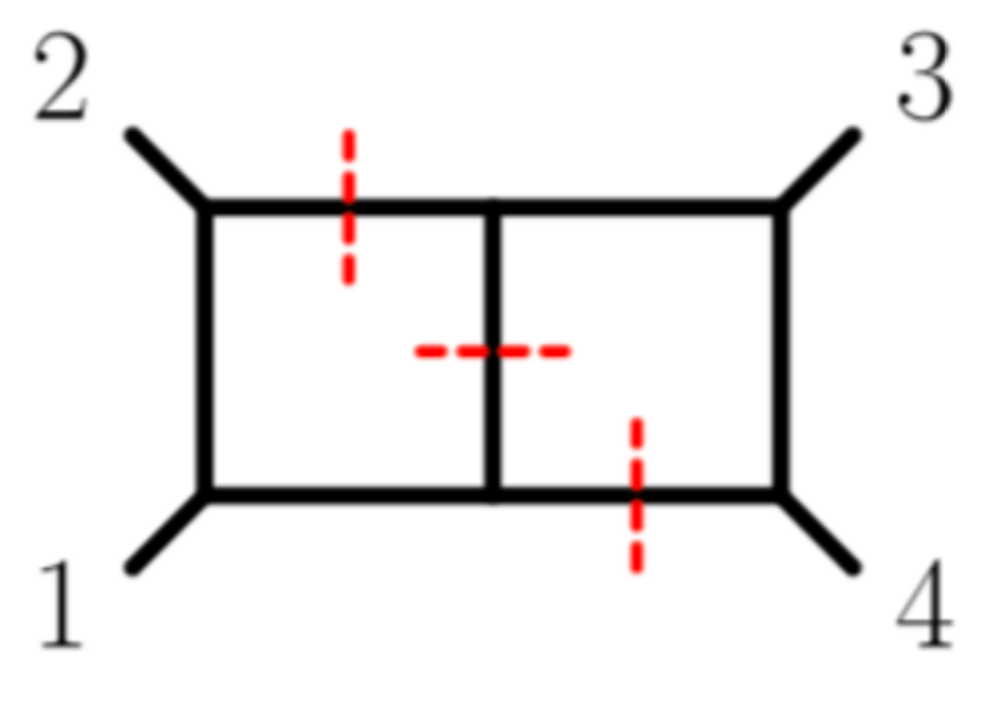}} \sim \frac{1}{z^4}
\end{align}
This procedure is only valid in $D=4$ because we critically use spinor helicity formalism and our ability to send only the holomorphic part of the cut loop momenta to infinity. It is likely that this cancelation at infinity is only valid for $D=4$ giving an evidence that the gravity integrands have special behavior only in this dimension. To substantiate this claim, one can perform a $D$-dimensional analysis of (\ref{fig:three_particle_cut}) by taking the Feynman integral expansion and plugging  three on-shell conditions $\ell^2_i=0$ as well as momentum conservation. With this prescription, one can express the cut (\ref{fig:three_particle_cut}), in terms of the following eight variables
\begin{align}
\big\{(\ell_1\cdot 1),(\ell_1\cdot 2),(\ell_1\cdot 3),(\ell_2\cdot 1),(\ell_2\cdot 2),(\ell_2\cdot 3),s,t\big\}\,.
\end{align}
If we analyze the cut  (\ref{fig:three_particle_cut}) for $(\ell_i\cdot j) \to z\, c_{ij}$ for some numerical constant $c_{ij}$, then we find that for $z\to\infty$ both the cut as well as individual integrals scale like $1/z^4$ so that there is no enhanced cancellation for $D$-dimensional kinematics, consistent with the analysis in \cite{Bern:2017lpv}. This shows that the four dimensional case is special as it allows to approach infinity in a way which leads to cancelations in gravity amplitudes.

\subsection{All-loop cut conjecture}
%
In the cut sequence discussed in subsec.~\ref{subsec:multi_unitarity_cut}, we can consider a general $L$-loop $(L+1)$-particle unitarity cut. This calculation would require sewing general graviton trees and performing the relevant state sums. In principle this is straightforward to do but can be a bit challenging in practice, especially because we want to make some all-loop statement. Here we pursue an alternative avenue and choose to cut additional propagators and consider a cut that can be easily computed to any loop order.  As should be familiar by now, we then compare the behavior at infinity of the on-shell function to the behavior of individual Feynman integrals. Naively this seems like a bad idea as cutting additional propagators deteriorates the large loop momentum scaling of the cuts. By definition the behavior of maximal cuts is as bad as that for individual diagrams. However, the logic here is as follows. If we see any improved behavior even for descendant cuts of the original $(L+1)$-particle unitarity cut discussed previously, it signals that the original $(L+1)$-cut itself, as well as the full uncut amplitude, if we are able to find good variables, have an improved UV behavior at infinity too (perhaps even further improved in comparison to the descendent cut due to additional cancellations).

There is an old example of a similar attempt to study the UV-structure of gravity scattering amplitudes in terms of multi-unitarity cuts by cutting an $L$-loop amplitude into a one-loop piece $\times$ a tree-level amplitude \cite{Bern:2006kd,Bern:2007xj} and potential cancellations were related to the "no-triangle property" of gravity at one-loop,

\begin{align}
\label{fig:one_loop_off_shell_cut}
\raisebox{-40pt}
{\includegraphics[scale=.48]{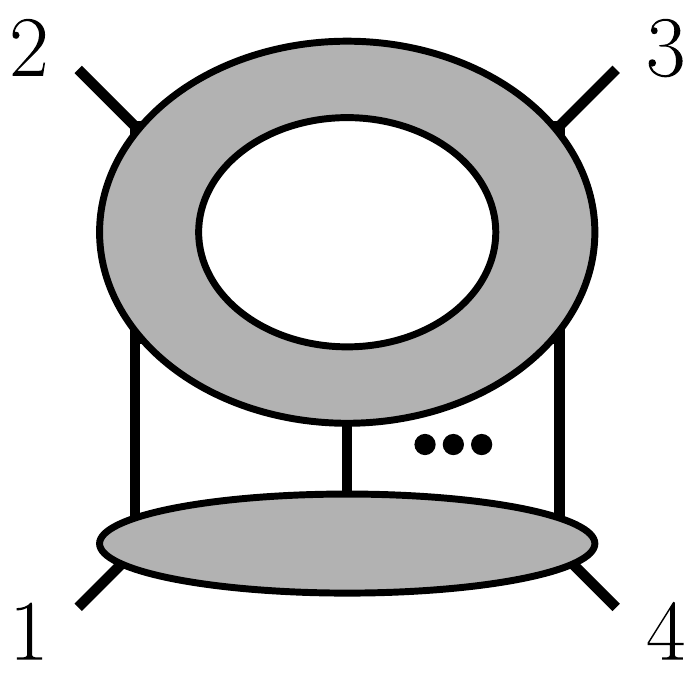}}\,,
\end{align}
which is to be compared to the scaling behavior of either the maximal cuts or iterated two-particle cuts. The original argument for cancellations given in \cite{Bern:2006kd,Bern:2007xj} starts from the observation, that the iterated two-particle cuts demand high powers of loop-momentum $\ell$ in the numerator of the associated local integrals,
\begin{align}
\raisebox{-50pt}
{\includegraphics[scale=.52]{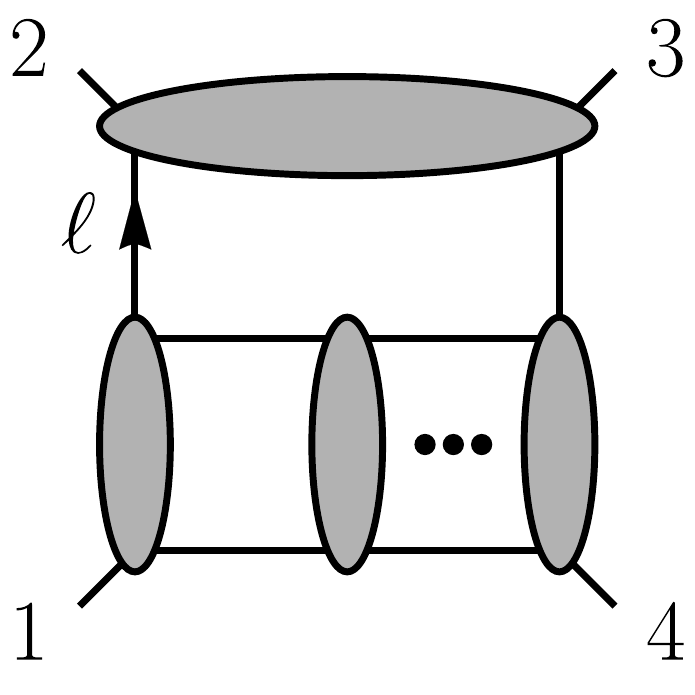}}
\Rightarrow 
N \sim stu \M^{(0)}_4(1234) \left[ t (\ell + p_1)^2\right]^{2(L-2)}\,.
\end{align}
When writing the $L$-loop amplitude as a sum of local integrals many different diagrams contribute on this cut including
\begin{align}
\label{fig:tennis_court_graph_high_loop}
\raisebox{-40pt}
{\includegraphics[scale=.43]{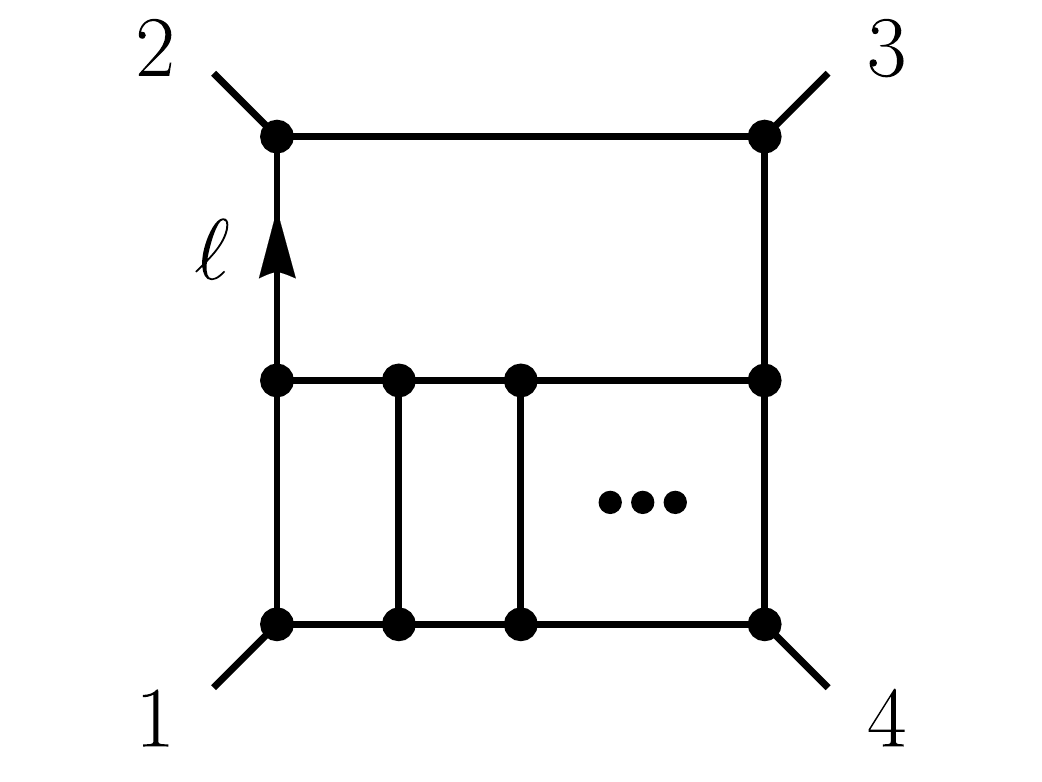}} 
\hskip -.5cm
\sim  \int \frac{d^4\ell \ \left[(\ell + p_1)^2\right]^{2(L-2)}}{\left[\ell^2\right]^{L+2}}\,,
\end{align}
which have high poles at infinity and violate the "no-triangle hypophysis". Based on this observation, it was argued that certain cancellations between diagrams had to occur. Note that $(\ell\cdot p_1)=(\ell+p_1)^2$ on the cut $\ell^2=0$ so either choice would lead to the same behavior. The difference between both numerator choices is their continuation off-shell. In this particular case, the worst behaved UV-terms can be pushed into contact terms in contrast to our later example (\ref{fig:uv_diagram}). The one-loop amplitude in (\ref{fig:one_loop_off_shell_cut}) has a box expansion and should therefore scale like $d^4\ell [\ell^2]^{-4}$ compared to the scaling of the higher-loop analog to the tennis-court integral of $d^4\ell [\ell^2]^{L-6}$. We see that, starting at $L=3$, there is a mismatch between the tennis-court integral and the one-loop box expansion. If one attempts to re-express the amplitude on the cut (\ref{fig:one_loop_off_shell_cut}) using a different set of Feynman integrals which manifest the box-type behavior of the uncut loop, one encounters some trouble. Even though the power-counting for the loop involving $\ell$ has been made manifest, this comes at a cost of spoiling the box-like UV-behavior on the other side of the diagram which can now involve triangles,

\vspace{-0.3cm}
\begin{align}
\label{fig:tennis_court_resummed}
\raisebox{-40pt}
{\includegraphics[scale=.43]{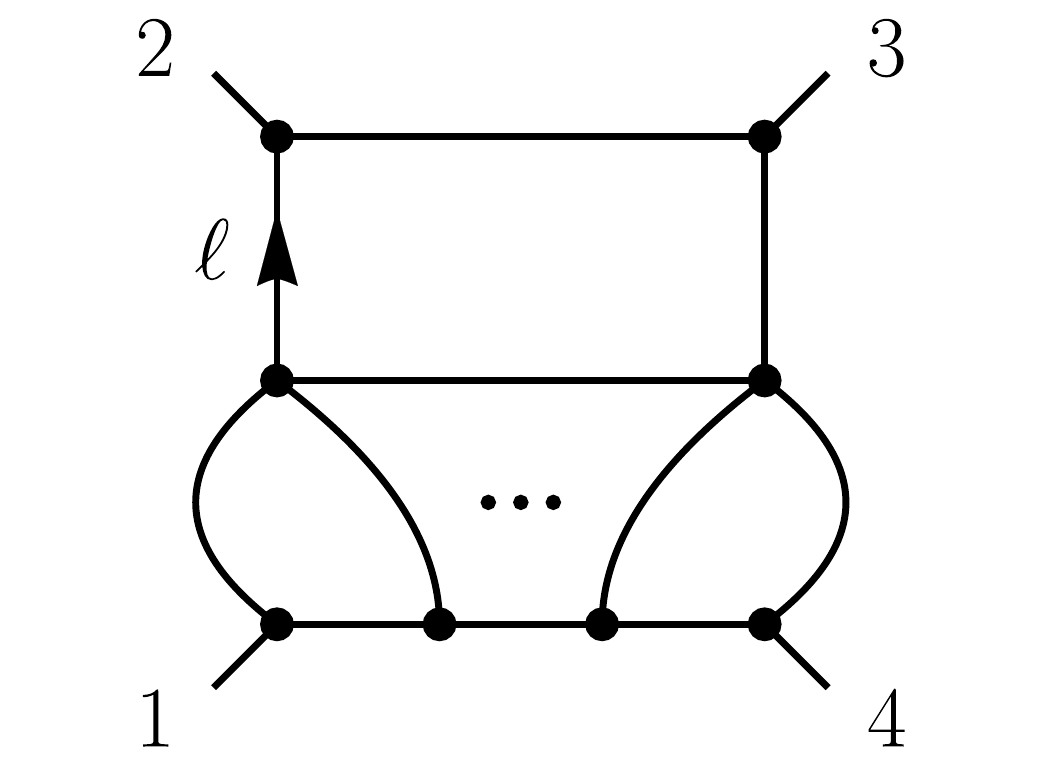}}\,.
\end{align}

There seems to be some irreducible problem which can be pushed back and forth between different loops. With a particular choice of representation, we can make the UV-behavior of an individual loop momentum manifest, but not all at the same time. Here, the problem is that we are approaching infinity via independent limits. The ultimate check if a given sum of Feynman integrals that contribute on a cut can show a global improvement of the UV-scaling is to send all loop momenta to infinity at the same time. Similar arguments apply to another cut where the bottom tree-level gravity amplitude in (\ref{fig:one_loop_off_shell_cut}) is replaced by another uncut loop. On this cut, the iterated two-particle analysis implies that some of the Feynman integrals need even higher tensor-power numerators with a worse UV-behavior. 

As mentioned before, at higher loops it becomes prohibitively complicated to determine the numerator structures of the Feynman integrals completely. Instead of summing all integrals which contribute on a given cut, we are going to focus directly on the on-shell function. It is the product of tree-level amplitudes which encodes the physical information independent of any particular integral expansion. We focus here on a cut that allows us to probe the simultaneous large loop-momentum scaling in all loops. At $L$-loops, the simplest cut to consider only involves $L+1$ three-point amplitudes whose helicity configuration is chosen (by picking a particular solution for the cut) such that the single $(L+3)$-point tree-level amplitude is forced to be in the MHV sector

\begin{align}
\label{fig:MHV_cut_sequence}
\raisebox{-45pt}{
\includegraphics[scale=.5]{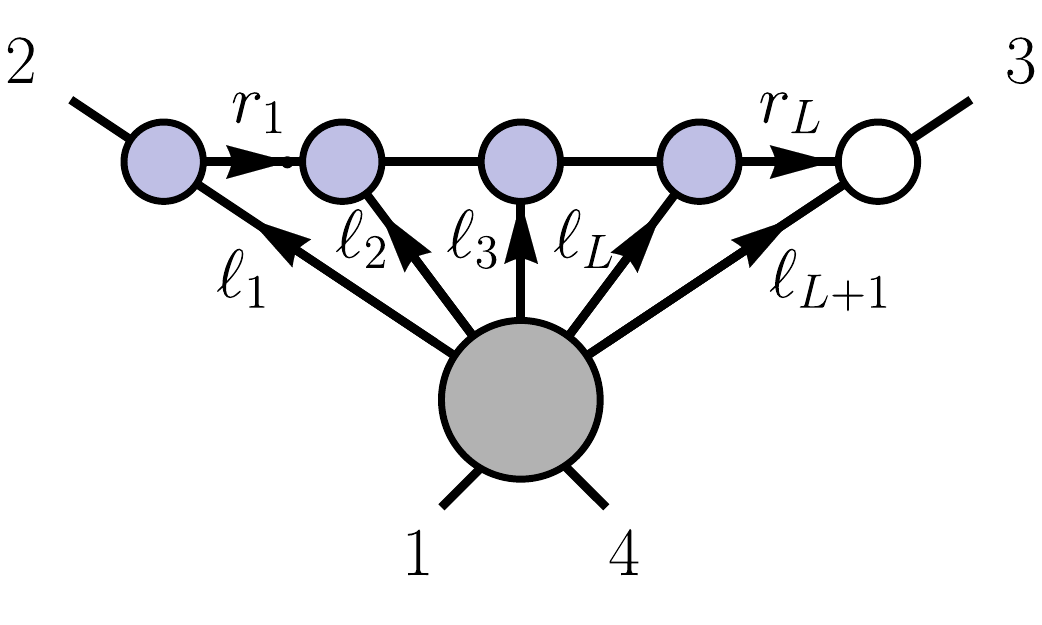}} \,.
\end{align}

This kind of cut is a natural $L$-loop generalization of the one-loop BCFW-cut (\ref{fig:one_loop_BCFW_cut}). Here we set $2L+1$ propagators on shell which leaves $2L-1$ unfixed degrees of freedom. These can be parameterized by $(L-1)$ two component spinors $\lam{x_i}\,,i=\{1,...,L-1\}$ and one additional parameter $\alpha$. The on-shell cut kinematics is therefore given by,
\begin{align}
\begin{split}
 \ell_i 	& = \lam{x_i}\lamt{2}	\,, \qquad  r_i = \sum^i_{j=1} \lam{x_j}\lamt{2} - \lam{2}\lamt{2}\,, \qquad\qquad\  i=\{1,...,L-1\} \\
 \ell_L 	& = \left(\lam{2}+\alpha \lam{3} - \sum^i_{j=1} \lam{x_j}\right)\lamt{2}\,, 
 			\quad r_L = \alpha \lam{3}\lamt{2} \,, 
			\quad \ell_{L+1} = \lam{3}(\alpha\lamt{2}-\lamt{3})\,.
\end{split}
\end{align}
In order to probe the UV-behavior of all loops simultaneously and in a uniform manner, we take $\alpha$ to be our scaling parameter and furthermore introduce a common reference spinor $\xi$ that allows us to shift all loops in the same direction,
\begin{align}
 \lam{x_i} = \lam{y_i} + \alpha\,\xi\,,
\end{align}
and ensures that all propagators are uniformly linear in $\alpha$ 
\begin{align}
 (\ell_i+\ell_j)^2\sim \alpha\,, \qquad  (\ell_i+p_k)^2\sim\alpha\,.
\end{align}
Equipped with this on-shell parameterization, it is now straightforward to compute the relevant on-shell function where the only nontrivial part is the $L+3$-point MHV tree amplitude. We are now in the position to compare the cut of the amplitude directly to the Feynman integral expansion in terms of the cubic graph representation in the known cases. The relevant $\alpha$-scaling as $\alpha \to \infty$ is summarized in table~\ref{tab:scaling_MHV_cut}. 
\begin{table}[ht!]
\centering
\begin{tabular}{|c|c|c|c|c|}
\hline
$L=$ 	& 2 & 3 & 4 & $L$ \\
\hline
\raisebox{-45pt}{
\includegraphics[scale=.4]{./figures/L_loop_MHV_unitarity_cut}} 		& $\alpha^{-2}$  &  $\alpha^{-3}$ &  $\alpha^{-4}$ &  $\alpha^{-L}$	\\
\hline
worst diagram	& $\alpha^{-2}$  &  $\alpha^{-1}$ &  $\alpha^{0} $ & 	$?$			\\
\hline
\end{tabular}
\caption{\label{tab:scaling_MHV_cut}Comparison of the UV-scaling of the gravity on-shell function (\ref{fig:MHV_cut_sequence}) and individual local integrals contributing on this cut as $\alpha\to\infty$. The on-shell parameterization is explained in the main text. We compare with the local BCJ-expressions; diagram $(h)$ at three loops from \cite{Bern:2010ue} and diagram $(35)$ of \cite{Bern:2010tq} are the worst behaved diagrams on this cut. The scalings are given without taking the overall Jacobian of the cut into account on both sides. }
\end{table}

In conclusion, we see that for higher loop orders, the cut sequence (\ref{fig:MHV_cut_sequence}) has an increasingly improved scaling behavior with $\alpha$, whereas individual integrals scale worse for higher loops. While we do not have control over all Feynman integrals to all loops the results in table \ref{tab:scaling_MHV_cut} demonstrate a clear trend showing the disparity between the cuts and individual integrals at infinity. In line with our previous discussions, we see that there must be comprehensive cancelations between Feynman integrals to match the correct $\alpha$ dependence of the on-shell function. We have also performed this cut analysis for pure gravity where it is necessary to specify the helicity configuration of external states. For this class of cuts there is a singlet- and non-singlet helicity configuration. In the singlet case, the exchanged internal helicity gravitons are the same in pure gravity as in $\N=8$ so that large $z$ scaling of pure gravity is identical to that of $\N=8$. For the non-singlet helicity configuration, there is a difference between the exchanged on-shell states and we have checked that maximal supersymmetry leads to the expected cancellation of eight powers of loop momentum. This result is in complete agreement with an earlier observation of Bern et al.~\cite{Bern:1998ug} of supersymmetry cancellations in certain MHV cuts.


We discuss one further all-loop cut which is a slight derivative of (\ref{fig:MHV_cut_sequence}),
\begin{align}
\label{fig:augmented_uv_cut}
\raisebox{-50pt}{\includegraphics[scale=.5]{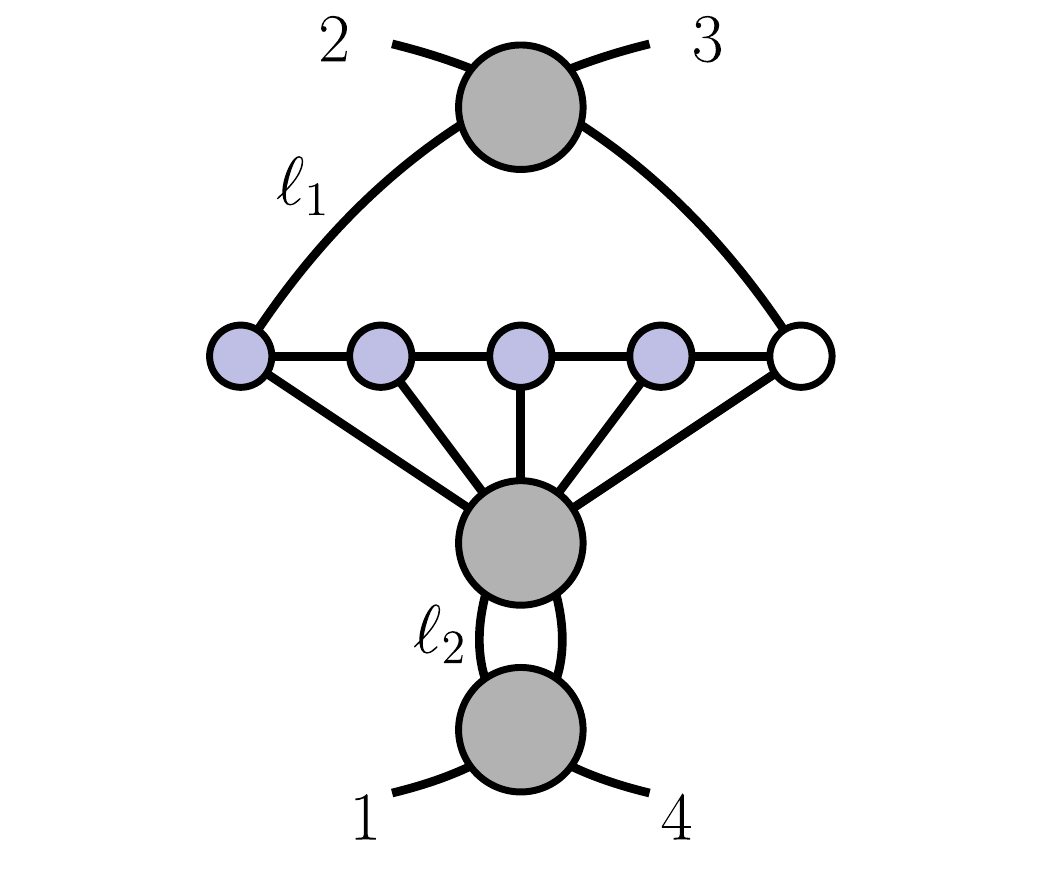}}\,.
\end{align}

The advantage of this cut is that we can compare it to particular integral topologies where the maximal cuts dictate the leading UV-behavior of the numerators to all loop orders in a simple iterative manner, see our discussion around (\ref{fig:os_func_uv_scaling_example}). 
Some of these topologies include the ones used in the literature as prototype UV-divergent integrals at higher loops.
\begin{align}
\label{fig:uv_diagram}
\raisebox{-53pt}{\includegraphics[scale=.5]{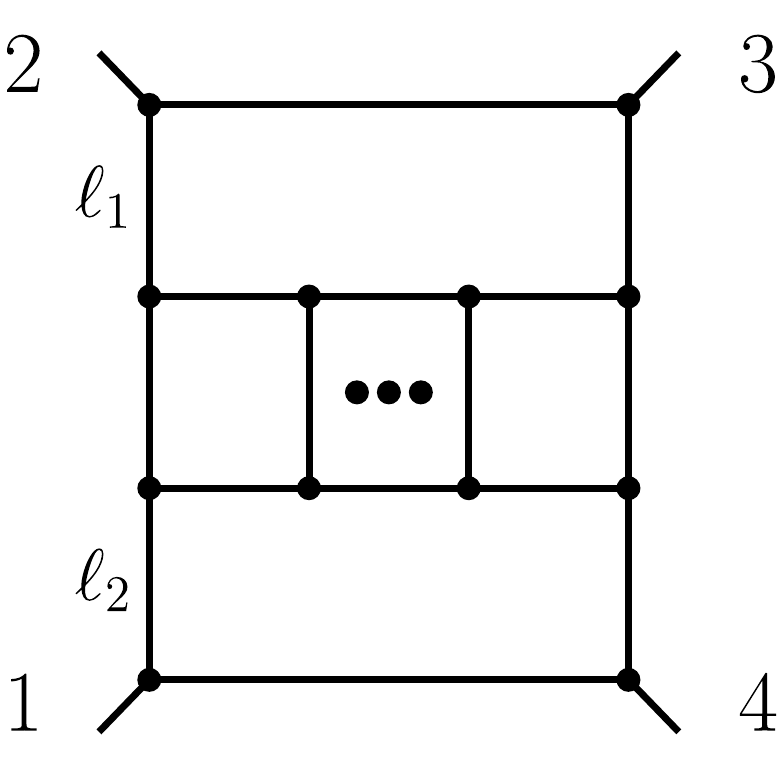}}
\quad \Leftrightarrow  \quad N \sim \left[(\ell_1\!+\!\ell_2)^2\right]^{2(L-3)}
\end{align}

From the point of view of poles at infinity, this integral has the same degree of pole as the one shown before (\ref{fig:tennis_court_graph_high_loop}) in the parameterization where all propagators scale uniformly. However, for integral (\ref{fig:tennis_court_graph_high_loop}) we can push the worst UV behaved piece into the contact term by rewriting the numerator. This is not possible for integral (\ref{fig:uv_diagram}) as any rewriting of the numerator by shuffling around contact terms still preserves the overall UV-scaling behavior of this integral. The numerator $\ell_1\cdot\ell_2$ represents an \emph{irreducible scalar product}, see e.g.~\cite{Grozin:2011mt}. 

\begin{table}[ht!]
\centering
\begin{tabular}{|c|c|c|c|c|}
\hline
$L=$ 	& 3 & 4 & 5 & $L$ \\
\hline
\raisebox{-35pt}{\includegraphics[scale=.3]{./figures/L_loop_MHV_unitarity_cut_augmented}} 		& $\alpha^{-10}$  	&  $\alpha^{-8}$ &  $\alpha^{-8}$ &  $\alpha^{-8}$				\\
\hline
\raisebox{-30pt}{\includegraphics[scale=.3]{./figures/bad_scaling_diag_label2}}					& $\alpha^{-5}$		&  $\alpha^{-4}$ &  $\alpha^{-3} $ &  $\alpha^{L-8}$				\\
\hline
\end{tabular}
\caption{\label{tab:scaling_MHV_uv_cut}Comparison of the UV-scaling of the gravity on-shell function (\ref{fig:augmented_uv_cut}) and the Feynman integrals (\ref{fig:uv_diagram}) contributing on this cut as $\alpha\to\infty$. The on-shell parameterization is explained in the main text. The scalings are given without taking the overall Jacobian of the cut into account on either side. }
\end{table}

In analyzing the UV-behavior of the on-shell function (\ref{fig:augmented_uv_cut}), we shift all the cut loop-momenta of the central part of the cut (which resembles (\ref{fig:MHV_cut_sequence})) proportional to $\alpha$ times some reference spinor $\xi$ which picks a certain direction. Parameterizing the cut loops associated to $\ell_1$ and $\ell_2$ we have some additional freedom. For the purpose of this analysis we parameterize the respective on-shell lines according to the appropriate relabelings of (\ref{eq:two_particle_cut_parameterization}) and identify $\beta\to\alpha,\ \delta\to\alpha$. In this parameterization $(\ell_1+\ell_2)^2\sim \alpha^2$ scales quadratically for large $\alpha$. Sending $\alpha\rightarrow \infty$ then sends all loop momenta of the core cut to infinity in the same direction as before and only the two special loops involving $\ell_1$ and $\ell_2$ are singled out. In this case, we summarize the relevant UV-scaling behavior in table~\ref{tab:scaling_MHV_uv_cut}. We again see that the behavior at infinity of the amplitude is better than for an individual Feynman integral. This improved UV-scaling of the cut then requires some elaborate cancelations between (\ref{fig:uv_diagram}) and other Feynman integrals. We schematically summarize all our findings in the table \ref{tab:summary_cuts}.

\begin{table}[h!]
\hspace{-.2cm}
\begin{tabular}{|c|c|c|c|} 
\hline
Object & Labels & Poles at infinity & Reality\\ \hline
\begin{tabular}{c}
Off-shell amplitude \\
\raisebox{-18pt}{\includegraphics[scale=.3]{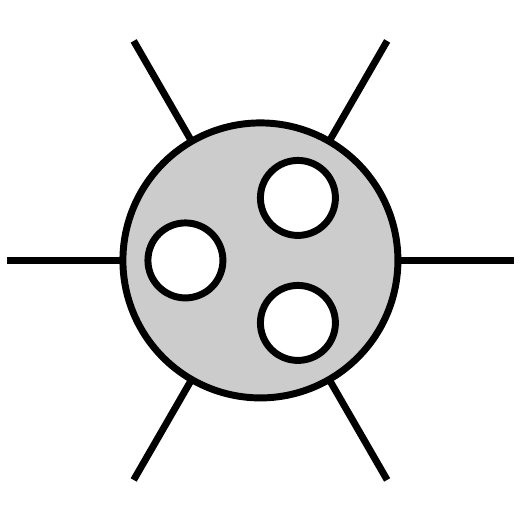}} 
\end{tabular}	& No & ??? & real \\ \hline
\begin{tabular}{c}
Multi-unitarity cut \\ 
\raisebox{-30pt}{\includegraphics[scale=.3,trim={0cm .5cm 0cm 1.1cm},clip]{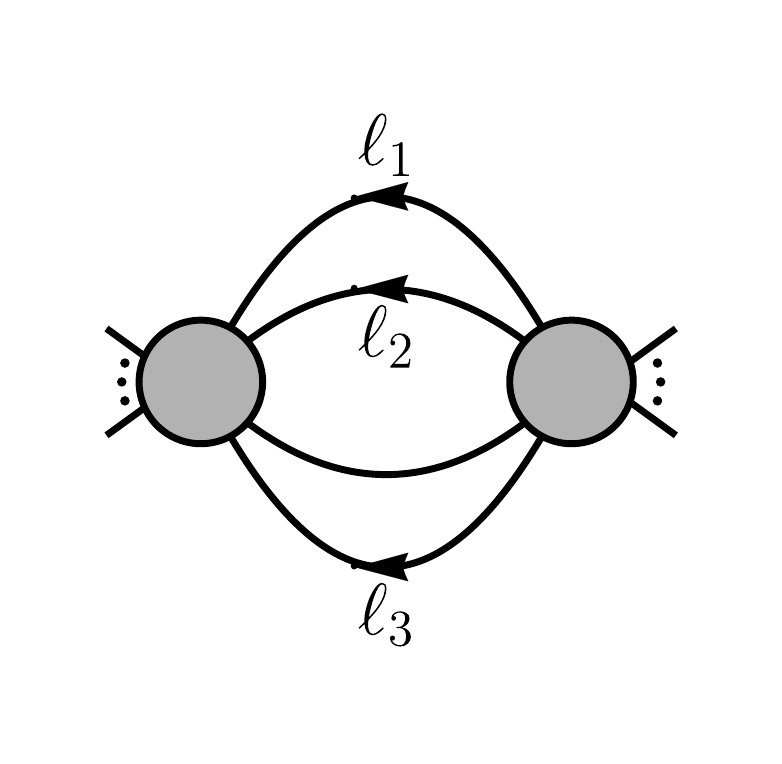}}
\end{tabular}					& Yes & \begin{tabular}{c} poles at infinity {\bf absent} \\ cancelations between diagrams \end{tabular} & real subregions 
\\ \hline
\begin{tabular}{c}
UV-probe cut \\
\raisebox{-25pt}{\includegraphics[scale=.3]{./figures/L_loop_MHV_unitarity_cut}}	
\end{tabular}	& Yes & \begin{tabular}{c} poles at infinity {\bf absent}\\ better scaling than diagrams \end{tabular} & complex \\ \hline
\begin{tabular}{c}
Maximal cut \\
\raisebox{-25pt}{\includegraphics[scale=.25]{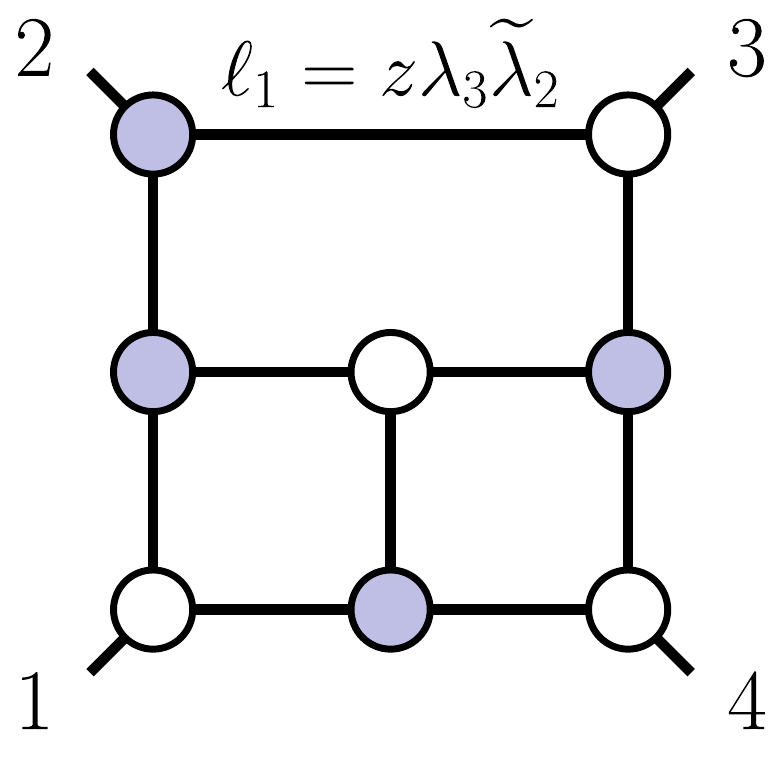}}	
\end{tabular}	& Yes & \begin{tabular}{c} poles at infinity {\bf present}\\ single diagram contributes \end{tabular} & complex \\ \hline
\end{tabular}
\caption{\label{tab:summary_cuts}Summary of various classes of on-shell functions discussed in our work. The second column marks whether or not a given cut defines unique loop labels. In the last column we indicate if the loop momenta are real or complex. For the full non-planar integrand the loop momenta are unconstrained and therefore can be chosen to be real. For the multi-unitarity cut we restrict them by on-shell conditions but there still exist real subregions and for the maximal cuts all loop momenta are complex. The speculation is that this might be important when relating the poles at infinity to UV structures in final amplitudes after integration (coefficient of divergencies etc).}
\end{table}

\subsection{Beyond the cuts}

In this subsection we want to provide evidence that the unexpected behavior is present even for the cases when not all loop momenta are cut. This seems contradictory to our earlier statement that all loop momenta should be cut in order to produce good variables. This of course still holds but we can use the trick: if we cut $L-1$ loop momenta the remaining object depends on one uncut loop momentum. In that case we can try to expand it using one-loop objects which are just boxes and pentagons and ``undo" the choice of variables by marginalizing over all labels of the uncut loop. The crucial part here is that we do not talk about the cut integrand when one of the loops is uncut -- that object is not well defined, but rather we talk about the box expansion of the cut of higher loop amplitude which is well defined. We will show it for the two-loop four- and five-point amplitudes. At higher loops one has to do a similar procedure or circumvent the problem by not choosing labels for the remaining loops and develop a symbolic diagrammatic expansion similar to the one that has been given for a particular collinear analysis in \cite{Herrmann:2016qea}.
\vskip -.9cm
\begin{align}
\label{fig:2loop_4pt_off_shell_triple_cut}
\raisebox{-45pt}
{\includegraphics[scale=.45]{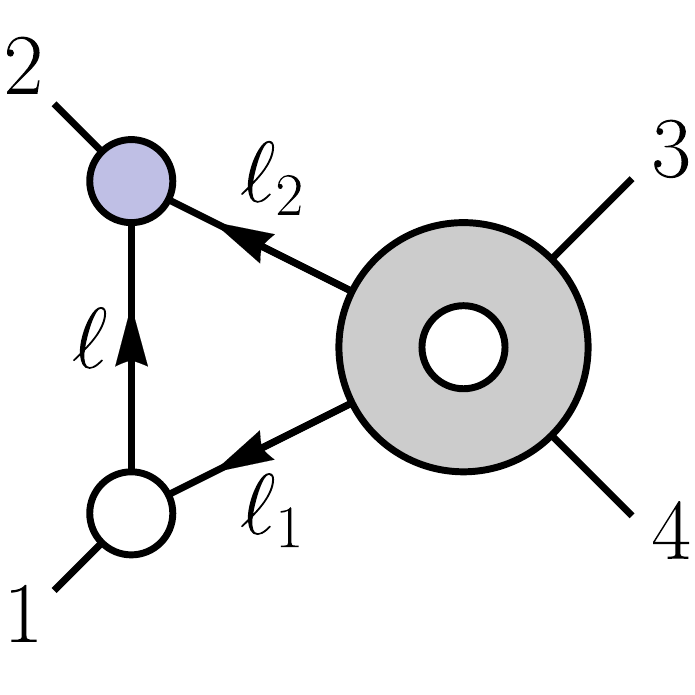}}
\end{align}
\vskip -.3cm
In this figure, we leave the second-loop momentum unspecified, but in a "would-be" box expansion, three classes of box-integrals can contribute,
\vskip -.7cm
\begin{align}
\raisebox{-40pt}
{\includegraphics[scale=.50]{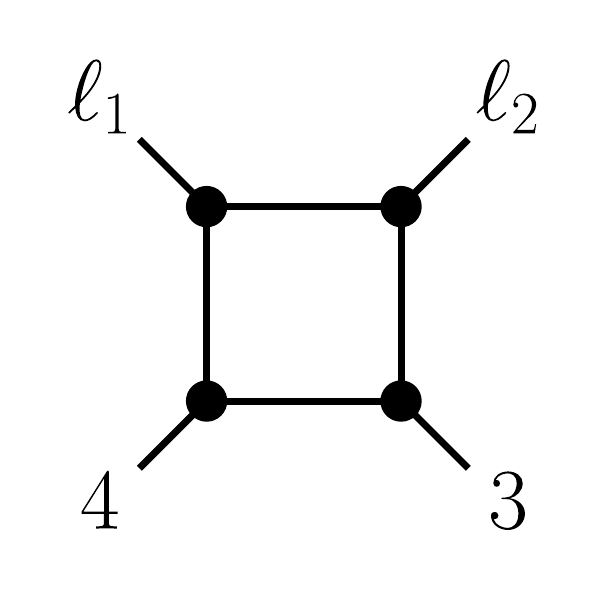}}\,,
\quad
\raisebox{-40pt}
{\includegraphics[scale=.50]{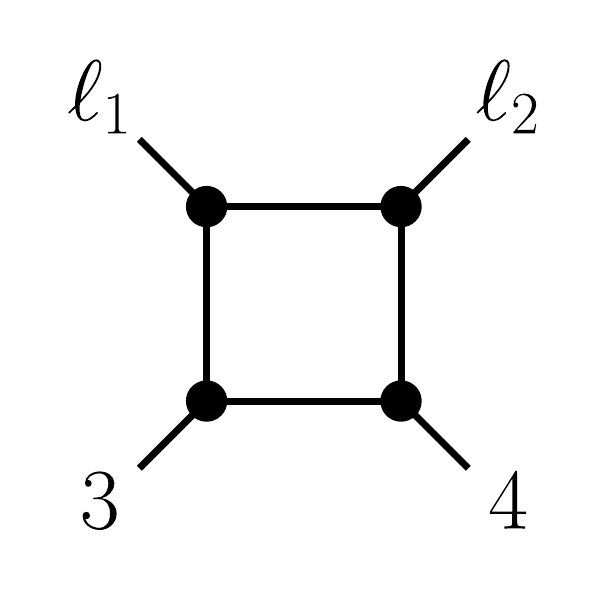}}\,,
\quad
\raisebox{-40pt}
{\includegraphics[scale=.50]{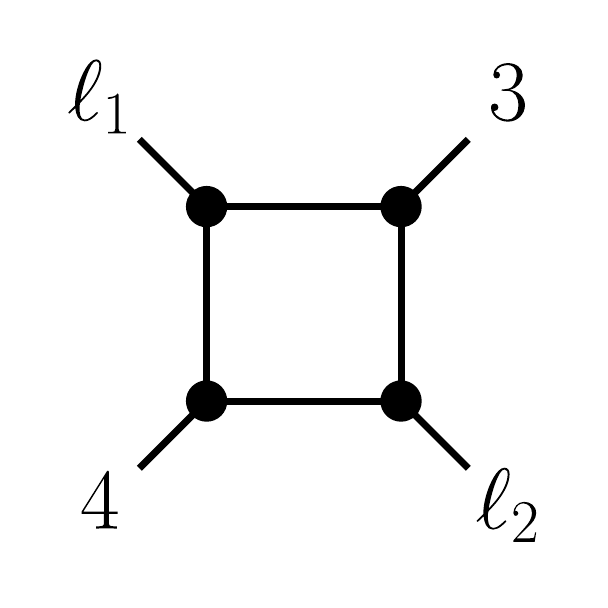}}\,.
\end{align}
\vskip -.4cm
The explicit solutions for the on-shell momenta $\ell_1 = \lam{1}(\lamt{1}+z \lamt{2})$ and $\ell_2 = (\lam{2}-z\lam{1})\lamt{2}$ are also z-dependent. In order to see the $1/z^3$ behavior, we introduce labels for these boxes in such a way that the unspecified loop-momentum $q$ is attached to all possible legs and in both possible directions. This gives a sum of 24 such boxes and we can write the one-loop subamplitude as
\vskip -.7cm
\begin{align}
\label{fig:one-loop-diagrammatic-sum}
 \M^{(1)}_4(\ell_1,\ell_2,3,4) = s \hat{t} \hat{u}\ \M^{(0)}_4 (\ell_1,\ell_2,3,4) \times \raisebox{-40pt}{\includegraphics[scale=.5]{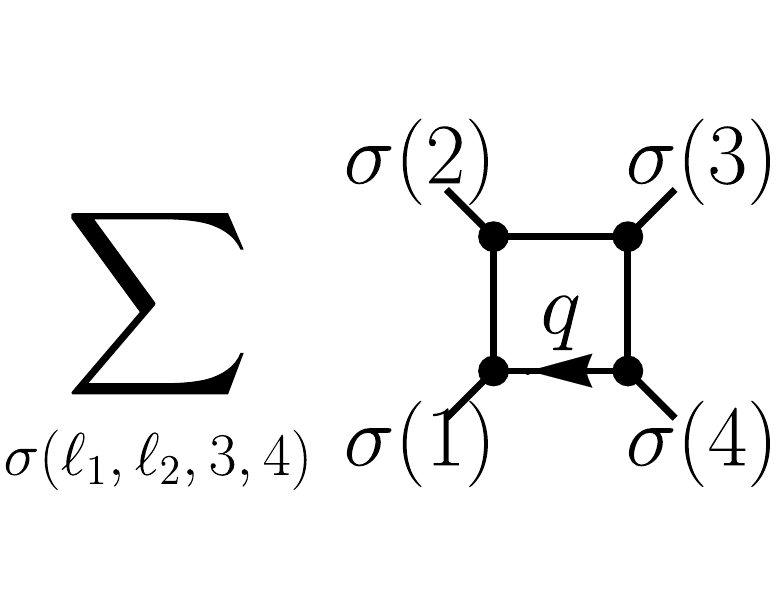}}\,,
\end{align}
where $\hat{t} = (\ell_1+p_4)^2$ and $\hat{u} = (\ell_1+p_3)^2$ and $s \hat{t}\hat{u} \M^{(0)}_4 (\ell_1,\ell_2,3,4)$ is totally crossing symmetric. Once we take into account the Jacobian of the cut, $\J_{\text{cut}}=1/(z\ s)$, evaluate the product of three-point functions and perform the supersymmetric state-sum the prefactor scales like $1/z$. Upon summing all box diagrams in (\ref{fig:one-loop-diagrammatic-sum}), one can check that the assembled cut indeed scales like $1/z^3$ while individual boxes scale like $1/z^2$,
\begin{align}
\raisebox{-45pt}
{\includegraphics[scale=.50]{./figures/2_loop_4pt_triple_cut_one_loop_uncut_4pt.pdf}}
= \int \frac{dz}{z} s^2 \hat{t} \hat{u}\ M^{(0)}_4(\ell_1,\ell_2,3,4)  \times \hskip -.4cm\raisebox{-40pt}{\includegraphics[scale=.5]{./figures/boxsum_triple_cut_2_loop.pdf}}\sim \frac{dz}{z^3}
\end{align}
This is the first nontrivial example that supports our more general claim of the $1/z^3$ scaling of triple cuts of $\N=8$ supergravity integrands.

We have also analyzed the two-loop five point amplitude on the triple-cut equivalent to (\ref{fig:2loop_4pt_off_shell_triple_cut})
\begin{align}
\label{fig:2loop_4pt_off_shell_triple_cut}
\raisebox{-50pt}
{\includegraphics[scale=.50]{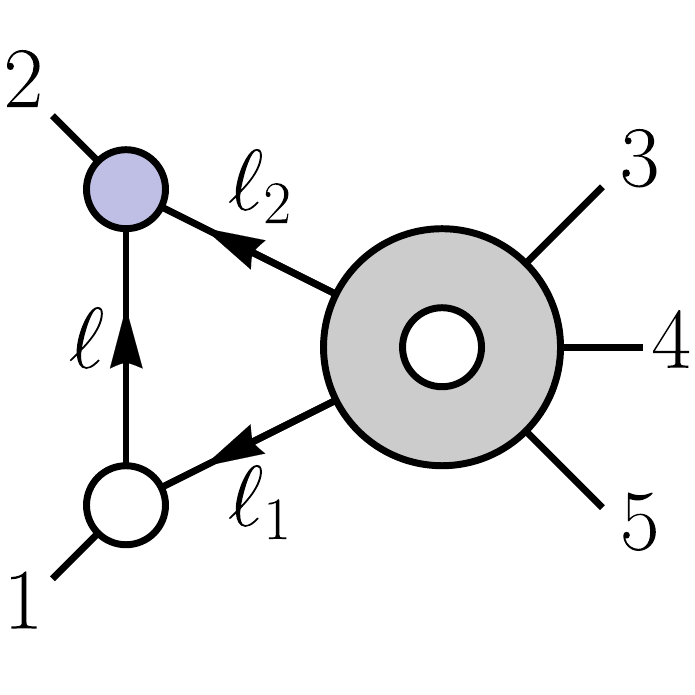}}\,,
\end{align}
where we require the one-loop expansion of the five-point ``integrand``. Here, the box expansion is insufficient as it only takes into account a parity-even combination of the full integrand. Instead, we use the representation given in \cite{Carrasco:2011mn} in terms of boxes and pentagons. If one symmetrizes over the possible choices of labelling the off-shell loop momentum $\ell_3$ in the one-loop five-point diagrams, one indeed obtains the claimed $1/z^3$ behavior of the triple cut in contrast to $1/z^2$ of individual boxes and pentagons.

\section{From IR to UV}
\label{sec:ir_to_uv}

In this section we review some facts about the IR properties of gravity integrands first observed in \cite{Herrmann:2016qea}, and study their interplay with the UV behavior of gravity on-shell functions discussed in the previous section. We will conjecture that combined the IR and UV constraints completely fix the gravity integrands making both constraints an important input for a possible geometric interpretation and promoting gravity to the equivalent level as nonplanar Yang-Mills theory.

\subsection{IR of gravity from cuts}

Due to the higher derivative nature of the gravity action the infrared divergences of gravity at loop level are very mild. For Yang-Mills scattering amplitudes, the leading IR-divergences of an $L$-loop amplitude calculated in dimensional regularization starts with the leading $1/\epsilon^{2L}$-term in the $\epsilon$-Laurent expansion. In contrast, the leading term in gravity is only 

\begin{equation}
\M^{(L)} \sim \frac{1}{\epsilon^L}\,. 
\label{IRdiv}
\end{equation}

The mild IR behavior of integrated gravity amplitudes can be nicely understood from properties of the on-shell functions at integrand level already. In order to draw this connection, we have to elaborate on the particular regions of loop-momentum integration where infrared divergencies can in principle arise. The first possibility for IR-divergencies comes from \emph{collinear regions} where the internal loop momentum is proportional to one of the external momenta, e.g. $\ell = \alpha p_1$. At the level of on-shell functions, this region is associated to cuts isolating a single massless external leg. 
\begin{align}
\label{fig:coll_cut_L_loop}
\underset{\ell^2=0=\sqb{\ell1}}{\text{Res}\ \I} = \raisebox{-31pt}{\includegraphics[scale=.55,trim={0cm .5cm 0cm .5cm},clip]{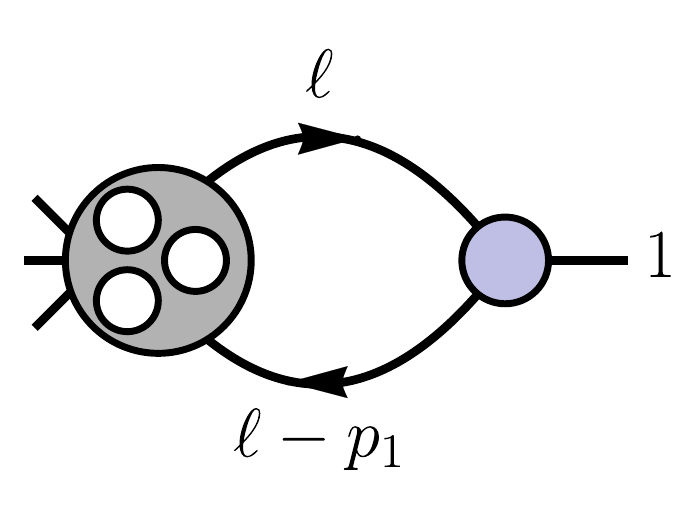}}
\end{align}
First, we put $\ell^2\!=\!0$ on shell where the loop momentum factorizes into a product of spinor-helicity variables $\ell = \lambda_\ell \widetilde{\lambda}_\ell$. Due to this factorization of $\ell$, if the following propagator-momentum differs from $\ell$ by a massless external momentum, say $p_1=\lam{1}\lamt{1}$, then $(\ell -p_1)^2$ also factorizes
\begin{equation}
(\ell - p_1)^2 = \ab{\ell 1} \sqb{\ell 1}\,.
\end{equation}
In order to approach the collinear region we have to set both factors to zero, $\ab{\ell 1} = \sqb{\ell 1} = 0$ which localizes $\ell = \alpha p_1$. Note that we are still cutting only two propagators $\ell^2$, $(\ell-p_1)^2$ but we impose three constraints. This residue can be thought of as cutting two propagators and a Jacobian. The relation between the residue of ${\cal I}$ on this cut and the IR divergence of the one-loop amplitude $\M^{(1)}$ is as follow: if $\I$ has a non-zero residue on a collinear cut $\ell = \alpha p_k$, \emph{and} there is an additional pole corresponding to a {\it soft-collinear} singularity, $\alpha=0$ or $\alpha=1$, for which either $\ell =0$ or $\ell-p_k = 0$, the combined IR-divergence of the amplitude is $\frac{1}{\epsilon^2}$. If the residue on $\ell=\alpha p_k$ is non-zero but there are no further poles for $\alpha=0$ or $\alpha=1$ there is only a collinear divergence $\frac{1}{\epsilon}$. For gravity, the cut integrand $\I_{cut}$ vanishes in the collinear region so that 
\begin{align}
\label{eq:coll_cut_approach}
 \I_{cut}=  \underset{\ell^2=\sqb{\ell 1}=\ab{\ell 1}=0}{\text{Res}\ \I}  = 0\,,
\end{align}
but there is still an IR divergence. This is a {\it soft} IR divergence which originates from a soft BCFW-type pole 
\vskip -1cm
\begin{align}
\raisebox{-48pt}{\includegraphics[scale=0.65, trim={0cm .5cm 0cm .8cm},clip]{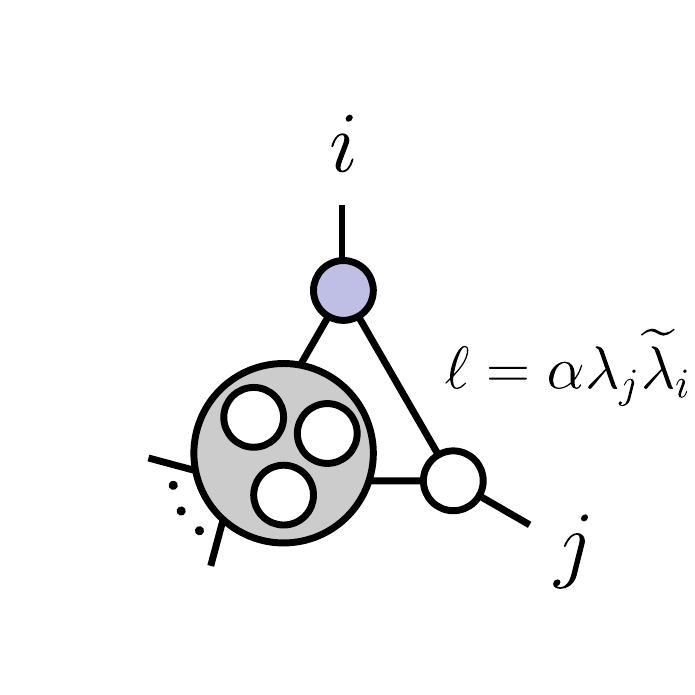}}
\hskip .1 cm
 = & \int\!\! \frac{d\alpha}{\alpha} f(\alpha) \,, \label{BCFWpole}
\end{align}
\vskip -.5cm
where the on-shell function has an explicit residue at $\alpha\!=\!0$ which corresponds to the soft region $\ell = 0$. If only such a pole is present at integrand level, upon integration this corresponds to a soft $\frac{1}{\eps}$ singularity of the amplitude. 

We can apply this logic to the one-loop four-point ${\cal N}=8$ supergravity amplitude (\ref{eq:one-loop-gravity-amp}) written in terms of three box integrals. On the collinear cut (\ref{eq:coll_cut_approach}) all box-integrals (\ref{eq:one-loop-gravity-amp}) contribute since each of them has a massless $p_1$-corner. The corresponding residue for one of the scalar boxes is:
\begin{align}
\raisebox{-50pt}{
\includegraphics[scale=.5]{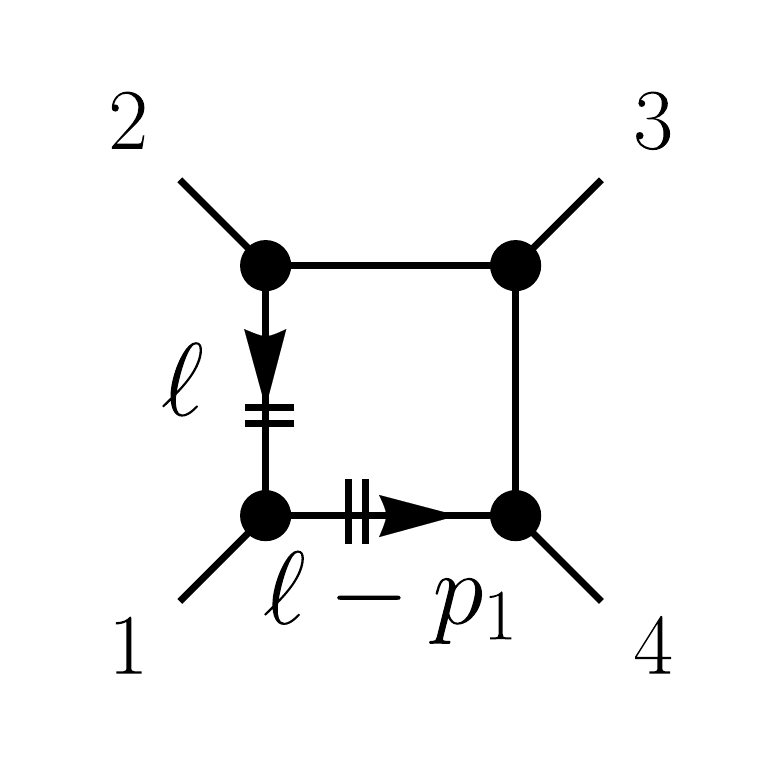}} \hskip -.8cm = \int\!\! \frac{d\alpha}{\alpha (1-\alpha)st} \,, \qquad \ell^\ast = \alpha p_1\,.
\end{align}
Analogous expressions for the other box integrals (\ref{eq:one-loop-gravity-amp}) can be obtained by appropriate relabelings of external momenta. Because of the presence of the $\alpha=1$ and $\alpha=0$ poles we conclude that all three integrals have soft-collinear singularities and are $\frac{1}{\epsilon^2}$ divergent. However, the combination of boxes which appears in the amplitude has a vanishing residue on the collinear pole at integrand level
\begin{align}
\underset{\ell^2=\sqb{\ell1}=\ab{\ell1}=0}{\large{\text{Res}}}\hskip -.4cm \I^{(1)}(1234) = \quad 0 \quad =  \hskip -.5cm
 \raisebox{-50pt}{
\includegraphics[scale=.5]{./figures/box1234_coll_cut_1}}
\hskip -.5cm + \text{perm}(234)\,,
\end{align}
where $+$perm$(234)$ instructs to sum over the $3!$ permutations  of the external legs 2,3, and 4. This explains why the $\frac{1}{\epsilon^2}$ divergence cancels as a consequence of the absence of the soft-collinear singularity. The subleading $\frac{1}{\epsilon}$ divergence then comes from the soft region --corresponding to the $\alpha=0$ residue-- where the on-shell function (\ref{BCFWpole}) has support. The generalization to higher loops is not straightforward as there are more intricate, possibly overlapping IR regions. However, the mild IR divergence (\ref{IRdiv}) indicates that the conspiracy in the collinear regions must still hold. 

In fact, we found in \cite{Herrmann:2016qea} based on the detailed analysis of gravity on-shell diagrams that the gravity loop integrands satisfy even more intricate collinear conditions. In particular, if the cut isolates three massless on-shell legs in one vertex and consequently sets $\lam{\ell_1}\sim \lam{\ell_2}\sim\lam{\ell_3}$, the on-shell function produces a factor $[\ell_1\ell_2]$ in the numerator
\begin{align}
\raisebox{-30pt}{
\includegraphics[scale=.35]{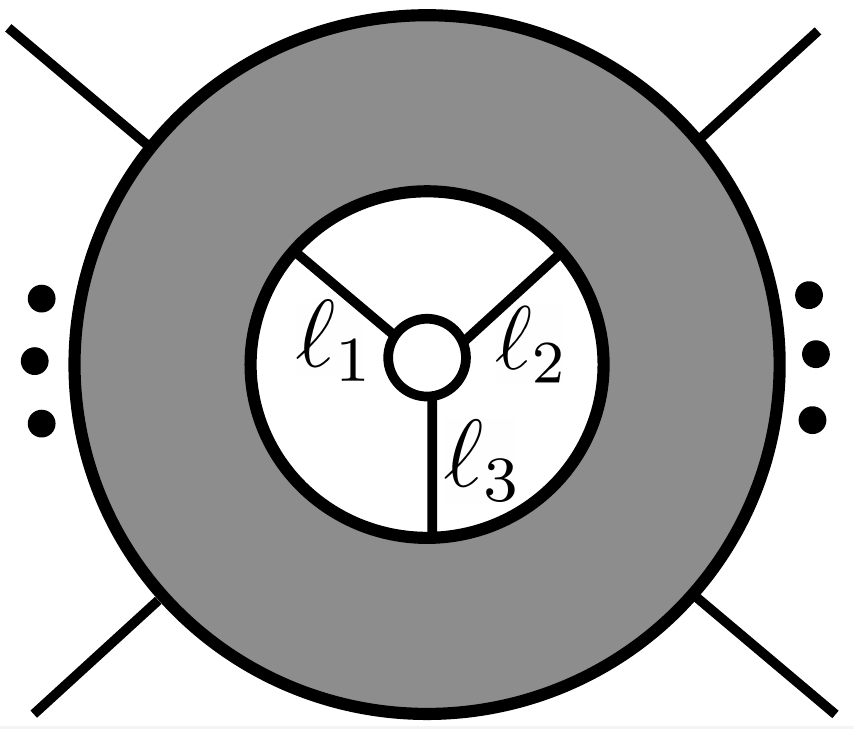}}
\sim \frac{\sqb{\ell_1\ell_2}}{\ab{\ell_1\ell_2}} \times \text{Regular}\,.
\label{IRcol}
\end{align}
An analogous statement also holds for the parity conjugate cut solution. The absence of collinear singularities discussed earlier is a special case when one of the legs is an external momentum $p_k$. The residue of the loop integrand on the cut $\ell^2=\ab{\ell 1}=0$ is
\begin{equation}
\underset{\ell^2=0=\ab{\ell1}}{\text{Res}}\M_n^{L-\text{loop}}  = [\ell 1] \times \text{Regular}
\end{equation}
Not only is there no further pole in $[\ell 1]$ but the integrand even vanishes for the kinematic configuration where both $\ab{\ell1}$ and $\sqb{\ell 1}$ are zero, i.e. $\ell=\alpha p_1$.  This collinear vanishing of the integrand goes beyond the absence of IR divergences discussed earlier. 

Another special case of the general setup (\ref{IRcol}) is when two of the legs are external momenta; then this reduces to the statement about collinear splitting functions in gravity \cite{Bern:1995ix,Bern:1998xc}, where the two collinear legs carry momentum $p_1=zP$ and $p_2=(1-z)P$. Here we give one example of the splitting function where two positive helicity gravitons split into a negative helicity graviton in an all outgoing convention
\begin{align}
 \text{Split}^{\text{gravity}}_{-} (z, 1^{+},2^{+}) & = \frac{-1}{z(1-z)} \frac{\sqb{12}}{\ab{12}}\,,
\end{align}
and again observe the characteristic $\sqb{12}/\ab{12}$ factor.

However, we would like to stress that our statement is much more general and also includes situations where all three on-shell momenta $\ell_1$, $\ell_2$, $\ell_3$ are internal. One simple example of this more general statement is the following on-shell function,
\begin{align}
\raisebox{-37pt}{\includegraphics[scale=.5]{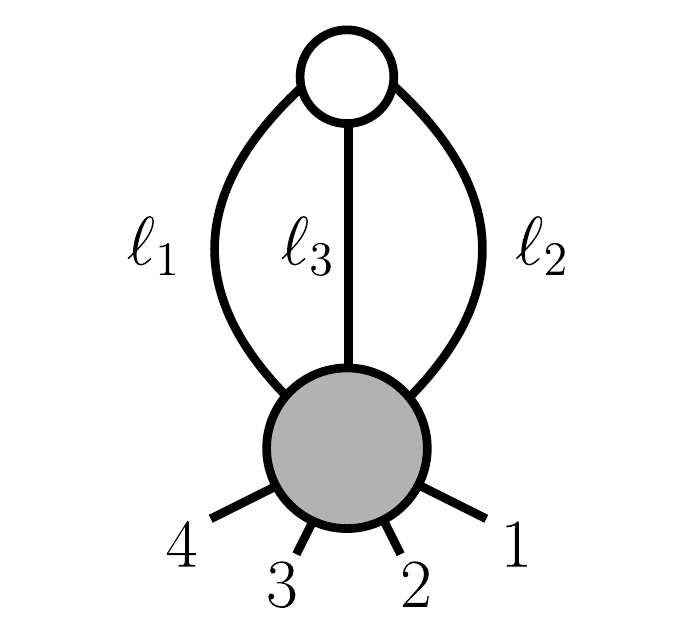}} \quad
\Leftrightarrow
\quad
\raisebox{-37pt}{\includegraphics[scale=.5]{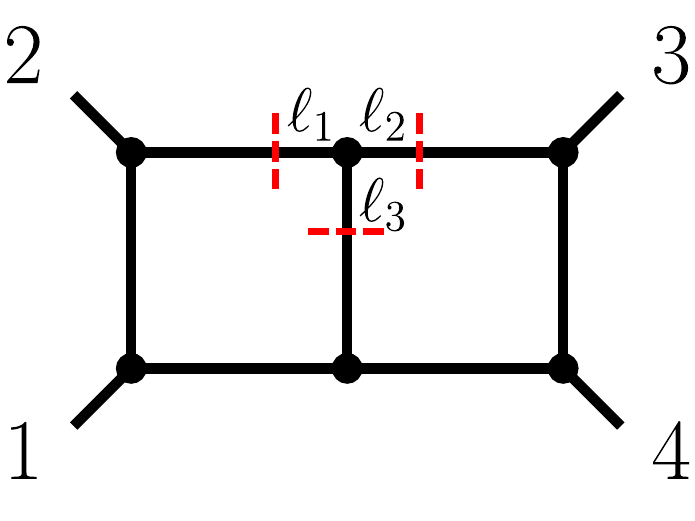}} + \cdots\,,
\end{align}
where we have also included an example of a Feynman integral that contributes to this cut for illustrative purposes.
While this collinear property was originally derived from on-shell diagrams, we now have a lot of evidence \cite{Herrmann:2016qea} that it also applies to more general on-shell functions.
\subsection{Daisy cut}
\label{subsec:daisy_cut}

In this subsection, we discuss a special class of cuts that ties both UV and IR properties together. In all these examples the IR property of the collinear conditions (\ref{IRcol}) force the on-shell function to vanish in the collinear limit which already requires intricate cancelations between contributing Feynman integrals in the local representation. On the other hand we also find nontrivial cancelations on the same cuts when one approaches the opposite large loop momentum limit. This exposes a fascinating tension between the UV and IR regions that go against one another. From the IR perspective, the numerators favor higher powers of loop momentum, whereas the absence of certain poles at infinity prefer lower powers of loop momentum. These two properties of gravity are therefore similar in spirit to the $\dlog$ and \emph{no poles at infinity} properties of Yang-Mills that also competed against one another to almost uniquely constrain the nonplanar loop integrand \cite{Bern:2015ple}. 

\vspace{-0.4cm}

\begin{align}
\raisebox{-60pt}{
\includegraphics[scale=.5]{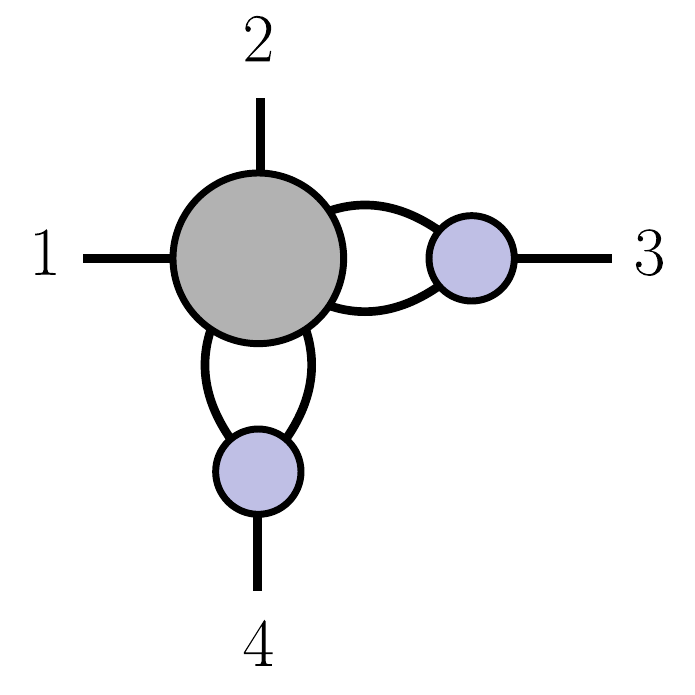}
\raisebox{1.5pt}{\includegraphics[scale=.5]{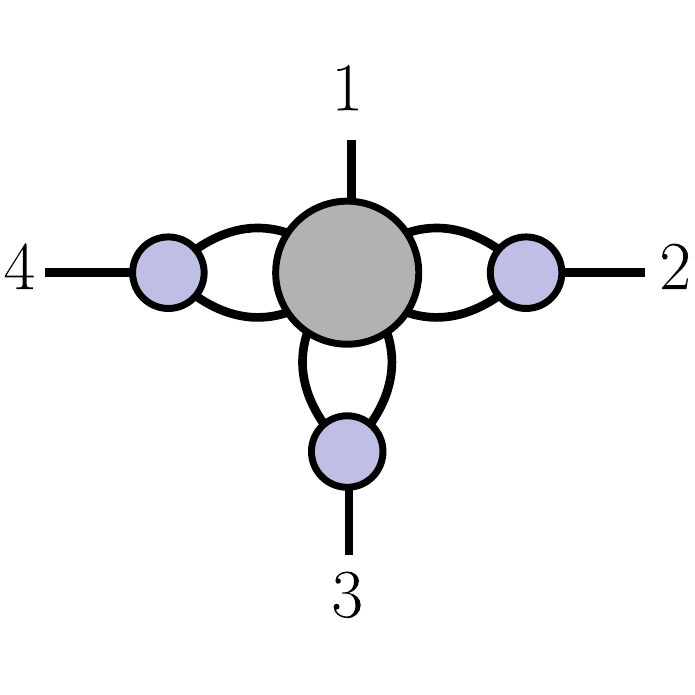}}
\raisebox{11pt}{\includegraphics[scale=.5]{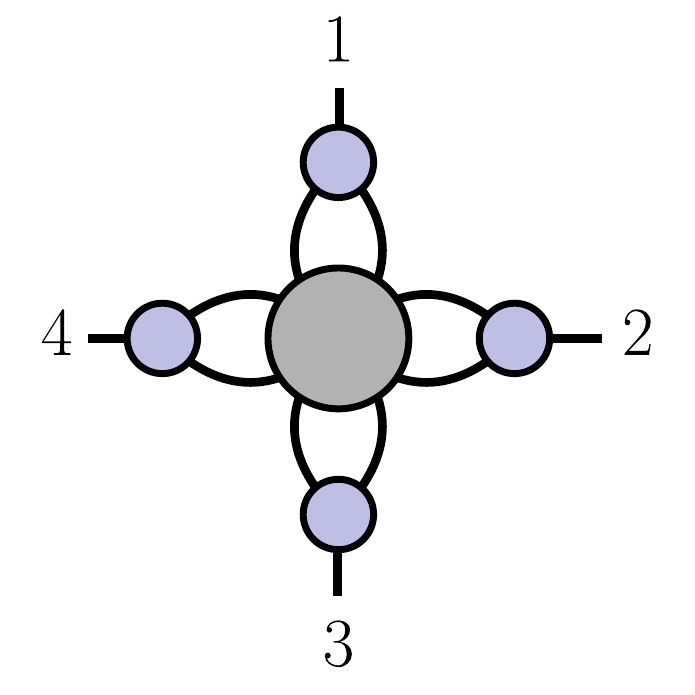}}}
\end{align}

In the following calculations we are not taking into account the respective Jacobians of the cuts. All formulae represent only the product of tree amplitudes evaluated on the on-shell kinematics as described. Since we always compare on-shell functions to the same cuts of the Feynman integral expansion of the amplitude, these Jacobian factors drop out and a comparison is meaningful. We only discuss two representative examples that clearly show the conflicting scaling requirements of both the IR- and UV regions. We have also worked out other examples but suppress them for the sake of brevity. 

In the two-loop case we parametrize the cut loop momenta to make the collinear vanishing properties of gravity manifest when $\beta_i\to0$.
\begin{align}
\label{figeq:MHV_bubble_cut_2}
\raisebox{-55pt}{\includegraphics[scale=.55]{./figures/bubble_cut_2_loop} } 
\hskip .5cm
\begin{cases}
\ell_1 = (\alpha_1 \lam{3} + \beta_1 \xi) \lamt{3} \\
\ell_2 = ((\alpha_1-1)\lam{3} + \beta_1 \xi) \lamt{3} \\
\ell_3 = (\alpha_2 \lam{4} + \beta_2 \xi) \lamt{4} \\
\ell_4 = ((\alpha_2-1)\lam{4} + \beta_2 \xi) \lamt{4} 
\end{cases}
\,.
\end{align}
For later convenience, we introduce a common reference spinor $\xi$. In this parametrization, the collinear vanishing condition (\ref{IRcol}) is reflected in the fact that the numerator of the on-shell function must depend quadratically on the $\beta$s (since we drop Jacobians), 
\begin{align}
 \text{cut} \sim \beta^2_1 \beta^2_2\  \overline{\text{cut}}\label{daisycut2}\,,
\end{align}
where $ \overline{\text{cut}}$ is regular for $\beta_i=0$. We can now approach the pole at infinity for all loop momenta by setting $\beta_i = \alpha x_i$ and sending $\alpha\rightarrow\infty$. In that case, we find the improved behavior of the on-shell function compared to individual integrals,
\begin{align}
 \raisebox{-30pt}{\includegraphics[scale=.3]{./figures/bubble_cut_2_loop} } \sim \frac{1}{\alpha^4} \qquad \text{vs.} \qquad \raisebox{-20pt}{\includegraphics[scale=.3]{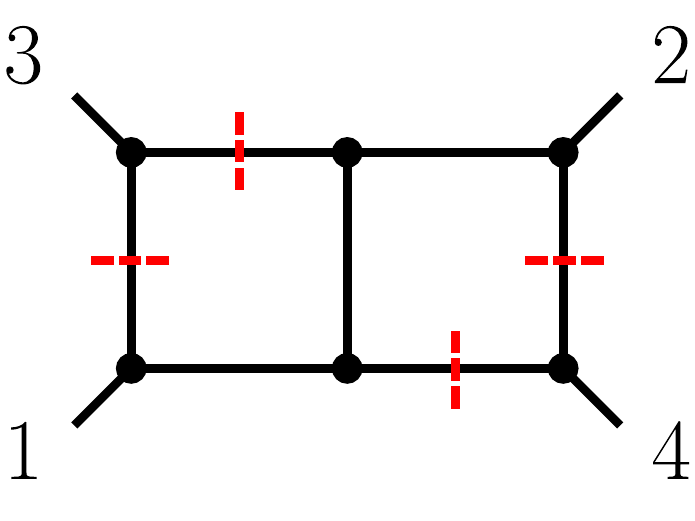}} \sim \frac{1}{\alpha^3}\,,
\end{align}
\vskip -.5cm
which requires cancelations of $1/\alpha^3$ pieces in the integral sum (\ref{eq:two_loop_four_point}). Note that this improved behavior is tightly connected to the presence of the $\beta_1^2\beta_2^2$ factor in (\ref{daisycut2}) originating from the collinear conditions (\ref{IRcol}). If there were higher powers of $\beta_1$ and $\beta_2$ no cancelations in UV would occur. On the other hand, if we had lower powers of $\beta_1$ and $\beta_2$ the collinear behavior would be violated. As a result, the combination of IR and UV conditions is extremely constraining for possible numerators in the Feynman integrals, fixing the the $\beta_i$ dependence on the cut completely.

In the four-loop case we parameterize the on-shell kinematics in complete analogy to the above example, and in the same limit the cut functions scales like $1/{\alpha^3}$. In contrast, one of the best behaved diagrams is the four-loop ladder which scales like $1/\alpha^5$. On the other end of the spectrum there are diagrams with loop-momentum containing numerators. Looking at the BCJ representation for diagram $(37)$ of \cite{Bern:2010tq}, the Yang-Mills BCJ numerator scales like $\sim \alpha^2$, so that the gravity numerator is proportional to $\alpha^4$ for large $\alpha$. Each of the five uncut propagators is linear in $\alpha$ and therefore the total scaling of the integral on the cut is proportional to $1/{\alpha}$.

\begin{align}
\label{figeq:MHV_bubble_cut_4}
\raisebox{-55pt}{\includegraphics[scale=.55]{./figures/bubble_cut_4_loop} } \hskip -.2cm \sim \frac{1}{\alpha^3},\quad
\raisebox{-50pt}{
\includegraphics[scale=.33]{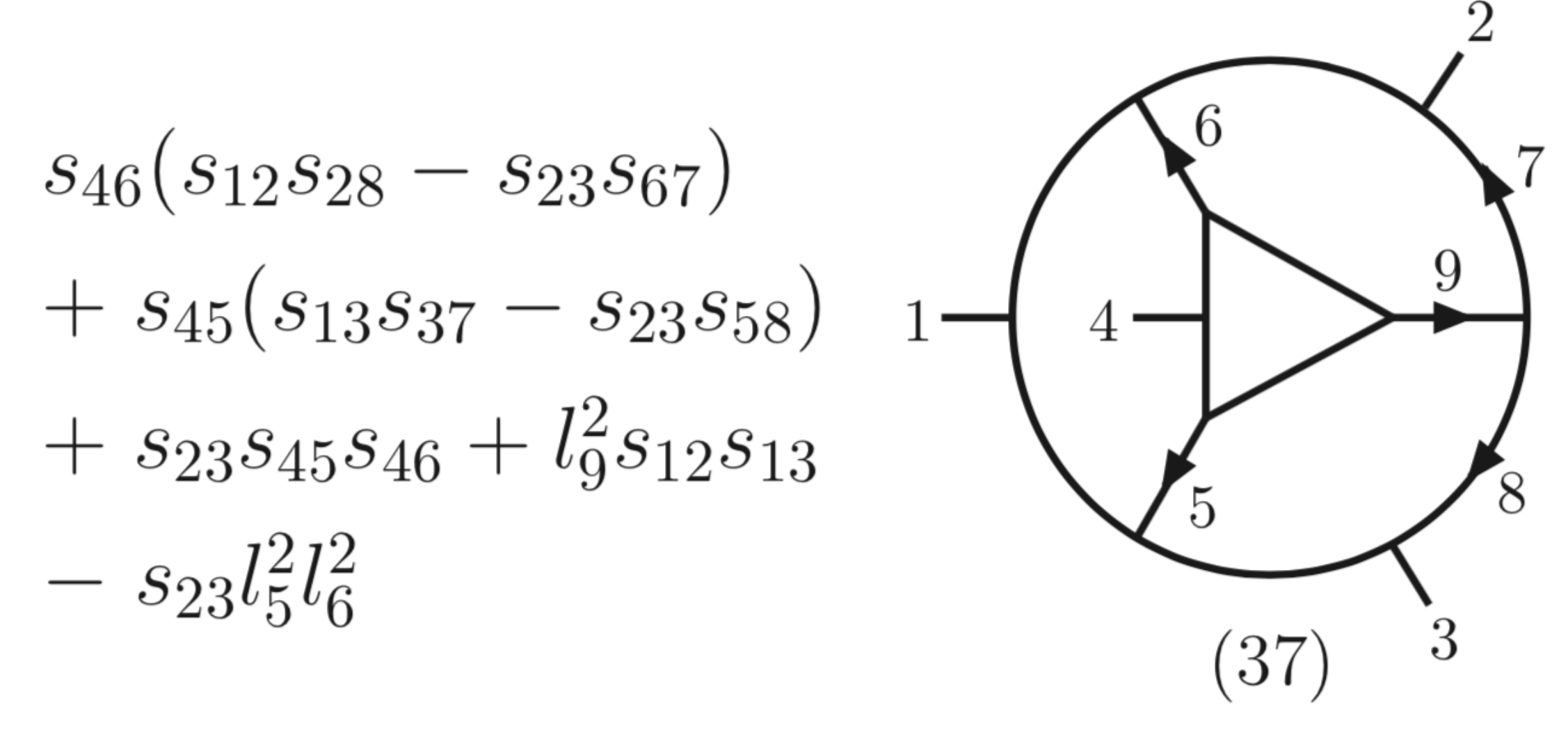}} \hskip -.1cm \sim \frac{1}{\alpha}
\end{align}

which requires cancelations of $1/\alpha$ and $1/\alpha^2$ terms between integrals in \cite{Bern:2010tq}. Again, turning things around, if we had not known the Feynman integral numerators, the combination of the improved UV behavior together with the IR collinear condition would impose very strong and constrains on a tentative ansatz.

\subsection{Uniqueness of gravity integrands}

As alluded to in the beginning of subsec.~\ref{subsec:daisy_cut}, from a specific perspective, the $\N=4$ sYM amplitude is rigidly defined by certain analytic properties. In the planar sector there exists a complete geometric formulation of the theory that predicts all of these analytic properties. In this formulation the loop integrand corresponds to a unique differential form with logarithmic singularities on the boundaries of the Amplituhedron \cite{Arkani-Hamed:2013jha}. Beyond planar $\N=4$ sYM theory, the equivalent analytic statements were translated to the language of Feynman integrals which allowed for the conjecture of an Amplituhedron-like object for the full non-planar $\N=4$ sYM amplitude \cite{Arkani-Hamed:2014via,Bern:2014kca,Bern:2015ple}. In the nonplanar case we can construct the basis of integrals subject to the following analytic constraints:
\begin{align}
\label{eq:YM_properties}
\mbox{(i) IR: logarithmic singularities:}\,\,\frac{dx}{x},\qquad \mbox{(ii) UV: no poles at infinity:}\,\, \ell\rightarrow\infty
\end{align}

These two conditions (\ref{eq:YM_properties}) are imposed term-by-term in the construction of the nonplanar integral basis. Combined, these constraints are very powerful and the basis of integrals satisfying both of them is relatively small. In order to completely fix all the coefficients in the expansion and uniquely specify the loop amplitude we need to impose only {\it homogeneous conditions}, i.e. specify on which cuts the amplitude should vanish rather than matching the non-trivial cut functions on the physical cuts. This all points towards the existence of a geometric picture for ${\cal N}=4$ sYM amplitudes beyond the planar limit \cite{Bern:2015ple} but the explicit construction requires the resolution of the nonplanar labeling problem. The special set of integrals found in \cite{Arkani-Hamed:2014via,Bern:2014kca,Bern:2015ple} were also used in attempts to generalize dual conformal symmetry beyond the planar limit \cite{Bern:2017gdk,Bern:2018oao,Chicherin:2018wes}. 

It is natural to explore the same uniqueness question for ${\cal N}=8$ supergravity amplitudes. We know that none of the Yang-Mills properties, (\ref{eq:YM_properties}), are valid as we have higher poles at infinity on maximal cuts. As argued in \cite{Bern:2014kca} if we only look at cuts in the IR (not poles at infinity) the singularities are still logarithmic as a consequence of the BCJ relations. Furthermore, there are strong collinear conditions (\ref{IRcol}) constraining the IR region. We showed in this paper that the ${\cal N}=8$ integrands have very non-trivial behavior in the UV region as well. We demonstrated these UV features on particular descendants of the multi-unitarity cut but the more general statement would be an improved behavior directly on the multi-unitarity cut (\ref{fig:multi_unitarity_cut}) in all possible channels.

The natural conjecture is that the combination of IR and UV conditions plus the homogeneous conditions on unphysical cuts uniquely selects the $\N=8$ supergravity amplitude. While the general philosophy is similar to the ${\cal N}=4$ sYM case --the amplitude is heavily constrained by the opposing IR and UV conditions which makes it unique-- the details in the expansion in terms of integrals is very different. In the ${\cal N}=4$ sYM case we were able to satisfy both IR and UV conditions term-by-term in individual integrals, and the homogeneous conditions tied them together leaving only one overall constant unfixed. We already saw before that in ${\cal N}=8$ supergravity neither IR nor UV conditions can be made manifest term-by-term in the integral expansion, and the non-trivial behavior of the amplitude requires cancelations between integrals. Therefore, there is no preferred integral basis for the ${\cal N}=8$ amplitude from the analytic point of view, the only constraint is power counting. All IR, UV and homogeneous constraints then lead to relations between the different basis coefficients and the conjecture is that the constraints uniquely fix them up to an overall factor. 

We tested this conjecture successfully on four point one loop and two loop amplitudes but this is obviously insufficient. More extensive checks are in progress, and we hope to resolve this problem including the precise definition of the UV constraints for all amplitudes in upcoming work \cite{wip_coll_const:2018}. 

\section{Conclusion}
\label{sec:conclusions}

In this paper we presented a series of observations that gravity loop integrands show surprising features in the UV region where loop momenta are large. In particular cases, we showed that certain poles at infinity are absent for gravity on-shell functions which requires massive cancelations between Feynman integrals contributing to the amplitude. These cancelations do not follow from any known symmetries or properties of gravity amplitudes. We spelled out the calculations in $\N=8$ supergravity but all statements are valid also in pure gravity, just the degree of the poles at infinity are different. In the last section of the paper we showed that these UV properties together with previously observed IR properties in the collinear regions heavily constrain the cut integrands and can possibly play the same role as the logarithmic singularities and absence of poles at infinity in $\N=4$ sYM theory. This all suggests that there might be some hidden mechanism or symmetry in $\N=8$ supergravity responsible for our observations.

In the future, we would like to improve our understand of this cancelation mechanism, it is likely to be related to some hidden properties of tree-level amplitudes. It would also be interesting to explore the role of BCJ relations and the connection of poles at infinity in on-shell functions to UV divergencies in this context. The overarching goal in this line of research is to attempt to formulate gravity integrands geometrically which ultimately requires to deal with the labeling problem and the definition of non-planar integrands. Before attacking these more fundamental questions, the first step in progress \cite{wip_coll_const:2018} is to test the conjecture that UV and IR constraints together with the absence of unphysical singularities indeed fix $\N=8$ amplitudes uniquely.

\section*{Acknowledgements} We would like to thank Zvi Bern, Lance Dixon, Alex Edison, Julio Parra-Martinez and J.~J. Stankowicz for useful discussions. The research of J.T. is supported in part by U.S. Department of Energy grant DE-SC0009999 and by the funds of University of California. E.H. was supported by the U.S. Department of Energy under contract DE-AC02-76SF00515. E.H. is grateful to the Mani L. Bhaumik Institute for Theoretical Physics at UCLA and QMAP at UC Davis for their kind hospitality.

\bibliographystyle{JHEP}
\phantomsection         
            \bibliography{References.bib}

\providecommand{\href}[2]{#2}\begingroup\raggedright\begin{thebibliography}{10}

\bibitem{tHooft:1974toh}
G.~'t~Hooft and M.~J.~G. Veltman, \emph{{One loop divergencies in the theory of
  gravitation}}, {\emph{Ann. Inst. H. Poincare Phys. Theor.} {\bfseries A20}
  (1974) 69}.

\bibitem{Goroff:1985sz}
M.~H. Goroff and A.~Sagnotti, \emph{{Quantum Gravity At Two Loops}},
  \href{https://doi.org/10.1016/0370-2693(85)91470-4}{\emph{Phys. Lett.}
  {\bfseries 160B} (1985) 81}.

\bibitem{Goroff:1985th}
M.~H. Goroff and A.~Sagnotti, \emph{{The Ultraviolet Behavior of Einstein
  Gravity}}, \href{https://doi.org/10.1016/0550-3213(86)90193-8}{\emph{Nucl.
  Phys.} {\bfseries B266} (1986) 709}.

\bibitem{vandeVen:1991gw}
A.~E.~M. van~de Ven, \emph{{Two loop quantum gravity}},
  \href{https://doi.org/10.1016/0550-3213(92)90011-Y}{\emph{Nucl. Phys.}
  {\bfseries B378} (1992) 309}.

\bibitem{Bern:2015xsa}
Z.~Bern, C.~Cheung, H.-H. Chi, S.~Davies, L.~Dixon and J.~Nohle,
  \emph{{Evanescent Effects Can Alter Ultraviolet Divergences in Quantum
  Gravity without Physical Consequences}},
  \href{https://doi.org/10.1103/PhysRevLett.115.211301}{\emph{Phys. Rev. Lett.}
  {\bfseries 115} (2015) 211301}
  [\href{https://arxiv.org/abs/1507.06118}{{\ttfamily 1507.06118}}].

\bibitem{Bjornsson:2010wm}
J.~Bjornsson and M.~B. Green, \emph{{5 loops in 24/5 dimensions}},
  \href{https://doi.org/10.1007/JHEP08(2010)132}{\emph{JHEP} {\bfseries 08}
  (2010) 132} [\href{https://arxiv.org/abs/1004.2692}{{\ttfamily 1004.2692}}].

\bibitem{Kallosh:2008rr}
R.~Kallosh and T.~Kugo, \emph{{The Footprint of E(7(7)) amplitudes of N=8
  supergravity}},
  \href{https://doi.org/10.1088/1126-6708/2009/01/072}{\emph{JHEP} {\bfseries
  01} (2009) 072} [\href{https://arxiv.org/abs/0811.3414}{{\ttfamily
  0811.3414}}].

\bibitem{Beisert:2010jx}
N.~Beisert, H.~Elvang, D.~Z. Freedman, M.~Kiermaier, A.~Morales and
  S.~Stieberger, \emph{{E7(7) constraints on counterterms in N=8
  supergravity}},
  \href{https://doi.org/10.1016/j.physletb.2010.09.069}{\emph{Phys. Lett.}
  {\bfseries B694} (2011) 265}
  [\href{https://arxiv.org/abs/1009.1643}{{\ttfamily 1009.1643}}].

\bibitem{Bern:2018jmv}
Z.~Bern, J.~J. Carrasco, W.-M. Chen, A.~Edison, H.~Johansson, J.~Parra-Martinez
  et~al., \emph{{Ultraviolet Properties of $N = 8$ Supergravity at Five
  Loops}},  \href{https://arxiv.org/abs/1804.09311}{{\ttfamily 1804.09311}}.

\bibitem{Bern:2007xj}
Z.~Bern, J.~J. Carrasco, D.~Forde, H.~Ita and H.~Johansson, \emph{{Unexpected
  Cancellations in Gravity Theories}},
  \href{https://doi.org/10.1103/PhysRevD.77.025010}{\emph{Phys. Rev.}
  {\bfseries D77} (2008) 025010}
  [\href{https://arxiv.org/abs/0707.1035}{{\ttfamily 0707.1035}}].

\bibitem{Bern:2012cd}
Z.~Bern, S.~Davies, T.~Dennen and Y.-t. Huang, \emph{{Absence of Three-Loop
  Four-Point Divergences in N=4 Supergravity}},
  \href{https://doi.org/10.1103/PhysRevLett.108.201301}{\emph{Phys. Rev. Lett.}
  {\bfseries 108} (2012) 201301}
  [\href{https://arxiv.org/abs/1202.3423}{{\ttfamily 1202.3423}}].

\bibitem{Bern:2017lpv}
Z.~Bern, M.~Enciso, J.~Parra-Martinez and M.~Zeng, \emph{{Manifesting enhanced
  cancellations in supergravity: integrands versus integrals}},
  \href{https://doi.org/10.1007/JHEP05(2017)137}{\emph{JHEP} {\bfseries 05}
  (2017) 137} [\href{https://arxiv.org/abs/1703.08927}{{\ttfamily
  1703.08927}}].

\bibitem{Arkani-Hamed:2013jha}
N.~Arkani-Hamed and J.~Trnka, \emph{{The Amplituhedron}},
  \href{https://doi.org/10.1007/JHEP10(2014)030}{\emph{JHEP} {\bfseries 10}
  (2014) 030} [\href{https://arxiv.org/abs/1312.2007}{{\ttfamily 1312.2007}}].

\bibitem{Arkani-Hamed:2013kca}
N.~Arkani-Hamed and J.~Trnka, \emph{{Into the Amplituhedron}},
  \href{https://doi.org/10.1007/JHEP12(2014)182}{\emph{JHEP} {\bfseries 12}
  (2014) 182} [\href{https://arxiv.org/abs/1312.7878}{{\ttfamily 1312.7878}}].

\bibitem{Arkani-Hamed:2017mur}
N.~Arkani-Hamed, Y.~Bai, S.~He and G.~Yan, \emph{{Scattering Forms and the
  Positive Geometry of Kinematics, Color and the Worldsheet}},
  \href{https://doi.org/10.1007/JHEP05(2018)096}{\emph{JHEP} {\bfseries 05}
  (2018) 096} [\href{https://arxiv.org/abs/1711.09102}{{\ttfamily
  1711.09102}}].

\bibitem{delaCruz:2017zqr}
L.~de~la Cruz, A.~Kniss and S.~Weinzierl, \emph{{Properties of scattering forms
  and their relation to associahedra}},
  \href{https://doi.org/10.1007/JHEP03(2018)064}{\emph{JHEP} {\bfseries 03}
  (2018) 064} [\href{https://arxiv.org/abs/1711.07942}{{\ttfamily
  1711.07942}}].

\bibitem{Frost:2018djd}
H.~Frost, \emph{{Biadjoint scalar tree amplitudes and intersecting dual
  associahedra}},  \href{https://arxiv.org/abs/1802.03384}{{\ttfamily
  1802.03384}}.

\bibitem{He:2018pue}
S.~He, G.~Yan, C.~Zhang and Y.~Zhang, \emph{{Scattering Forms, Worldsheet Forms
  and Amplitudes from Subspaces}},
  \href{https://arxiv.org/abs/1803.11302}{{\ttfamily 1803.11302}}.

\bibitem{Arkani-Hamed:2015bza}
N.~Arkani-Hamed and J.~Maldacena, \emph{{Cosmological Collider Physics}},
  \href{https://arxiv.org/abs/1503.08043}{{\ttfamily 1503.08043}}.

\bibitem{Arkani-Hamed:2017fdk}
N.~Arkani-Hamed, P.~Benincasa and A.~Postnikov, \emph{{Cosmological Polytopes
  and the Wavefunction of the Universe}},
  \href{https://arxiv.org/abs/1709.02813}{{\ttfamily 1709.02813}}.

\bibitem{Arkani-Hamed:2017vfh}
N.~Arkani-Hamed, H.~Thomas and J.~Trnka, \emph{{Unwinding the Amplituhedron in
  Binary}}, \href{https://doi.org/10.1007/JHEP01(2018)016}{\emph{JHEP}
  {\bfseries 01} (2018) 016}
  [\href{https://arxiv.org/abs/1704.05069}{{\ttfamily 1704.05069}}].

\bibitem{Arkani-Hamed:2014via}
N.~Arkani-Hamed, J.~L. Bourjaily, F.~Cachazo and J.~Trnka, \emph{{Singularity
  Structure of Maximally Supersymmetric Scattering Amplitudes}},
  \href{https://doi.org/10.1103/PhysRevLett.113.261603}{\emph{Phys. Rev. Lett.}
  {\bfseries 113} (2014) 261603}
  [\href{https://arxiv.org/abs/1410.0354}{{\ttfamily 1410.0354}}].

\bibitem{Bern:2014kca}
Z.~Bern, E.~Herrmann, S.~Litsey, J.~Stankowicz and J.~Trnka, \emph{{Logarithmic
  Singularities and Maximally Supersymmetric Amplitudes}},
  \href{https://doi.org/10.1007/JHEP06(2015)202}{\emph{JHEP} {\bfseries 06}
  (2015) 202} [\href{https://arxiv.org/abs/1412.8584}{{\ttfamily 1412.8584}}].

\bibitem{Bern:2015ple}
Z.~Bern, E.~Herrmann, S.~Litsey, J.~Stankowicz and J.~Trnka, \emph{{Evidence
  for a Nonplanar Amplituhedron}},
  \href{https://doi.org/10.1007/JHEP06(2016)098}{\emph{JHEP} {\bfseries 06}
  (2016) 098} [\href{https://arxiv.org/abs/1512.08591}{{\ttfamily
  1512.08591}}].

\bibitem{ArkaniHamed:2012nw}
N.~Arkani-Hamed, J.~L. Bourjaily, F.~Cachazo, A.~B. Goncharov, A.~Postnikov and
  J.~Trnka, \emph{{Grassmannian Geometry of Scattering Amplitudes}}. Cambridge
  University Press, 2016, [\href{https://arxiv.org/abs/1212.5605}{{\ttfamily
  1212.5605}}].

\bibitem{Herrmann:2016qea}
E.~Herrmann and J.~Trnka, \emph{{Gravity On-shell Diagrams}},
  \href{https://doi.org/10.1007/JHEP11(2016)136}{\emph{JHEP} {\bfseries 11}
  (2016) 136} [\href{https://arxiv.org/abs/1604.03479}{{\ttfamily
  1604.03479}}].

\bibitem{Heslop:2016plj}
P.~Heslop and A.~E. Lipstein, \emph{{On-shell diagrams for $ \mathcal{N} $ = 8
  supergravity amplitudes}},
  \href{https://doi.org/10.1007/JHEP06(2016)069}{\emph{JHEP} {\bfseries 06}
  (2016) 069} [\href{https://arxiv.org/abs/1604.03046}{{\ttfamily
  1604.03046}}].

\bibitem{ArkaniHamed:2010kv}
N.~Arkani-Hamed, J.~L. Bourjaily, F.~Cachazo, S.~Caron-Huot and J.~Trnka,
  \emph{{The All-Loop Integrand For Scattering Amplitudes in Planar $N=4$
  SYM}}, \href{https://doi.org/10.1007/JHEP01(2011)041}{\emph{JHEP} {\bfseries
  1101} (2011) 041} [\href{https://arxiv.org/abs/1008.2958}{{\ttfamily
  1008.2958}}].

\bibitem{Green:1982sw}
M.~B. Green, J.~H. Schwarz and L.~Brink, \emph{{N=4 Yang-Mills and N=8
  Supergravity as Limits of String Theories}},
  \href{https://doi.org/10.1016/0550-3213(82)90336-4}{\emph{Nucl. Phys.}
  {\bfseries B198} (1982) 474}.

\bibitem{Baadsgaard:2015twa}
C.~Baadsgaard, N.~E.~J. Bjerrum-Bohr, J.~L. Bourjaily, S.~Caron-Huot, P.~H.
  Damgaard and B.~Feng, \emph{{New Representations of the Perturbative
  S-Matrix}}, \href{https://doi.org/10.1103/PhysRevLett.116.061601}{\emph{Phys.
  Rev. Lett.} {\bfseries 116} (2016) 061601}
  [\href{https://arxiv.org/abs/1509.02169}{{\ttfamily 1509.02169}}].

\bibitem{Geyer:2015jch}
Y.~Geyer, L.~Mason, R.~Monteiro and P.~Tourkine, \emph{{One-loop amplitudes on
  the Riemann sphere}},
  \href{https://doi.org/10.1007/JHEP03(2016)114}{\emph{JHEP} {\bfseries 03}
  (2016) 114} [\href{https://arxiv.org/abs/1511.06315}{{\ttfamily
  1511.06315}}].

\bibitem{Geyer:2016wjx}
Y.~Geyer, L.~Mason, R.~Monteiro and P.~Tourkine, \emph{{Two-Loop Scattering
  Amplitudes from the Riemann Sphere}},
  \href{https://doi.org/10.1103/PhysRevD.94.125029}{\emph{Phys. Rev.}
  {\bfseries D94} (2016) 125029}
  [\href{https://arxiv.org/abs/1607.08887}{{\ttfamily 1607.08887}}].

\bibitem{Cachazo:2008vp}
F.~Cachazo, \emph{{Sharpening The Leading Singularity}},
  \href{https://arxiv.org/abs/0803.1988}{{\ttfamily 0803.1988}}.

\bibitem{ArkaniHamed:2010gh}
N.~Arkani-Hamed, J.~L. Bourjaily, F.~Cachazo and J.~Trnka, \emph{{Local
  Integrals for Planar Scattering Amplitudes}},
  \href{https://doi.org/10.1007/JHEP06(2012)125}{\emph{JHEP} {\bfseries 06}
  (2012) 125} [\href{https://arxiv.org/abs/1012.6032}{{\ttfamily 1012.6032}}].

\bibitem{Bourjaily:2015jna}
J.~L. Bourjaily and J.~Trnka, \emph{{Local Integrand Representations of All
  Two-Loop Amplitudes in Planar SYM}},
  \href{https://doi.org/10.1007/JHEP08(2015)119}{\emph{JHEP} {\bfseries 08}
  (2015) 119} [\href{https://arxiv.org/abs/1505.05886}{{\ttfamily
  1505.05886}}].

\bibitem{Bourjaily:2017wjl}
J.~L. Bourjaily, E.~Herrmann and J.~Trnka, \emph{{Prescriptive Unitarity}},
  \href{https://doi.org/10.1007/JHEP06(2017)059}{\emph{JHEP} {\bfseries 06}
  (2017) 059} [\href{https://arxiv.org/abs/1704.05460}{{\ttfamily
  1704.05460}}].

\bibitem{Bern:2007ct}
Z.~Bern, J.~J.~M. Carrasco, H.~Johansson and D.~A. Kosower, \emph{{Maximally
  supersymmetric planar Yang-Mills amplitudes at five loops}},
  \href{https://doi.org/10.1103/PhysRevD.76.125020}{\emph{Phys. Rev.}
  {\bfseries D76} (2007) 125020}
  [\href{https://arxiv.org/abs/0705.1864}{{\ttfamily 0705.1864}}].

\bibitem{Benincasa:2015zna}
P.~Benincasa, \emph{{On-shell diagrammatics and the perturbative structure of
  planar gauge theories}},  \href{https://arxiv.org/abs/1510.03642}{{\ttfamily
  1510.03642}}.

\bibitem{ArkaniHamed:2008gz}
N.~Arkani-Hamed, F.~Cachazo and J.~Kaplan, \emph{{What is the Simplest Quantum
  Field Theory?}}, \href{https://doi.org/10.1007/JHEP09(2010)016}{\emph{JHEP}
  {\bfseries 09} (2010) 016} [\href{https://arxiv.org/abs/0808.1446}{{\ttfamily
  0808.1446}}].

\bibitem{Bern:1998xc}
Z.~Bern, L.~J. Dixon, M.~Perelstein and J.~S. Rozowsky, \emph{{One loop n point
  helicity amplitudes in (selfdual) gravity}},
  \href{https://doi.org/10.1016/S0370-2693(98)01397-5}{\emph{Phys. Lett.}
  {\bfseries B444} (1998) 273}
  [\href{https://arxiv.org/abs/hep-th/9809160}{{\ttfamily hep-th/9809160}}].

\bibitem{Bern:2010ue}
Z.~Bern, J.~J.~M. Carrasco and H.~Johansson, \emph{{Perturbative Quantum
  Gravity as a Double Copy of Gauge Theory}},
  \href{https://doi.org/10.1103/PhysRevLett.105.061602}{\emph{Phys. Rev. Lett.}
  {\bfseries 105} (2010) 061602}
  [\href{https://arxiv.org/abs/1004.0476}{{\ttfamily 1004.0476}}].

\bibitem{Bern:1998ug}
Z.~Bern, L.~J. Dixon, D.~C. Dunbar, M.~Perelstein and J.~S. Rozowsky, \emph{{On
  the relationship between Yang-Mills theory and gravity and its implication
  for ultraviolet divergences}},
  \href{https://doi.org/10.1016/S0550-3213(98)00420-9}{\emph{Nucl. Phys.}
  {\bfseries B530} (1998) 401}
  [\href{https://arxiv.org/abs/hep-th/9802162}{{\ttfamily hep-th/9802162}}].

\bibitem{Bern:2006kd}
Z.~Bern, L.~J. Dixon and R.~Roiban, \emph{{Is N = 8 supergravity ultraviolet
  finite?}}, \href{https://doi.org/10.1016/j.physletb.2006.11.030}{\emph{Phys.
  Lett.} {\bfseries B644} (2007) 265}
  [\href{https://arxiv.org/abs/hep-th/0611086}{{\ttfamily hep-th/0611086}}].

\bibitem{Bern:2010tq}
Z.~Bern, J.~J.~M. Carrasco, L.~J. Dixon, H.~Johansson and R.~Roiban, \emph{{The
  Complete Four-Loop Four-Point Amplitude in N=4 Super-Yang-Mills Theory}},
  \href{https://doi.org/10.1103/PhysRevD.82.125040}{\emph{Phys. Rev.}
  {\bfseries D82} (2010) 125040}
  [\href{https://arxiv.org/abs/1008.3327}{{\ttfamily 1008.3327}}].

\bibitem{Grozin:2011mt}
A.~G. Grozin, \emph{{Integration by parts: An Introduction}},
  \href{https://doi.org/10.1142/S0217751X11053687}{\emph{Int. J. Mod. Phys.}
  {\bfseries A26} (2011) 2807}
  [\href{https://arxiv.org/abs/1104.3993}{{\ttfamily 1104.3993}}].

\bibitem{Carrasco:2011mn}
J.~J. Carrasco and H.~Johansson, \emph{{Five-Point Amplitudes in N=4
  Super-Yang-Mills Theory and N=8 Supergravity}},
  \href{https://doi.org/10.1103/PhysRevD.85.025006}{\emph{Phys. Rev.}
  {\bfseries D85} (2012) 025006}
  [\href{https://arxiv.org/abs/1106.4711}{{\ttfamily 1106.4711}}].

\bibitem{Bern:1995ix}
Z.~Bern and G.~Chalmers, \emph{{Factorization in one loop gauge theory}},
  \href{https://doi.org/10.1016/0550-3213(95)00226-I}{\emph{Nucl. Phys.}
  {\bfseries B447} (1995) 465}
  [\href{https://arxiv.org/abs/hep-ph/9503236}{{\ttfamily hep-ph/9503236}}].

\bibitem{Bern:2017gdk}
Z.~Bern, M.~Enciso, H.~Ita and M.~Zeng, \emph{{Dual Conformal Symmetry,
  Integration-by-Parts Reduction, Differential Equations and the Nonplanar
  Sector}}, \href{https://doi.org/10.1103/PhysRevD.96.096017}{\emph{Phys. Rev.}
  {\bfseries D96} (2017) 096017}
  [\href{https://arxiv.org/abs/1709.06055}{{\ttfamily 1709.06055}}].

\bibitem{Bern:2018oao}
Z.~Bern, M.~Enciso, C.-H. Shen and M.~Zeng, \emph{{Dual Conformal Structure
  Beyond the Planar Limit}},
  \href{https://arxiv.org/abs/1806.06509}{{\ttfamily 1806.06509}}.

\bibitem{Chicherin:2018wes}
D.~Chicherin, J.~M. Henn and E.~Sokatchev, \emph{{Implications of nonplanar
  dual conformal symmetry}},
  \href{https://arxiv.org/abs/1807.06321}{{\ttfamily 1807.06321}}.

\bibitem{wip_coll_const:2018}
A.~Edison, E.~Herrmann, J.~Parra-Martinez and J.~Trnka, \emph{{Uniqueness of
  gravity integrands}}, {\emph{work in progress} (2018) }.

\end{thebibliography}\endgroup
            \clearpage

\end{document}